\documentclass[numberedappendix,twocolappendix]{emulateapj}

\pdfoutput=1

\usepackage{apjfonts} 
\usepackage{amsmath} 
\usepackage{color} 
\usepackage[colorlinks=true,linkcolor=blue,urlcolor=blue,citecolor=blue]{hyperref}

\shorttitle{Stellar Evolution and the mass function}
\shortauthors{F.R.N.~Schneider et al.}

\newcommand{\msun}{{\rm M}_{\odot}}
\newcommand{\lsun}{{\rm L}_{\odot}}
\newcommand{\rsun}{{\rm R}_{\odot}}

\bibpunct{(}{)}{;}{a}{}{,}

\begin{document}

\title{Evolution of mass functions of coeval stars through wind mass loss and binary interactions}

\email{fabian.schneider@physics.ox.ac.uk}
\author{
	F.R.N.~Schneider\altaffilmark{1,2},
	R.G.~Izzard\altaffilmark{1,3},
	N.~Langer\altaffilmark{1} and
	S.E.~de~Mink\altaffilmark{4}
}
\altaffiltext{1}{Argelander-Institut f{\"u}r Astronomie der Universit{\"a}t Bonn, Auf dem H{\"u}gel 71, 53121 Bonn, Germany}
\altaffiltext{2}{Department of Physics, University of Oxford, Denys Wilkinson Building, Keble Road, Oxford OX1 3RH, UK}
\altaffiltext{3}{Institute of Astronomy, University of Cambridge, Madingley Road, Cambridge CB3 0HA, UK}
\altaffiltext{4}{Astronomical Institute 'Anton Pannekoek', Amsterdam University, Science Park 904, 1098 XH, Amsterdam, The Netherlands}

\begin{abstract}
Accurate determinations of stellar mass functions and ages of stellar
populations are crucial to much of astrophysics. We analyse the evolution
of stellar mass functions of coeval main sequence stars including
all relevant aspects of single- and binary-star evolution. We show
that the slope of the upper part of the mass function in a stellar
cluster can be quite different to the slope of the initial mass function.
Wind mass loss from massive stars leads to an accumulation of stars which is visible as a 
peak at the high mass end of mass functions, thereby flattening the mass function slope.
Mass accretion and mergers in close
binary systems create a tail of rejuvenated binary products. These
blue straggler stars extend the single star mass function by up to
a factor of two in mass and can appear up to ten times 
younger than their parent stellar cluster. Cluster
ages derived from their most massive stars that are close to the turn-off
may thus be significantly biased. 
To overcome such difficulties, we
propose the use of the binary tail of stellar mass functions
as an unambiguous clock to derive the cluster age because the location of
the onset of the binary tail identifies the cluster turn-off mass.
It is indicated by a pronounced jump in the mass function of old stellar
populations and by the wind mass loss peak in young stellar populations.
We further characterise the binary induced blue straggler population
in star clusters in terms of their frequency, binary fraction and
apparent age.
\end{abstract}


\section{Introduction}\label{sec:introduction}

Stellar mass functions are
important for population studies, both nearby and at high redshift.
Mass functions are not static but change their shape when the 
frequencies of stars or their masses are altered over time \citep{1986FCPh...11....1S,2013pss5.book..115K}.
Possible causes include the evaporation of stars and mass segregation
in star clusters \citep{2002MNRAS.331..245D,2008ApJ...679.1272M,2010MNRAS.409..628H,2013A&A...556A..26H} 
and mass loss in the course of stellar evolution \citep{2012ARA&A..50..107L}.
These mechanisms leave characteristic fingerprints in
mass functions from which insights into the evolutionary status of
stellar populations and the mechanisms themselves can be gained.

Massive stars are subject to strong stellar wind mass loss, which decreases their masses
already on the main sequence (MS), directly affecting mass functions. Furthermore,
it emerges that binary stars play an important role in stellar populations
of various ages and even dominate the evolution of massive stars \citep{2012Sci...337..444S}.
The multiplicity fraction, i.e.\ the number of multiple stars divided
by the total number of stellar systems, is larger for higher
masses \citep[e.g.][]{2010ARA&A..48..339B,2013ARA&A..51..269D}: it
exceeds 40\% for solar like stars (F, G and K stars) and 70\%
for the most massive stars (O-stars). In close binaries, mass is exchanged
between the binary components during Roche lobe overflow (RLOF) or
in stellar mergers, directly affecting stellar masses and hence the
mass function. The mass gainers are rejuvenated and can appear much younger than they really are.
Some mass gainers may also be visible as blue straggler stars
\citep{1995A&A...297..483B,2007MNRAS.376...61D,2011Natur.478..356G}.

The determination of stellar ages is a fundamental task in stellar astrophysics \citep{2010ARA&A..48..581S} 
that can be biased by rejuvenated binary products.
Various methods are used to determine stellar ages.
The surface properties of individual stars can be compared
with evolutionary model predictions, e.g., luminosity, surface temperature or surface gravity
\citep{2007A&A...475..519H,2014A&A...570A..66S}, rotation
rates in low mass stars \citep{2007ApJ...669.1167B} or surface nitrogen abundances
in massive stars \citep{2012A&A...544A..76K}. In star clusters, the most widely used
age determination method compares the main sequence turn-off with theoretical isochrones
\citep[e.g.][]{2006MNRAS.373.1251N,2010A&A...516A...2M}. Close binary evolution such as
mass transfer and stellar mergers leads to spurious or inaccurate results in all of 
these methods. To derive unambiguous age estimates one must be able to distinguish between 
rejuvenated binary products and genuinely single stars.

In this paper we investigate how the modulation of stellar mass functions 
by single and binary star evolution can be used both to identify binary products 
and to derive unambiguous stellar ages. 
More generally, we explore what can be learned about stellar evolution and 
the evolutionary status of whole stellar populations from observed mass functions. 
We also investigate quantitatively how single
and binary star evolution influence the determination of initial 
mass functions. To that end, we perform detailed 
population synthesis calculations of coeval stellar 
populations using a rapid binary evolution code.

We describe our method in Sec.~\ref{sec:method} and present mass functions
of coeval single and binary star populations of ages ranging from $3\,\mathrm{Myr}$ to
$1\,\mathrm{Gyr}$ in Sec.~\ref{sec:results}. Through binary interactions, blue straggler stars 
are formed in our models. We characterise their binary fraction and ages, and compare 
their predicted frequencies to those found in Galactic open star clusters in 
Sec.~\ref{sec:blue-stragglers}. Blue straggler stars predominantly populate the high mass
end of mass functions and may bias determinations of stellar cluster ages. We show how to
use mass functions to overcome such biases when determining cluster ages 
in Sec.~\ref{sec:determination-cluster-age} and conclude in Sec.~\ref{sec:conclusions}.

\section{Method}\label{sec:method}

We compute the evolution of single and close, interacting binary
stars and construct, at predefined ages, mass functions by
counting how many stars of certain masses exist. This approach ensures
that we factor in all the relevant single and binary star physics
to investigate their influence on the present-day mass function (PDMF).

The initial parameter space of binaries is large compared to that
of single stars, which essentially consists of the initial mass.
In binaries, there are two masses, the orbital separation, the eccentricity of the
orbit and the relative orientation of the spin axis of both stars.
We apply some standard simplifications to reduce this huge parameter space:
we impose circular orbits and that both stellar spins are aligned
with the orbital angular momentum. All our models are calculated for
a metallicity $Z=0.02$ (unless stated otherwise). Furthermore
we focus on the main sequence because stars spend typically
$90\%$ of their life in this evolutionary stage.
Still, we need to follow the evolution of a large number of stellar systems
to sample the remaining binary parameter space and to resolve effects
at the high mass end of the PDMF. Hence, we work with a rapid binary
evolution code.

\subsection{Rapid binary evolution code}\label{sec:bse-code}

We use the binary population and nucleosynthesis code of 
\citet{2004MNRAS.350..407I,2006A&A...460..565I,2009A&A...508.1359I} with modifications due to \citet{2013ApJ...764..166D}
which is based on the rapid binary evolution code of \citet{2002MNRAS.329..897H}.
This code uses analytic formulae \citep{2000MNRAS.315..543H} fitted
to detailed single star evolutionary models \citep{1998MNRAS.298..525P}
to approximate the evolution of single stars for a wide range of masses and metallicities.

The fitting formulae of \citet{2000MNRAS.315..543H} are based on
detailed stellar model sequences of stars with mass up to $50\,\msun$ 
\citep{1998MNRAS.298..525P}. The evolution of stars with mass in excess 
of $50\,\msun$ is thus based on extrapolations of the original fitting formulae. 
The MS lifetime, $\tau_{\mathrm{MS}}$, is particularly inaccurately extrapolated 
by the appropriate fit, so we replace it with a logarithmic tabular interpolation 
of the MS lifetimes taken directly from the models of \citet{1998MNRAS.298..525P} 
in the mass range $20\leq M\leq50\,\msun$. More massive than this we extrapolate 
the final two masses in the grid of detailed models \citep{1998MNRAS.298..525P}.
This results in a reduction of the MS lifetime of e.g. a $100\,\msun$ star at $Z=0.02$ from
$3.5$ to $2.9\,\mathrm{Myr}$, i.e.\ in a reduction of $17\%$,
which is in agreement with state-of-the-art non-rotating detailed stellar 
models of \citet{2011A&A...530A.115B} and \citet{2012A&A...537A.146E}.

Stellar wind mass loss for stars with luminosities
$L>4000\,\lsun$ is given by \citet{1990A&A...231..134N}. This recipe
is modified by a factor $Z^{0.5}$ according to \citet{1989A&A...219..205K}
to mimic the impact of the metallicity $Z$ on wind mass loss rates.

Binary stars can exchange mass either by RLOF,
wind mass transfer or merging.
RLOF occurs when one star (hereafter the donor) fills its Roche lobe
and transfers mass to its companion (hereafter the accretor) through
the inner Lagrangian point. Depending on the physical state of the
donor star one distinguishes between Case A, B and C mass transfer. 
Following the definition of \citet{1967ZA.....65..251K},
Case A mass transfer occurs during core hydrogen burning, Case B after
the end of core hydrogen burning and Case C, defined by \citet{1970A&A.....7..150L},
after the end of core helium burning.

Our binary evolution code differs from \citet{2002MNRAS.329..897H}
in its treatment of RLOF. For stable mass transfer it is expected
that the stellar radius $R$ adjusts itself to the Roche lobe radius
$R_{\mathrm{L}}$, i.e.\ $R\approx R_{\mathrm{L}}$. Whenever RLOF
occurs ($R>R_{\mathrm{L}}$) we remove as much mass as needed to shrink
the donor star back into its Roche lobe.
The resulting mass transfer and mass accretion rates are capped by
the thermal timescales of the donor and accretor, respectively. 

We follow \citet{2002MNRAS.329..897H} to determine the occurrence of common envelope
evolution and contact phases that lead to stellar mergers. MS mergers are expected 
either if the initial orbital separation is so small that both stars 
fill their Roche lobes and thus come into physical contact
or if the mass ratio of the accretor to the donor star falls below
a certain limit at the onset of mass transfer, $q=M_{2}/M_{1}<q_{\mathrm{crit}}$,
which drives the accretor out of thermal equilibrium and hence results in a contact system 
\citep[e.g.][]{1976ApJ...206..509U,1977A&A....54..539K,1977PASJ...29..249N,2001A&A...369..939W}. 
The critical mass ratio is approximately $q_{\mathrm{crit,MS}}=0.56$ for MS stars \citep{2007A&A...467.1181D},
$q_{\mathrm{crit,HG}}=0.25$ if the donor star is a Hertzsprung gap
star and is given by a fitting formula if the donor star has a deep
convective envelope \citep{2002MNRAS.329..897H}.

Mass transfer because of either stable RLOF or during a stellar merger makes the mass gainers
appear younger than they really are \citep{1995A&A...297..483B,1998A&A...334...21V,2007MNRAS.376...61D}.
Such rejuvenated stars may stand out as blue stragglers in Hertzsprung--Russell (HR) diagrams.
Rejuvenation is handled following \citet{1997MNRAS.291..732T} and
\citet{2002MNRAS.329..897H} but with improvements as described in \citet{2008A&A...488.1017G} and
\citet{2013ApJ...764..166D}. The apparent age $T$ of a star is given by 
the amount of burnt fuel compared to the total available. For a 
star with MS lifetime $\tau_\mathrm{MS}$ we therefore have (in a linear approximation)
$T=f_\mathrm{burnt} \tau_\mathrm{MS}$ where $f_\mathrm{burnt}=M_\mathrm{burnt} / 
M_\mathrm{available}$ is the mass ratio of the burnt to totally available fuel.
After mass transfer onto a MS star with a convective core, the mass of the already burnt
fuel is given by the fraction of burnt material, $f_\mathrm{burnt}$, times the 
convective core mass before mass transfer, $M_\mathrm{c}$ (primes 
indicate quantities after mass transfer). After mass transfer,
the convective core and hence the available fuel of the accretor grow in mass, 
i.e.\ $M'_\mathrm{c}>M_\mathrm{c}$, because the total stellar mass increases 
(and vice versa for the donor star). Thus, the fraction of burnt fuel after mass
transfer is $f'_\mathrm{burnt}=f_\mathrm{burnt}M_\mathrm{c}/M'_\mathrm{c}$
and the apparent age, $T'=f'_\mathrm{burnt} \tau_\mathrm{MS}'$, is
\begin{equation}
T'=\frac{M_{\mathrm{\mathrm{c}}}}{M_{\mathrm{c}}'}\frac{\tau_{\mathrm{MS}}'}{\tau_{\mathrm{MS}}}T.\label{eq:rlof-rejuvenation}
\end{equation}
This equation holds for the accretor and also for the donor when 
setting $M_{\mathrm{c}}=M_{\mathrm{c}}'$ (no burnt fuel is mixed out 
of the core upon mass loss from the stellar surface) and shows 
that the accreting stars rejuvenate upon mass transfer ($T'<T$ because $M'_\mathrm{c}\geq M_\mathrm{c}$
and $\tau_\mathrm{MS}'<\tau_\mathrm{MS}$) while the donor stars age
($T'>T$ because $\tau_\mathrm{MS}'>\tau_\mathrm{MS}$).
The accretor and donor will have burnt less (more) 
fuel than a single star of the same mass that did not accrete (lose) mass.
If stars do not have convective cores, e.g., stars with initial masses in the range
$0.3\leq M/\msun\leq1.3$ or Hertzsprung gap stars which have radiative
cores, we set $M_{\mathrm{c}}=M_{\mathrm{c}}'$ because no fresh fuel
is expected to be mixed into their cores. 

To model the rejuvenation of MS mergers we follow \citet{2013ApJ...764..166D}: first, we assume
that a fraction $f_\mathrm{loss}$ of the total mass $M_{3}=M_{1}+M_{2}$ is lost during
the merger; we adopt $f_\mathrm{loss}=0.1$. Second, we approximate the core mass fraction $f_{\mathrm{c}}$
of MS stars according to fitting functions \citep{2008A&A...488.1017G}
and estimate the apparent age $T_{3}$ of the newly formed merged
star from,
\begin{equation}
T_{3}=\tau_{\mathrm{MS,3}}\cdot\frac{f_{\mathrm{c,1}}\cdot\frac{T_{1}}{\tau_{\mathrm{MS,1}}}+f_{\mathrm{c,2}}\cdot\frac{T_{2}}{\tau_{\mathrm{MS,2}}}}{f_{\mathrm{c,3}}^{\mathrm{eff}}},\label{eq:merger-rejuvantion}
\end{equation}
where $T_{1}$ and $T_{2}$ are the effective ages, $f_{\mathrm{c,1}}$
and $f_{\mathrm{c,2}}$ are the core mass fractions of the progenitor
stars and $\tau_{\mathrm{MS}}$ denotes the MS lifetime of the corresponding
star. The denominator contains the effective core mass fraction 
$f_{\mathrm{c,3}}^{\mathrm{eff}}=f_{\mathrm{c,3}}+f_\mathrm{mix}\cdot(1-f_{\mathrm{c,3}})$
of the merger product which is given by its core mass fraction
modified by an additional mixing of $f_\mathrm{mix}=10\%$ of the hydrogen-rich envelope. 
The resulting rejuvenation is less than in the original \citet{2002MNRAS.329..897H} code,
where it is assumed that the whole star is mixed, and closer to that seen
in hydrodynamic and smoothed particle hydrodynamics simulations of stellar mergers 
\citep[e.g.][]{1995ApJ...445L.117L,1997ApJ...487..290S,2001ApJ...548..323S,2008A&A...488.1017G,2013MNRAS.434.3497G}.

\subsection{Initial distribution functions}\label{sec:pop-syn}

We set up a grid of stellar systems to cover the parameter space of single and
binary stars and assign each stellar system $j$ a probability
of existence $\delta p_{j}$. The
probabilities $\delta p_{j}$ are calculated from $\delta p_{j}=\Psi \delta \ln V$ 
where $\Psi$ is a distribution function
of the initial masses and the initial orbital periods and $\delta \ln V$ is the volume 
of the parameter space filled by the stellar system $j$. 
The initial distribution function, $\Psi$, reads, 
\begin{equation}
\Psi=\begin{cases}
\psi(\ln m) & \text{single stars,}\\
\psi(\ln m_{1})\,\phi(\ln m_{2})\,\chi(\ln P) & \text{binary stars,}
\end{cases}
\end{equation}
where $m_{1}$ and $m_{2}$ are the initial masses of the primary
and secondary star in binaries, respectively, and $P$ is the
initial orbital period. The functions $\psi(\ln m_{1})$, $\phi(\ln m_{2})$
and $\chi(\ln P)$ are the IMFs of the primary
and secondary star and the distribution function of the initial orbital
period, respectively. Stellar masses $m$ are given in solar
masses.

Single stars and also the primaries of binary stars, i.e.\ the initially
more massive component, are distributed according to a Salpeter IMF
\citep{1955ApJ...121..161S}. The IMF $\psi(\ln m_{1})$
is then 
\begin{equation}
\psi(\ln m_{1})\equiv\frac{\mathrm{d}p}{\mathrm{d}\ln m_{1}}=m_{1}\frac{\mathrm{d}p}{\mathrm{d}m_{1}}=Am_{1}^{\Gamma}\label{eq:psi}
\end{equation}
with $\Gamma=-1.35$ the slope of the mass function and $A$ the normalization
constant such that 
\begin{equation}
\int_{m_{{\rm l}}}^{m_{{\rm u}}}\,\frac{{\rm d}p}{{\rm d}m_{1}}\,{\rm d}m_{1}=1.
\end{equation}
The lower and upper mass limits are chosen such that we do not exceed
the validity of the fitting functions used in the code, hence $m_{l}=0.1$
and $m_{u}=100$.

\citet{2013ARA&A..51..269D} review stellar multiplicity
(multiplicity fractions, mass ratio distributions and orbital separations
distributions) and its dependence on primary mass and environment.
A complete picture is yet lacking (e.g., for MS primary stars
in the mass range $8$--$16\,\msun$) but it seems that a flat
mass ratio distribution, 
\begin{equation}
\phi(\ln m_{2})=\frac{\mathrm{d}p}{\mathrm{d}\ln m_2}=q\frac{\mathrm{d}p}{\mathrm{d}q}\propto q,
\end{equation}
is reasonable maybe except at the lowest primary masses ($\lesssim1\,\msun$).
We adopt this as the distribution function of the initial mass ratios,
meaning that all mass ratios $q$ are equally probable ($\mathrm{d}p/\mathrm{d}q=\mathrm{const.}$).
We adopt a minimum mass ratio, $q_{\mathrm{min}}=0.1/m_{1}$.

In terms of orbital separations, we are only interested in close binaries,
i.e.\ binaries that can interact by mass exchange at some point during
their life. Our binary systems therefore have initial orbital separations
$a$ between $3\,\rsun$ and $10^{4}\,\rsun$ ($\sim46\,\mathrm{AU}$).
In practice it turns out that only binaries with initial orbital separations
less than about $3\times10^{3}\,\rsun\approx15\,\mathrm{AU}$
interact. Initially wider binaries are effectively single stars. The
distribution function $\chi(\ln P)$ of the initial orbital periods
$P$ is given by \citet{2012Sci...337..444S} for binaries
with O-type companions (i.e.\ $m_{1}\geq15$) and mass ratios $q>0.1$
and by a flat distribution in $\ln P$ for all other binaries, i.e.\ 
$f(P)\,\mathrm{d}P\propto\mathrm{d}P/P$ \citep{1924PTarO..25f...1O},
\begin{equation}
\chi(\ln P)\propto \begin{cases}
\left(\log P\right)^{-0.55}, & 0.15\leq\log P/{\rm d}\leq3.5\\
\text{const.} & \text{otherwise.}
\end{cases}
\end{equation}
The lower boundary of the
initial orbital separations, $a_{l}=3\,\rsun$, is increased if a
star fills its Roche lobe on the zero-age main sequence (ZAMS)
such that stars cannot interact immediately by RLOF. 

For our mass function calculations, we evolve a total of $2,500,000$ 
binary and $250,000$ single stars. The overall mass range of $1$--$100\,\msun$ 
is subdivided in $10$ equally spaced mass intervals. Each of the $10$ mass
intervals covers $2500$ single stars of different initial masses and $100\times 50 \times 50$ 
($m_1 \times q \times a$) binaries. Masses and orbital separations are distributed equidistantly
on a logarithmic grid while the mass ratios are distributed equidistantly on a linear grid.

\subsection{Binary parameter space}\label{sec:binary-parameter-space}

To understand the influence of binary interactions
on the PDMF, we need to know quantitatively how much mass is
transferred and accreted during binary mass exchange. Given an initial
primary mass $m_{1}$, we investigate the binary parameter space spanned
by the initial secondary mass $m_{2}$ ($<m_{1}$)
and initial orbital separation $a$ to calculate the amount of transferred
and accreted mass because of RLOF ($\Delta M_{\mathrm{trans}}$ and $\Delta M_{\mathrm{acc}}$
respectively) during the MS evolution of the secondary star.
We further compute the mass transfer efficiency,
\begin{equation}
\beta=\frac{\Delta M_{\mathrm{acc}}}{\Delta M_{\mathrm{trans}}},\label{eq:beta-code-practice}
\end{equation}
during the different mass transfer Cases~A, B and C. 
$\Delta M_{\mathrm{acc}}$ and $\Delta M_{\mathrm{trans}}$ are then the
accreted and the transferred mass, respectively, during the different mass transfer cases. 

\begin{figure}
\center
\includegraphics[width=0.46\textwidth]{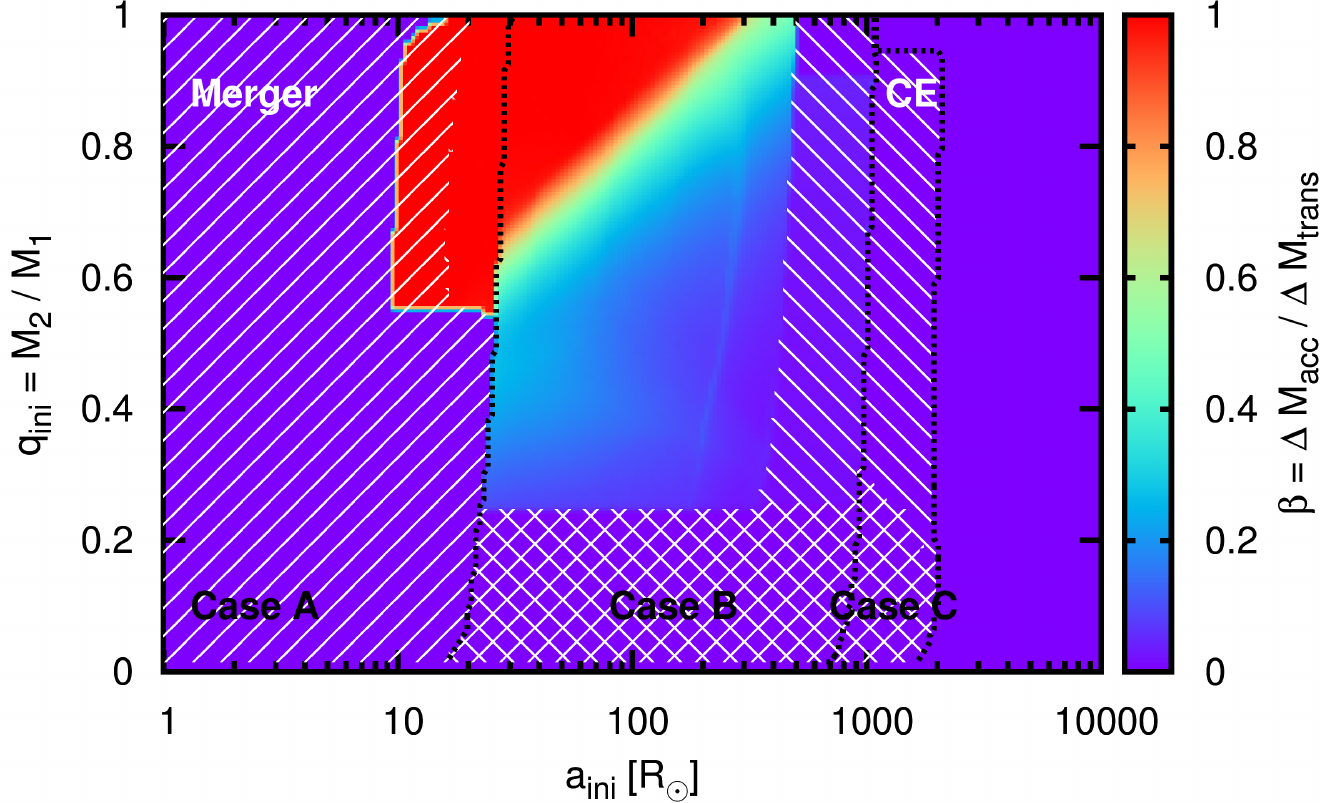}
\caption{Mass transfer efficiency $\beta=\Delta M_{\mathrm{acc}}/\Delta M_{\mathrm{trans}}$
in binaries with $M_{1}=10\,\msun$ primary stars as a function of
initial orbital separation $a_{{\rm ini}}$ and initial
mass ratio $q_{{\rm ini}}=m_{2}/m_{1}$. The black, dotted
lines indicate the boundaries between the three different mass transfer
Cases~A,~B and~C. The white area hatched from bottom-left to top-right 
shows binaries that result in stellar mergers whereas those hatched from
bottom-right to top-left show binaries that
go through at least one common envelope phase during their
evolution. Binary evolution is followed until the secondary stars leave the MS.}
\label{fig:m1-10-beta-no-op}
\end{figure}

In Fig.~\ref{fig:m1-10-beta-no-op} we show the mass transfer efficiency
$\beta$ (Eq.~\ref{eq:beta-code-practice}) in binaries with
$10\,\msun$ primary stars as a function of initial mass ratio, $q_\mathrm{ini}$, and
initial orbital separation, $a_\mathrm{ini}$. 
Binaries interact by Case~A mass transfer if the initial orbital separation
is less than about $20$-$30\,\rsun$. They first interact by
Case~B mass transfer if the initial orbital separation is longer
than the boundary for Case~A mass transfer and shorter than $800$--$1000\,\rsun$.
They first interact by Case~C mass transfer if the initial orbital
separation is longer than $800$--$1000\,\rsun$. They do not interact
by RLOF at all if the initial orbital separation is longer than
about $2000\,\rsun$. The boundaries depend
on the initial mass ratio: if the mass ratio is larger (for a fixed orbital separation),
the primary overflows its Roche lobe earlier in its evolution because 
the Roche lobe is smaller. The boundaries between the different mass
transfer cases therefore shift to longer initial orbital separations for larger
mass ratios.

There is a small zone/gap above the Case~C
mass transfer region for large initial mass ratios ($q\gtrsim0.95$)
where stars interact by Case~C
mass transfer, but only after the secondary has left the MS, hence
the gap. 

In Fig.~\ref{fig:m1-10-beta-no-op}, we also mark those binaries
as mergers which start RLOF on the ZAMS (we do not treat these binaries in our simulations). 
We assume that stars enter a
contact phase if their mass ratio at the onset of Case~A mass transfer
is less than $0.56$ (Sec.~\ref{sec:bse-code}). 
Only the products of these Case~A mergers are MS stars and only these
contribute to our analysis of the PDMF. 

There are two more critical
mass ratios visible in Fig.~\ref{fig:m1-10-beta-no-op}: we assume
that stars enter common envelope evolution at the onset of Case~B
RLOF if the donor star is a Hertzsprung gap star and the mass ratio
is below $0.25$. For giant-like donor stars, i.e.\
donor stars with a deep convective envelope (e.g.\ red supergiants),
we use a formula (Eq.~57 in \citealp{2002MNRAS.329..897H})
to calculate the critical mass ratio.
Mass loss from a star with a deep convective envelope leads to an
increase of its radius, hence to even more mass transfer and is thus
dynamically unstable. The giant-like donor star engulfs its
companion and the binary enters common envelope evolution.
If all these criteria do not apply for the binary star at the onset
of RLOF we use the critical mass ratio $q_{\mathrm{c}}=0.33$.

\begin{figure}
\center
\includegraphics[width=0.46\textwidth]{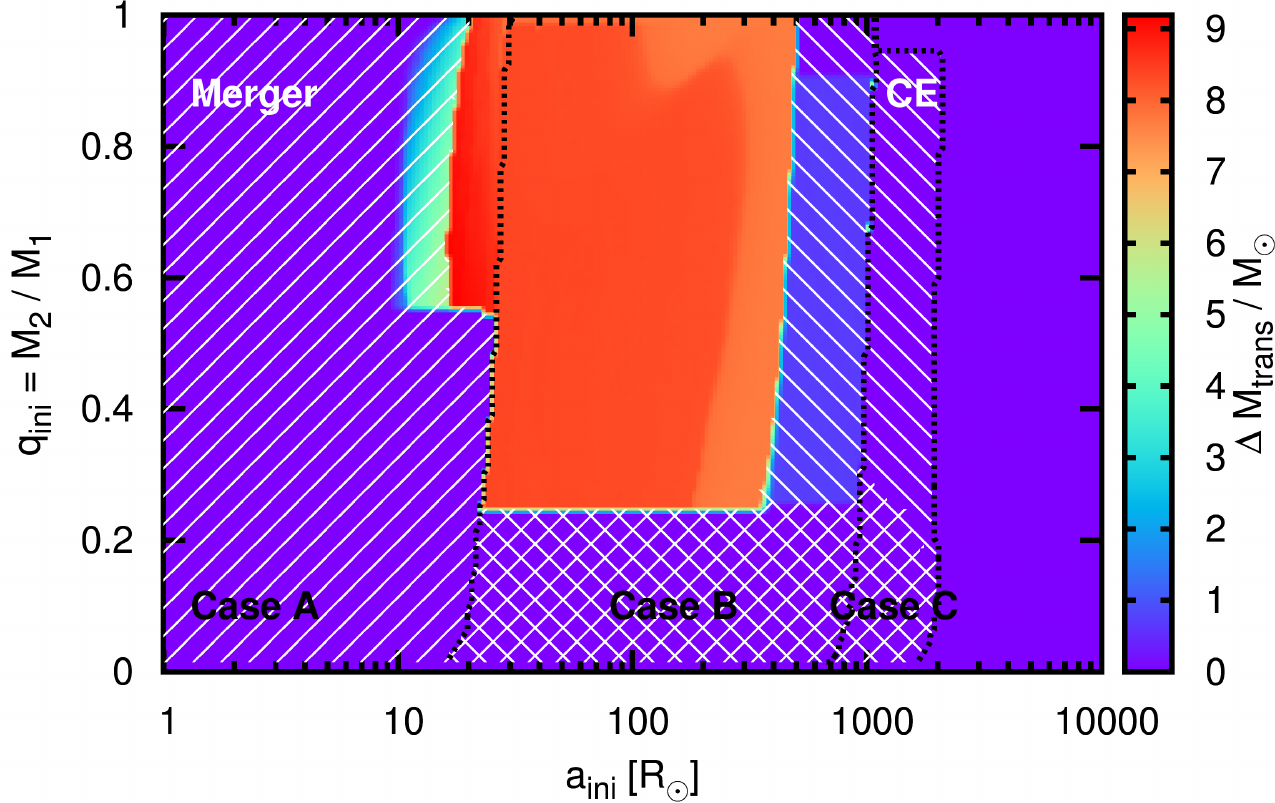}
\caption{The mass $\Delta M_{\mathrm{trans}}$ transferred from $10\,\msun$
donor stars by RLOF during the MS evolution of the secondary star
as a function of the initial mass ratio and the initial orbital separation.
The hatched regions have the same meaning as in Fig.~\ref{fig:m1-10-beta-no-op}.}
\label{fig:m1-10-dMtrans-nop-tot}
\end{figure}

\begin{figure}
\center
\includegraphics[width=0.46\textwidth]{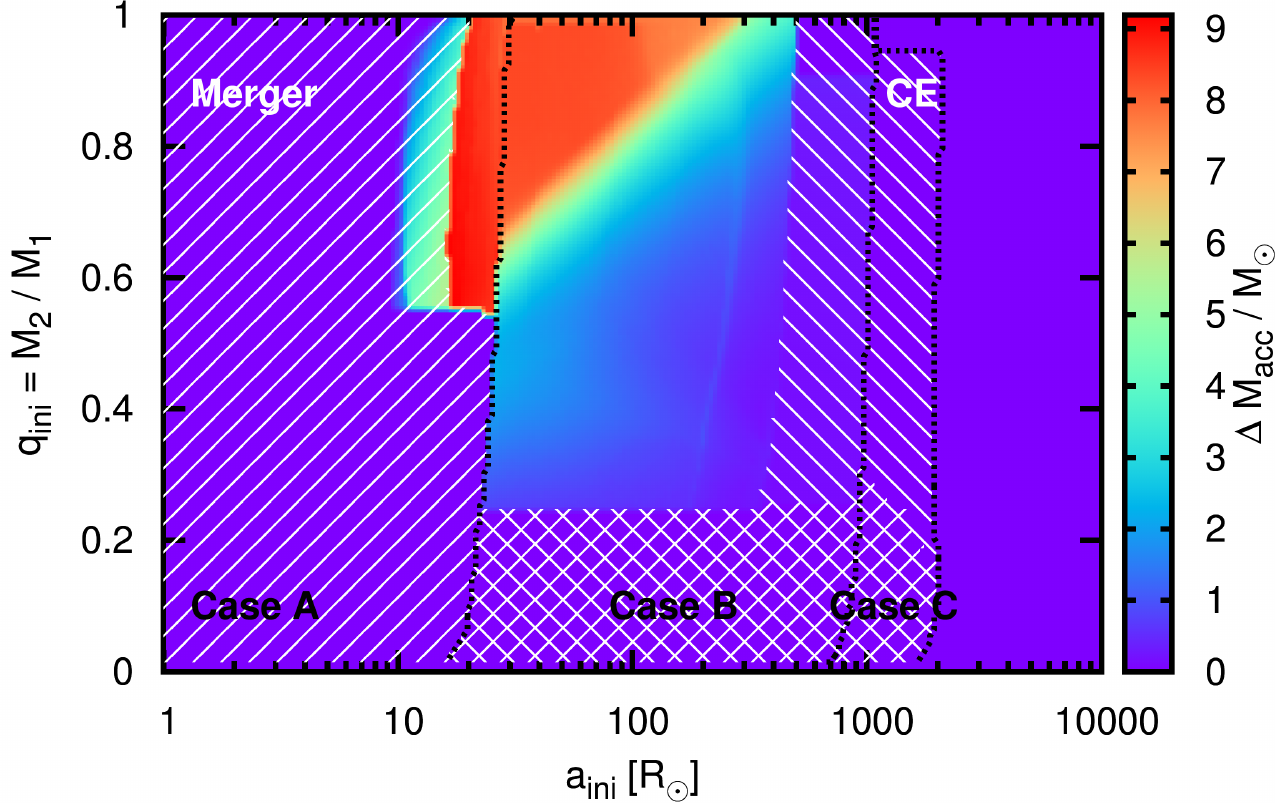}
\caption{As Fig.~\ref{fig:m1-10-dMtrans-nop-tot} but for the total
mass accreted by the secondary star, $\Delta M_{\mathrm{acc}}$,
during RLOF.}
\label{fig:m1-10-dMacc-nop}
\end{figure}

The transferred mass $\Delta M_{\mathrm{trans}}$ from a $10\,\msun$
primary star during RLOF is shown in Fig.~\ref{fig:m1-10-dMtrans-nop-tot}.
The transferred
mass is very similar in all interacting binaries that do not enter
a contact phase 
Case~A mass transfer is caused by increasing
stellar radii due to nuclear evolution during core hydrogen burning.
MS stars with masses larger than $1.25\,\msun$ have radiative envelopes
in our models. Mass transfer from stars with radiative envelopes is
stable, i.e.\ such stars shrink back into their Roche lobe as a reaction
to mass loss. Case~A RLOF starts earlier and lasts longer for smaller
initial orbital separations, hence the smaller the initial orbital
separation the larger the mass lost by the donor, i.e.\ the transferred
mass $\Delta M_{\mathrm{trans}}$. During Case~B and~C mass transfer,
RLOF is driven by the expansion of the stellar envelope. Stars stop
overfilling their Roche lobes only after losing (nearly) their whole
envelope. The envelope mass of the $10\,\msun$ donor stars of the
Case~B binaries which do not enter a CE evolution is always about
the same, hence is the total transferred mass in Fig.~\ref{fig:m1-10-dMtrans-nop-tot}.

Figure~\ref{fig:m1-10-dMacc-nop} shows how much of the mass transferred
from the $10\,\msun$ donor stars is accreted by the secondary
stars during their MS evolution. The only limit on mass accretion
in our models is the thermal timescale of the accretor which ensures
that the accretors remain in thermal equilibrium, an implicit assumption when using
fitting functions for single star evolution. The thermal (Kelvin-Helmholtz)
timescale is given by
\begin{equation}
\tau_{{\rm KH}}\approx10^7 \frac{M/\msun M_\mathrm{env}/\msun}{R/\rsun L/\lsun}\,\mathrm{yr},\label{eq:thermal-timescale}
\end{equation}
where $M$ is the total mass, $M_\mathrm{env}$ the mass of the 
envelope ($M_\mathrm{env}=M$ for MS stars), $R$ the
radius and $L$ the luminosity of the star. Nearly all transferred mass
is accreted during Case~A mass transfer.
Mass transfer during the Hertzsprung gap, i.e.\ Case~B mass transfer,
proceeds on the thermal timescale of the primary. At longer initial orbital 
separations, the thermal timescale of the primary is shorter because the primaries have
larger radii when they overfill their Roche lobes (mass and luminosity are nearly constant in the
Hertzsprung gap). The thermal timescale of the MS secondary at the same time 
is inversely proportional to a power of its mass ($\tau_{\mathrm{KH}}\propto M^{-x}\,,x>0$).
We therefore
expect less accretion at longer initial orbital separations (because of shorter thermal timescales of
the donor stars and hence faster mass transfer) and for smaller initial secondary masses (because
of longer thermal timescales of less massive accretors). This causes
the gradient visible in the Case~B region of Figs.~\ref{fig:m1-10-beta-no-op}
and~\ref{fig:m1-10-dMacc-nop}.

Combining the results of the transferred and accreted mass during
RLOF reveals the mass transfer efficiency as shown in Fig.~\ref{fig:m1-10-beta-no-op}.
Mass transfer is nearly conservative during Case~A, i.e.\ almost all transferred
mass is accreted ($\beta\approx1$), and becomes non-conservative
during Case~B and~C mass transfer.

\begin{figure}
\center
\includegraphics[width=0.46\textwidth]{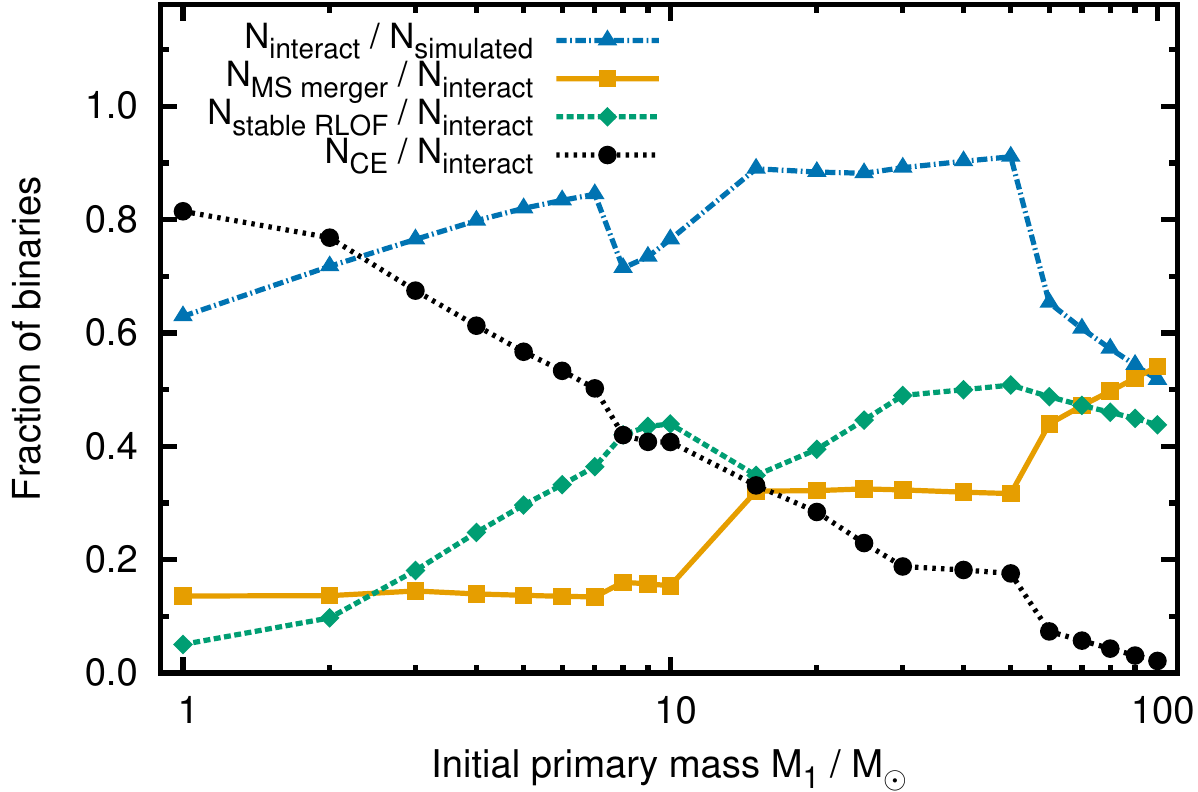}
\caption{Fraction of simulated binaries that interact by RLOF during the MS of
the secondary stars and fraction of interacting binaries that merge on the MS,
that transfer mass stably by RLOF and that go through a common envelope (CE) phase as a
function of the mass of the primary stars. The number of simulated binaries is
given by $N_\mathrm{simulated}$, the number of interacting binaries by $N_\mathrm{interact}$,
the number of binaries that merge on the MS by $N_\text{MS merger}$, the number of
binaries that transfer mass stably by RLOF by $N_\text{stable RLOF}$ and the number 
of binaries that go through a CE phase by $N_\mathrm{CE}$. By interacting binaries, we mean
that the primary overfills its Roche lobe resulting in 
mass transfer or a merger.}
\label{fig:phase-space-analysis}
\end{figure}

Next, we analyse the binary parameter space for different primary
masses. Figures~\ref{fig:binary-parameter-space-2msun}--\ref{fig:binary-parameter-space-100msun}
in Appendix~\ref{sec:appendix-binary-parameter-space-cont} contain the mass 
transfer efficiency, transferred and accreted mass
as in Figs.~\ref{fig:m1-10-beta-no-op}, \ref{fig:m1-10-dMtrans-nop-tot}
and~\ref{fig:m1-10-dMacc-nop} but for primary masses of $2$, $5$,
$20$, $50$, $70$ and $100\,\msun$. Binaries in which the primary star once filled its
Roche lobe are called interacting binaries. From the data of Fig.~\ref{fig:m1-10-beta-no-op},
we compute the ratios of the number of binaries that go through a common
envelope phase, merge on the MS and transfer mass stably by RLOF to the
number of interacting binaries and the ratio of the number of interacting to simulated
binaries. We plot these ratios as a function of primary mass in Fig.~\ref{fig:phase-space-analysis}
taking into account the initial distribution functions of binaries discussed in Sec.~\ref{sec:pop-syn}.
The overall trend is that the more massive the primary star the larger
the fraction of stars that transfer mass stably by RLOF, the larger the fraction
of stars that merge on the MS and hence the smaller the fraction of binaries
that go through a common envelope phase.

The number of interacting binaries decreases rapidly around $50\,\msun$
because more massive stars cross the Humphreys--Davidson limit \citep{1979ApJ...232..409H}
after their MS evolution and are subject to strong wind mass loss.
The mass loss widens the orbits such that binaries cannot interact 
by RLOF \citep[cf.][]{1991A&A...252..159V}. Consequently, the fraction of MS mergers 
increases because it is normalized by the number of interacting binaries
while the number of Case~B binaries that go through a common envelope phase 
($q_\mathrm{ini}<0.25$) decreases. 

The fraction of interacting binaries shows a kink around $8\,\msun$
because this is the mass above which stars explode as supernovae. 
Our $7\,\msun$ model reaches a maximum radius that is larger than that of
e.g.\ our $8\,\msun$ model because the latter
star explodes before reaching a similarly large radius. The parameter space
for interaction is therefore smaller in binaries with $8\,\msun$ primary stars.
From thereon, the number of interacting binaries gradually increases with primary mass 
because more massive primary stars reach larger maximum radii. 
Above $\sim13\,\msun$, the maximum radius is larger than that of a 
$7\,\msun$ star and so is the fraction of interacting binaries.

Between $2$ and $10\,\msun$ the fraction
of binaries that transfer mass stably by RLOF increases while the fraction of systems that
go through a common envelope phase decreases. Stars more massive than about $2\,\msun$
expand significantly while crossing the Hertzsprung-Russell diagram 
and the chance of interaction by Case~B
mass transfer is therefore larger. Less massive stars are cooler and
develop convective envelopes, ascending the giant branch (GB), after little expansion.
Mass transfer from such stars is dynamically
unstable leading to a common envelope phase. The more massive stars
are, the larger the Hertzsprung gap, the larger the number of stars
that interact by Case~B mass transfer while having a radiative envelope
and the fewer the number of binaries that enter a common envelope phase.

Stars with $M\gtrsim13\,\msun$ ignite helium in the core during the
Hertzsprung gap in our models and only slightly climb the GB, meaning
there is only a limited range of initial orbital separations that
leads to stars that interact as giants and thus enter a common envelope phase. 
Contrarily, the fraction of binaries
that transfer mass stably by RLOF plateaus for primary masses greater
than $10\,\msun$ because the available range of separations to
transfer mass from a star without a fully convective envelope does
not change significantly.

Binaries in which the primary star is more massive than about $22\,\msun$
do not experience Case~C mass transfer because the primary stars lose their
envelopes through strong stellar winds during core helium burning which widens the orbit and
prevents the star from further expansion. As a consequence, the star cannot 
overfill its Roche lobe after core helium burning and there is no Case~C mass transfer.

Binaries with O-type primaries, i.e.\ with $M_{1}\geq15\,\msun$
are distributed according to the initial orbital period distributions of \citet{2012Sci...337..444S}.
This enhances the number of stellar mergers on the MS and the occurrence 
of Case~A mass transfer because there are more binaries with short orbital periods according to
the orbital period distribution of \citet{2012Sci...337..444S} than for a distribution that is
flat in $\log P$ as used for less massive stars. The change of the period 
distribution leads to the abrupt increase in the fraction of MS
mergers around $15\,\msun$ and the decrease of the fraction of
binaries transferring mass stably by RLOF because the emphasis is now
on binaries in the closest orbits which preferentially merge (see Figs.~\ref{fig:m1-10-beta-no-op}
and \ref{fig:binary-parameter-space-2msun}--\ref{fig:binary-parameter-space-100msun}).

We discuss uncertainties in our models regarding single and binary star evolution
and the initial distribution functions in Appendix~\ref{sec:uncertainties-models}.

\subsection{Construction of mass functions}\label{sec:construct-mf}

Constructing the mass function of a population of single stars is
straightforward: we bin the masses and count the number of stars 
per mass bin. In the case of binary stars we construct three different
mass functions according to:
\begin{enumerate}
\item The primary masses only. 
\item The secondary masses only. 
\item An \emph{observed} mass reconstructed from the total stellar luminosity. 
\end{enumerate}
In the latter case we derive masses from the total luminosity of binaries:
we add the luminosities of both components and numerically invert
the mass-luminosity (ML) relation given by our fitting formula
(Eq.~12 of \citealp{2000MNRAS.315..543H}). We call mass functions
constructed in this way observed mass functions, because it comes
closest to mass functions from photometry where luminosities are directly
translated into masses and binaries are unresolved. We emphasize again
that we only consider MS stars in our analysis.

\section{Modulation of mass functions by stellar evolution}\label{sec:results}

\subsection{Single star populations}\label{sec:single-star-populations}

\begin{figure}
\center
\includegraphics[width=0.46\textwidth]{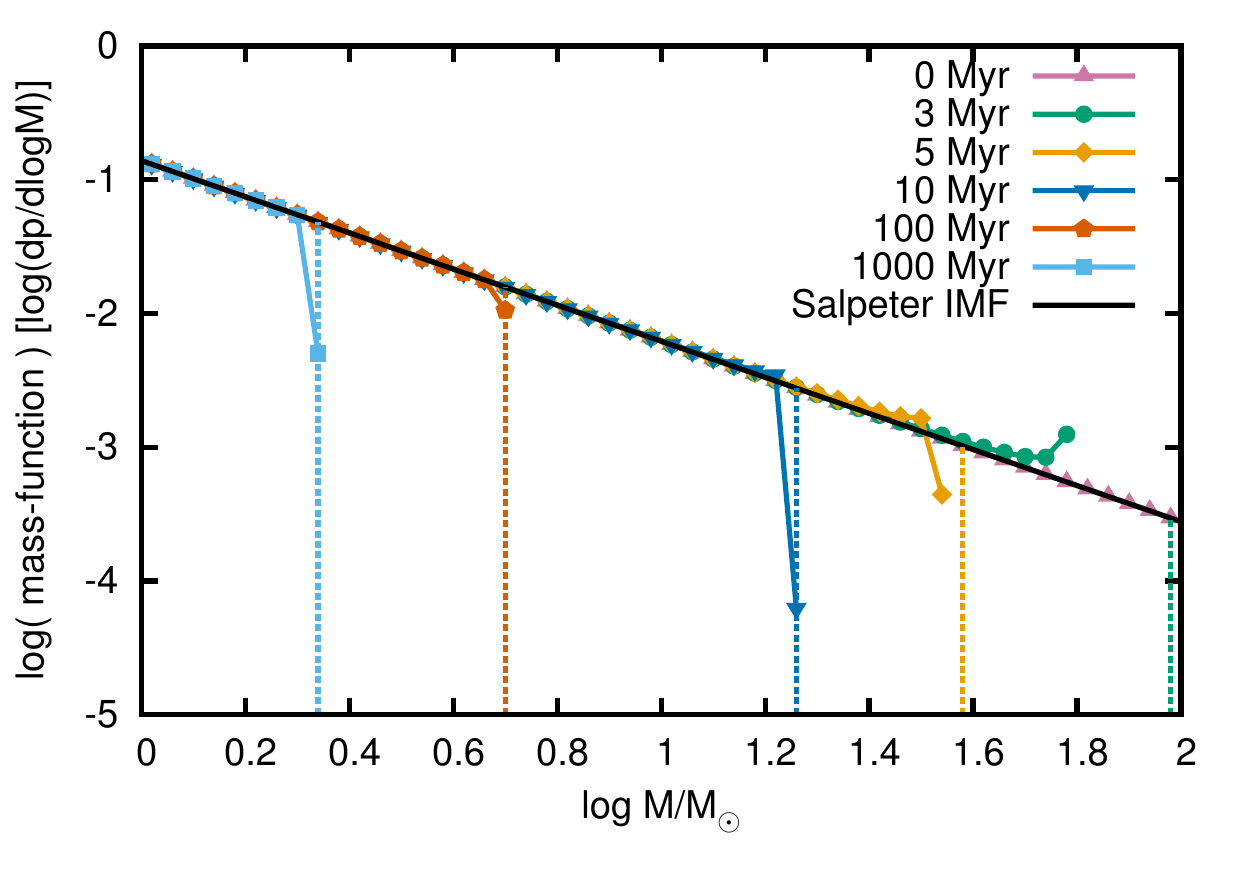}
\caption{Present-day mass functions of coeval single star populations of different
ages. Dashed vertical lines correspond to the mass of stars with a
MS lifetime equal to the age of the population. The most massive stars
have the shortest MS lifetime and hence disappear first. Stellar wind
mass loss creates the peaks at the high mass end most prominently 
seen in the $3.0\,{\rm Myr}$ old population.}
\label{fig:histo-singlestars-star1}
\end{figure}

Two phenomena influence the mass functions of a
population of single stars. First, stars have shorter lives the more massive they are. 
When stars leave the MS, they disappear from our mass functions. Second,
massive stars lose mass through stellar winds. This reduces stellar
masses and hence changes the mass function. In Fig.~\ref{fig:histo-singlestars-star1}
we show PDMFs of coeval single star populations of different ages.
The mass functions are constructed from the theoretically known stellar
masses (approach~1 in Sec.~\ref{sec:construct-mf}). The dashed vertical
lines mark the initial mass of stars that have a MS lifetime equal
to the age of the population. 
Over-plotted in black is the initial distribution of
stellar masses, i.e.\ the \citet{1955ApJ...121..161S} IMF.

The most massive stars leave the MS first
which causes a truncation close to the dashed vertical lines.
From here on we call this truncation the turn-off because stars at the truncation
of the mass function are located at the turn-off in the Hertzsprung--Russell (HR) diagram of a star cluster.
Stars at the turn-off are called ``turn-off stars'' and their mass the ``turn-off mass''.
Younger than about $10\,{\rm Myr}$ the turn-off
and the dashed lines do not coincide, they are displaced from each
other. The displacement is caused by stellar wind mass loss
and amounts to $\sim40\,\msun$ in the $3.0\,{\rm Myr}$ PDMF
for our adopted mass loss recipes.

We use a maximum initial stellar mass of $100\,\msun$. Such
stars have MS lifetimes of about $3.0\,\mathrm{Myr}$, so 
the upper mass boundary plays no role in populations 
older than $3.0\,\mathrm{Myr}$ because all stars with
initially masses larger than $100\,\msun$ have left the MS.
Our results are influenced by boundary effects at ages younger than $3.0\,\mathrm{Myr}$.

A prominent feature in the PDMF is the accumulation of stars
close to the turn-off because of mass loss through stellar winds. The more massive a
star the more mass is lost during its MS evolution. Stars with masses
close to the turn-off mass are the currently most massive stars in the cluster and will
soon leave the MS, so the greatest accumulation of stars in the PDMF
is found close to the turn-off. In our calculations, stars initially less massive than $9\,\msun$ at $Z=0.02$
($\tau_{\mathrm{MS}}\gtrsim30\,\mathrm{Myr}$) lose less than $1\%$ of their initial mass
via winds during the MS --- the accumulation of stars
in the PDMF disappears totally and the initial and the present-day turn-off masses 
are the same (the dashed vertical lines coincide with the turn-off). 
Note that we do not see a sharp truncation as indicated by the dashed
vertical lines because of the finite size of our mass bins.

\begin{figure*}
\center
\includegraphics[width=0.7\textwidth]{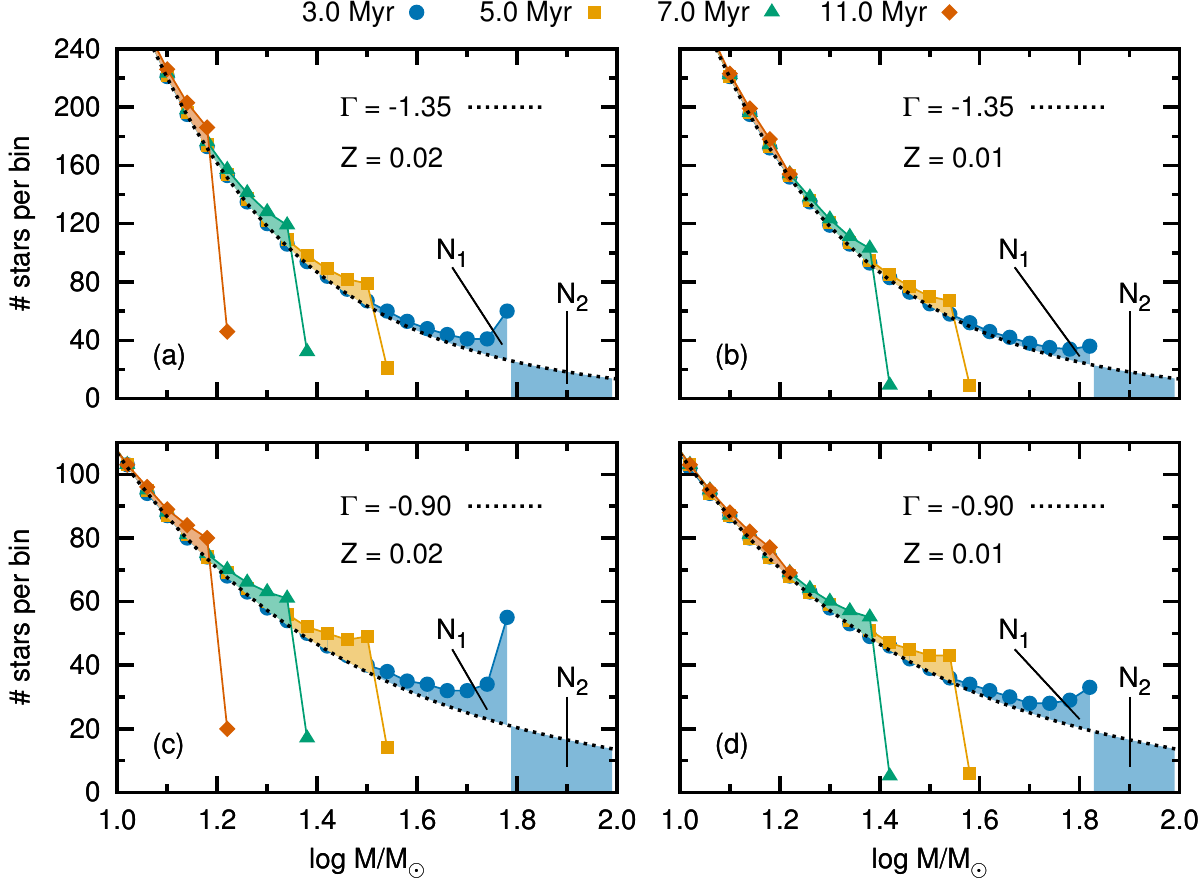}
\caption{High mass end ($\geq 10\,\msun$) of mass functions of coeval stellar populations
with IMF slopes $\Gamma$ and metallicities $Z$.
The panels show (a) a population drawn from a Salpeter IMF $\Gamma=-1.35$ at solar metallicity
$Z=0.02$, (b) as (a) with $Z=0.01$, (c) as (a) with
$\Gamma=-0.9$ and (d) as (c) with $Z=0.01$. The black dotted
line shows the IMF and the filled regions labelled $N_{1}$ indicate
the accumulation of stars because of stellar wind mass loss at an
age of $3\,{\rm Myr}$. Because the area labelled $N_1$ contains those stars that depopulated
the mass function for masses beyond the mass function peak,
it contains the same number of stars as the area labelled
$N_{2}$. The wind mass loss peaks are stronger the flatter, i.e.\ the 
more positive, the IMF slope and the stronger the wind mass loss (i.e.\ the higher 
the metallicity).}
\label{fig:pdmf-bump-imf-z}
\end{figure*}

The magnitude of the wind mass loss peak in the PDMF depends on how many stars are shifted
to lower masses and by how much. It therefore depends
on the strength of stellar wind mass loss, i.e.\
also on the metallicity, and the slope of the IMF. In Fig.~\ref{fig:pdmf-bump-imf-z} we show
this dependence at the high mass end of PDMFs ($\geq10\,\msun$)
on a linear scale where the size of the bump is more apparent: the higher the metallicity,
i.e.\ the stronger the wind mass loss, the bigger the peak. Similarly, the flatter
the mass function, i.e.\ the more high mass stars exist compared to
lower mass stars, the bigger the peak. The number
of stars in the highest-mass bin of the $3\,{\rm Myr}$ mass function
is increased compared to the IMF by $127\%$ for $\Gamma=-1.35$
and by $162\%$ for $\Gamma=-0.9$ at a metallicity $Z=0.02$ and
by about $55\%$ for $\Gamma=-1.35$ and $70\%$ for $\Gamma=-0.9$
at $Z=0.01$. 

The wind mass loss peak allows us
to determine the overall mass lost by stars during their MS
evolution. In practice, it may be necessary to model luminosity functions
rather than mass functions with different mass loss recipes to match
observed luminosity functions. Only by doing so can one overcome the 
inherent problem that masses derived from observed luminosities rely
implicitly on the mass loss prescription used to derive the applied
ML relations.

Figures~\ref{fig:histo-singlestars-star1} and~\ref{fig:pdmf-bump-imf-z} show that the wind 
mass loss peak is only visible in sufficiently young clusters in which stellar winds are strong.
We now investigate the age and metallicity range of stellar populations whose 
mass functions are likely to show the wind mass loss peak. 
Let $M_{{\rm to,p}}$ be the present-day turn-off mass in the star cluster and $N_{1}$ the number
of excess stars in the peak compared to the IMF
which is a power-law with slope $\Gamma$ (Eq.~\ref{eq:psi}). We
can then redistribute these stars such that the peak is removed by
filling up the IMF from the top end, thereby obtaining the
initial mass of the turn-off stars, $M_{{\rm to,i}}$. Let $\Delta M$ be the mass 
lost by turn-off stars over their MS evolution, i.e.\ $\Delta M = M_\mathrm{to,i}-M_\mathrm{to,p}$.
Because the number of excess stars $N_{1}$ equals the number of stars $N_{2}$ 
between $M_{{\rm to,p}}$ (Fig.~\ref{fig:pdmf-bump-imf-z}), we have, 
\begin{eqnarray}
N_{1}=N_{2}&=\int_{M_{{\rm to,p}}}^{M_{{\rm to,p}} + \Delta M}\,\psi(M)\,{\rm d}M=N_2 \nonumber \\
&= \frac{A}{\Gamma} M_\mathrm{to,p}^\Gamma \left[ \left(1 + \frac{\Delta M}{M_\mathrm{to,p}}\right)^\Gamma -1 \right]. 
\label{eq:excess-stars-def}
\end{eqnarray}
To judge whether the accumulation of $N_1$ stars is enough to see a wind mass loss peak,
we have to compare $N_1$ to the number of stars expected from the IMF in a mass range 
that is slightly less massive than the turn-off, $N_\mathrm{imf}$. 
We define $N_\mathrm{imf}$ to be the number of stars in the
mass range from $f M_\mathrm{to,p}$ to $M_\mathrm{to,p}$ where $f<1$. For $f=0.75$, we 
require that $N_1$ has to exceed 10\% of $N_\mathrm{imf}$ to be able to see the wind mass loss peak.
The 10\% requirement and the value for $f$ are chosen such that
a visible wind mass loss peak is predicted in the $7\,\mathrm{Myr}$ but 
not $11\,\mathrm{Myr}$ stellar populations at 
$Z=0.01$ and $Z=0.02$ (cf. Fig.~\ref{fig:pdmf-bump-imf-z}). Analogously to Eq.~(\ref{eq:excess-stars-def}), we 
compute $N_\mathrm{imf}$ and find that,
\begin{equation}
\frac{N_1}{N_\mathrm{imf}} = \frac{\left(1+\Delta M/M_\mathrm{to,p}\right)^\Gamma - 1}{1-f^\Gamma} > 10\%\;\mathrm{for}\;f=0.75
\label{eq:mdot-peak-visibility-criterion}
\end{equation}
in order for the wind mass loss peak to be visible. This criterion translates into a relative wind mass loss of,
\begin{equation}
\frac{\Delta M}{M_\mathrm{to,p}} \geq \left[ \frac{N_1}{N_\mathrm{imf}} \left( 1 - f^\Gamma\right)+1\right]^{1/\Gamma}-1.
\label{mdot-peak-visibility-criterion-mass-loss} 
\end{equation}
According to our criterion, the minimum relative mass loss is 3.7\% 
for $\Gamma=-1.35$ and 3.3\% for $\Gamma=-0.7$.

\begin{figure}
\center
\includegraphics[width=0.46\textwidth]{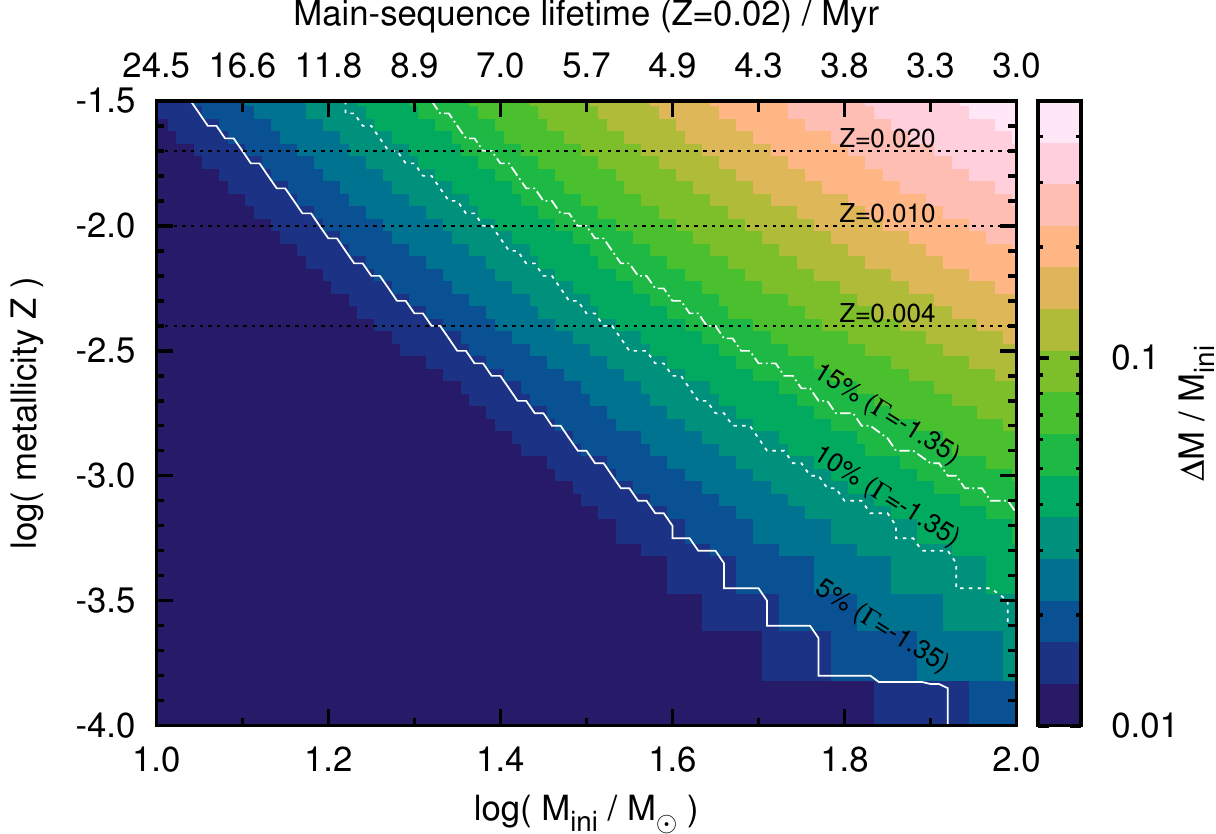}
\caption{Fractional initial mass loss, $\Delta M/M_\mathrm{ini}$, of stars during 
the main sequence as a function of initial mass and metallicity $Z$. The solid and dotted 
lines show the criterion of the visibility of the wind mass loss peak
in mass functions from Eq.~(\ref{eq:mdot-peak-visibility-criterion})
for $N_1/N_\mathrm{imf}=5$, 10 and 15\%, respectively, and an IMF slope of $\Gamma=-1.35$. 
According to our criterion, a wind mass loss peak is likely visible in stellar 
populations to the right of the solid, 10\% line (see text for more details).}
\label{fig:visibility-mdot-peak}
\end{figure}

In Fig.~\ref{fig:visibility-mdot-peak}, 
we show how much mass is lost by stars through stellar winds
during their MS evolution as a function of initial mass and metallicity. 
We further indicate the corresponding MS lifetimes for $Z=0.02$ on the top and plot the criterion 
from Eq.~\ref{eq:mdot-peak-visibility-criterion} for $N_1/N_\mathrm{imf}=5$, 10 and 15\%.
The mass functions of stellar populations of ages and metallicities to the right of 
the $N_1/N_\mathrm{imf}=10\%$ line are likely to show the wind mass loss peak. 
As a rule of thumb, the mass functions of stellar populations younger than about $10\,\mathrm{Myr}$
at $Z=0.02$ are likely to show a visible wind mass loss peak.
Note that our criterion is a rather rough estimate and depends on the
wind mass loss prescription and its metallicity scaling.

\subsection{Binary star populations}\label{sec:binary-star-populations}

Two processes additionally influence the mass functions of binary star populations
compared to those of single stars. First, stars
can merge during their evolution and, second, stars can accrete mass by RLOF.
In what follows we discuss three different mass functions to disentangle
these effects: we present mass functions of primary stars,
of secondary stars and those constructed from the ML inversion method
(observed mass functions) as discussed in Sec.~\ref{sec:construct-mf}.

\begin{figure}
\center
\includegraphics[width=0.46\textwidth]{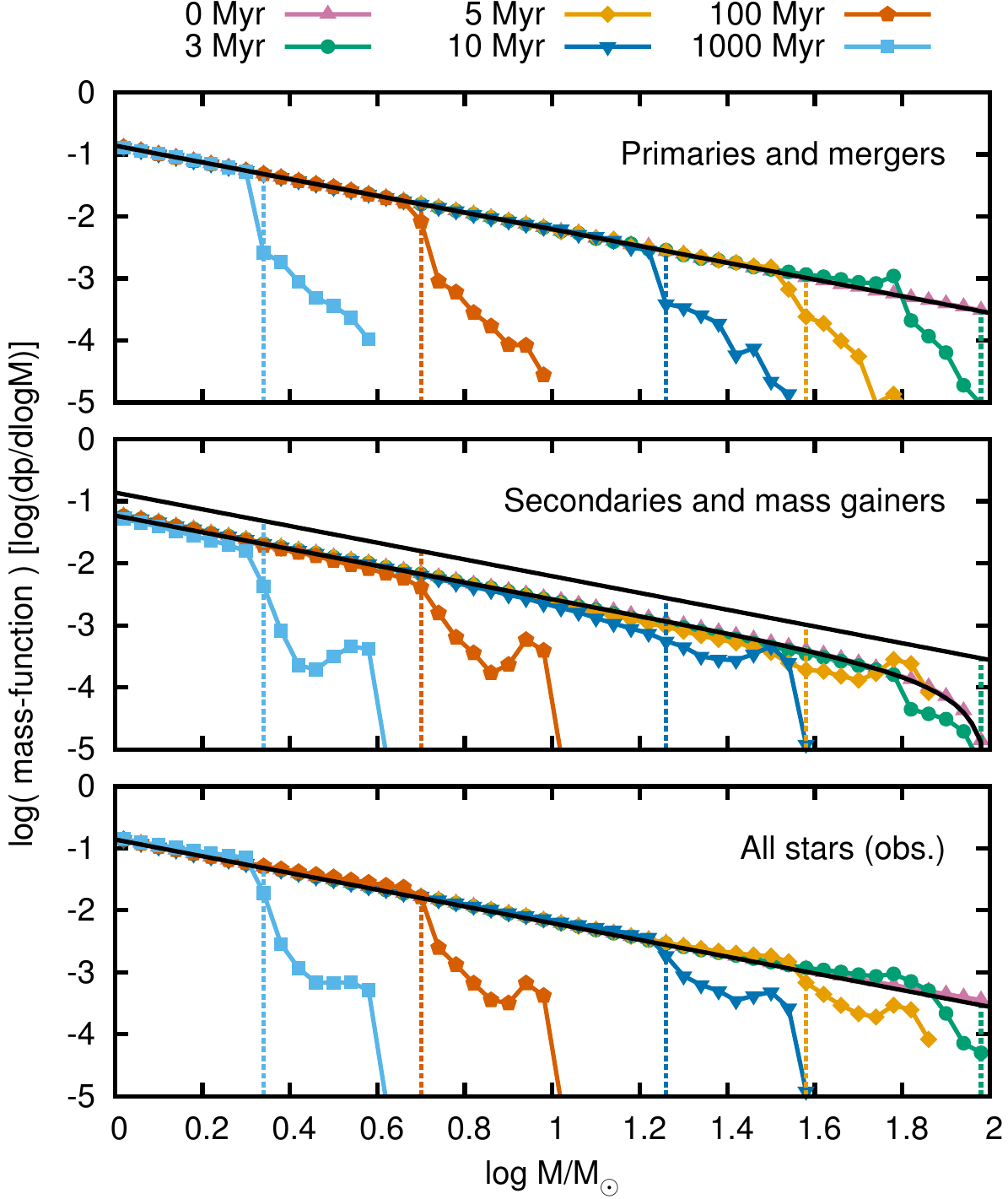}
\caption{Temporal evolution of PDMFs of all primary stars and mergers (top panel), all secondary stars
(middle panel) and for observed masses for all stars (bottom panel; see Sec.~\ref{sec:construct-mf}).
By construction, merged stars appear in the PDMFs in the top panel
and mass gainers of RLOF in the PDMFs in the middle panel. The vertical dashed lines indicate
the mass of stars having a MS lifetime equal to the age of the population, i.e.\
they indicate the initial mass of the turn-off stars.
Contrary to a population made only of single stars, stars more massive than the
initial mass of the turn-off stars exist. The black solid
lines are the Salpeter IMF and the initial distribution function of
the secondary stars (Eq.~\ref{eq:imf-star2-analytic}) respectively.}
\label{fig:histo-binaries-no-overspin-protection}
\end{figure}

The primary star is initially more massive than the secondary star,
so it evolves faster and is the donor star during RLOF. Analogously,
the secondary star is the mass gainer. 
In the top panel of Fig.~\ref{fig:histo-binaries-no-overspin-protection}
we show the mass functions of primary stars, i.e.\ stars that have not yet interacted, that have
lost mass by mass transfer or that have merged. The mass functions in the middle
panel of Fig.~\ref{fig:histo-binaries-no-overspin-protection} show secondary stars that have not interacted
yet and that have gained mass by RLOF. The observed mass function 
in the bottom panel of Fig.~\ref{fig:histo-binaries-no-overspin-protection}
is a combination of all stellar evolutionary effects.

The primary stars
are initially distributed according to the Salpeter IMF (Eq.~\ref{eq:psi})
and the secondary stars according to
\begin{eqnarray}
\frac{{\rm d}p}{{\rm d}\ln m_{2}} & = & \int_{\ln m_{2}}^{\ln100}\psi(\ln m_{1})\phi(\ln m_{2})\,\mathrm{d}\ln m_{1}\cdot\underbrace{\int_{\ln a}\chi(\ln a)\,\mathrm{d}\ln a}_{=1}\nonumber \\
 & \approx & \frac{A}{\Gamma-1}m_{2}\left(100^{\Gamma-1}-m_{2}^{\Gamma-1}\right),\label{eq:imf-star2-analytic}
\end{eqnarray}
where $\psi(\ln m_{1})$, $\phi(\ln m_{2})$ and $\chi(\ln a)$ are the
initial distribution functions of the primary and secondary masses and 
the orbital separations as defined in Sec.~\ref{sec:pop-syn} 
(we assume $q_\mathrm{min}\approx0.0$ in the last step). Compared to
the IMF of primary stars, the initial mass distribution of 
secondary stars is lowered by a factor $(1-\Gamma)^{-1}$ ($\approx0.43$). The decline of the secondary IMF
at the high mass end is caused by the maximum initial mass
(cf.\ the first term on the right-hand side of Eq.~\ref{eq:imf-star2-analytic}:
the maximum initial mass in our calculations is $100\,\msun$). The
overall slope is $\Gamma=-1.35$ again because of the flat mass ratio
distribution.

In single stars, no star is more 
massive than the turn-off. Including binary stars
it is possible to populate a tail of stars that extends the high mass end of single star
PDMFs by about a factor of 2 in mass.

In the top panel of Fig.~\ref{fig:histo-binaries-no-overspin-protection}
only stellar mergers populate the PDMF tail (i.e.\ the PDMF on the right-hand
side of the vertical dashed lines). The relative number of mergers in the tail
is lower the older the population as expected from
our analysis of the binary parameter space in which the fraction of
binary stars that merge on the MS increases with the initial mass
of the primary star (Fig.~\ref{fig:phase-space-analysis}).

The maximum mass of stellar mergers is reached by initially equal-mass binaries
in which the final mass is the total mass of the binary minus a fraction of 
10\% that we assume is lost during a merger (Sec.~\ref{sec:bse-code}). 
This corresponds to a mass increase by
a factor of $1.8$ or a shift of about $0.26\,\mathrm{dex}$ on the
logarithmic mass scale. The PDMF at $10\,\mathrm{Myr}$
extends slightly further because of the rejuvenation of the merged stars. The mass
of the merged star is not increased by more than a factor of $1.8$
relative to the primary mass, but fresh hydrogen is mixed into its
core. This decreases the fraction of burnt fuel and hence the apparent age
of the star. Compared to genuine single
stars of comparable mass, i.e.\ stars which have not interacted, the
stellar merger has more available fuel and thus stays longer on
the MS. Rejuvenated stars can thus appear to be shifted by more than $0.26\,\mathrm{dex}$
because the turn-off mass decreases simultaneously.

The PDMF tails in the middle panel of Fig.~\ref{fig:histo-binaries-no-overspin-protection}
contain secondary stars that have accreted mass. The PDMFs again extend to slightly larger masses
than expected from mass accretion alone because of rejuvenation. Mass accretion
in young stellar populations (ages $\lesssim10\,\mathrm{Myr}$) forms PDMF tails
that even exceed the initial distribution of the secondary stars.
At later times, the relative number of stars in the tail is less because the number of interacting
binaries that transfer mass stably by RLOF decreases for initially
less massive primary stars (Fig.~\ref{fig:phase-space-analysis}) as does 
the overall mass transfer efficiency, $\beta$, which is coupled
to the thermal timescale of the mass gainers during Case~B mass transfer. The thermal timescales
of stars become more comparable for larger masses because the ML
and mass-radius relations are less steep the larger the mass.

In populations younger than $3.0\,{\rm Myr}$, the effect of mass
accretion on the PDMF is modest
because the most massive stars in our models have just left the MS and hence there
is no contribution from Case~B mass transfer to the PDMF (Case~C
does not occur for primary masses larger than about
$22\,\msun$, see Sec.~\ref{sec:binary-parameter-space}). The number
of interacting binaries drops strongly for primary masses larger than
about $50\,\msun$ because of the Humphreys--Davidson limit (Sec.~\ref{sec:binary-parameter-space}).
This is important in stellar populations younger than $\sim 4.3\,\mathrm{Myr}$
which is the MS lifetime of $50\,\msun$ stars in our models.

In the bottom panel of Fig.~\ref{fig:histo-binaries-no-overspin-protection}
we use the ML inversion method (approach~3 in Sec.~\ref{sec:construct-mf})
to construct the PDMFs. Both, mergers and accretors show up in the PDMF tails.
The star with the highest mass in each population is formed by RLOF.
However, the difference between the highest mass achieved by stellar
mergers and by RLOF is small --- at most one bin-width, i.e.\
$\Delta\log m=0.04\,{\rm dex}$ --- and depends on the assumed 
mass loss in stellar mergers and the mass transfer efficiency. Only at early times ($\lesssim3.5\,\mathrm{Myr}$)
does the star with the highest mass originate in a stellar merger.
Unresolved binaries also contribute to the tail of the observed
PDMFs but their contribution is small (at most $30\%$ of the tail stars are unresolved binaries in
young, $\lesssim3$--$4\,\mathrm{Myr}$ populations) and greatest
for the largest masses because of a flatter ML relation
of high mass stars compared to lower masses.
Unresolved binaries extend the single star mass function by at most $\sim20\%$ in mass
for $1\dots2\,\msun$ stars and by at most $\sim50\%$ in mass for $60\dots100\,\msun$ stars
(cf. Eq.~\ref{eq:mobs} in Sec.~\ref{sec:quantifying-evol-effects}).

\begin{figure}
\center
\includegraphics[width=0.46\textwidth]{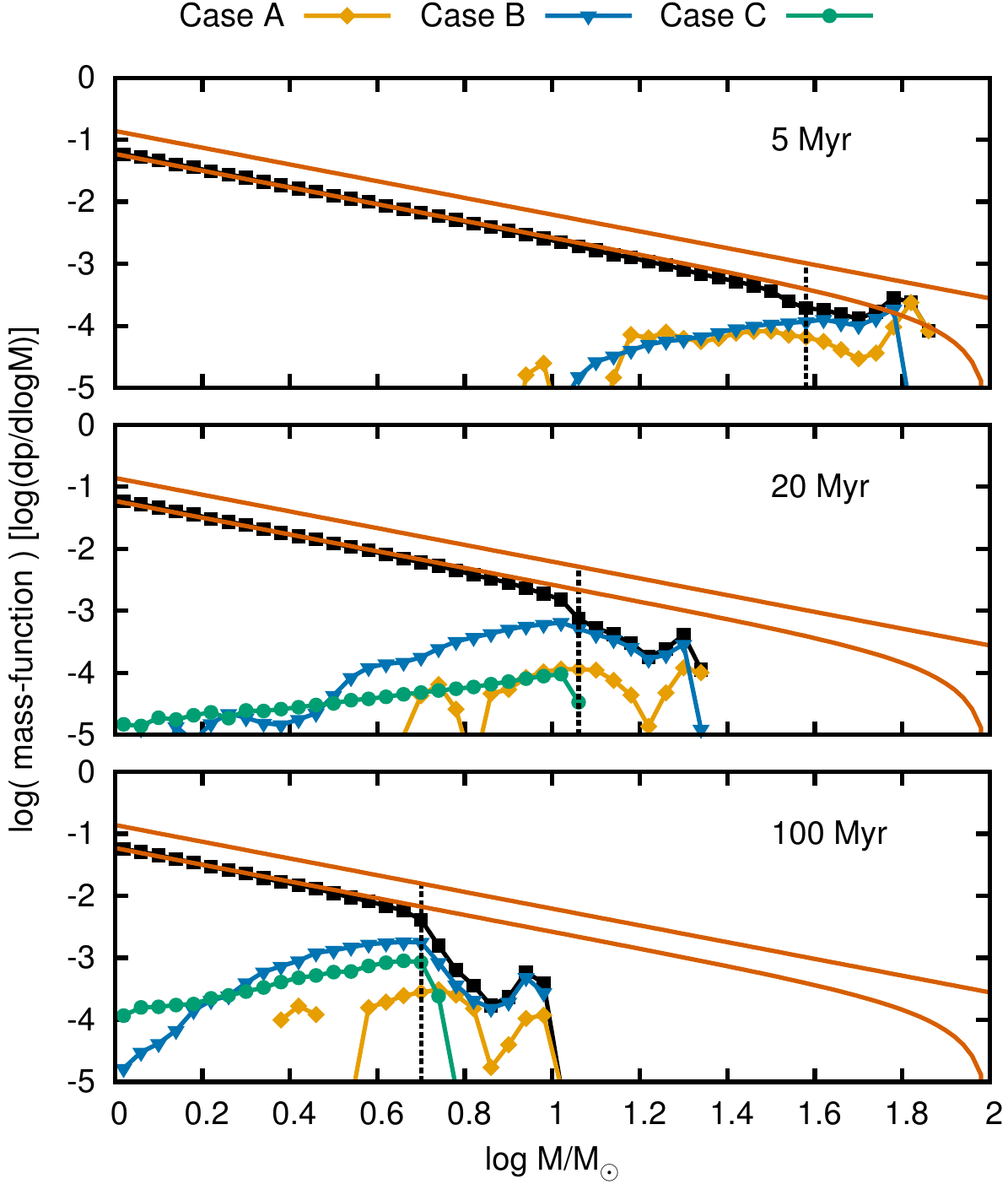}
\caption{Mass functions at $5$, $20$ and $100\,\mathrm{Myr}$ for secondary stars, i.e.\ mass accretors.
The curves show the contributions of the mass transfer
Cases~A,~B and~C (as defined in Sec.~\ref{sec:bse-code}) to the
total PDMF of secondaries (black squares). Case~B mass transfer is the
dominant contribution to the tail stars for $\gtrsim3\,\mathrm{Myr}$,
hence the mass loser is not a MS star and therefore not present in
the PDMFs of primaries in the top panel of Fig.~\ref{fig:histo-binaries-no-overspin-protection}.
This is why the mass gainers are seen clearly in the PDMFs whereas the mass losers
are not.}
\label{fig:mt-cases}
\end{figure}

In Fig.~\ref{fig:mt-cases} we show the PDMF of secondary stars at three different times --- $5$, $20$
and $100\,\mathrm{Myr}$. 
Case~B mass transfer is the dominant contribution after $3\,\mathrm{Myr}$,
therefore the mass losing star has left the MS and is no longer included
in the PDMFs of the primary stars in the top panel of Fig.~\ref{fig:histo-binaries-no-overspin-protection}.
The Case~A mass losers are difficult to find. Nevertheless some of them can
be found by comparing the PDMFs of populations of single (Fig.~\ref{fig:histo-singlestars-star1})
and binary stars only (top panel of Fig.~\ref{fig:histo-binaries-no-overspin-protection}).
The mass losers are easiest to spot by the difference in the magnitude of
the accumulation of stars because of wind mass loss e.g. in
the PDMFs of the $5\,{\rm Myr}$ old populations. Populations made
only of single stars show a larger accumulation of stars than those made of
binary stars, because some donor stars that would appear in the wind mass loss
peak lose additional mass during Case~A mass transfer.
In practice it is very hard to detect the primaries after they 
have lost mass \citep{2014ApJ...782....7D}.

Case~C mass transfer is typically highly non-conservative in our models
(Sec.~\ref{sec:binary-parameter-space} and Appendix~\ref{sec:appendix-binary-parameter-space-cont}). 
The secondary stars of such systems therefore do not gain enough mass to make 
any significant contribution to the tail.

The stars in the tail of the mass functions are rejuvenated binary products
and hence appear younger than the real age of the population --- they are 
blue stragglers. We further characterise these blue straggler stars in Sec.~\ref{sec:blue-stragglers}
in terms of their binary fraction and apparent ages and compare their frequencies to observations.

\subsection{Stellar populations with varying binary fractions}\label{sec:star-cluster}

\begin{figure}
\center
\includegraphics[width=0.46\textwidth]{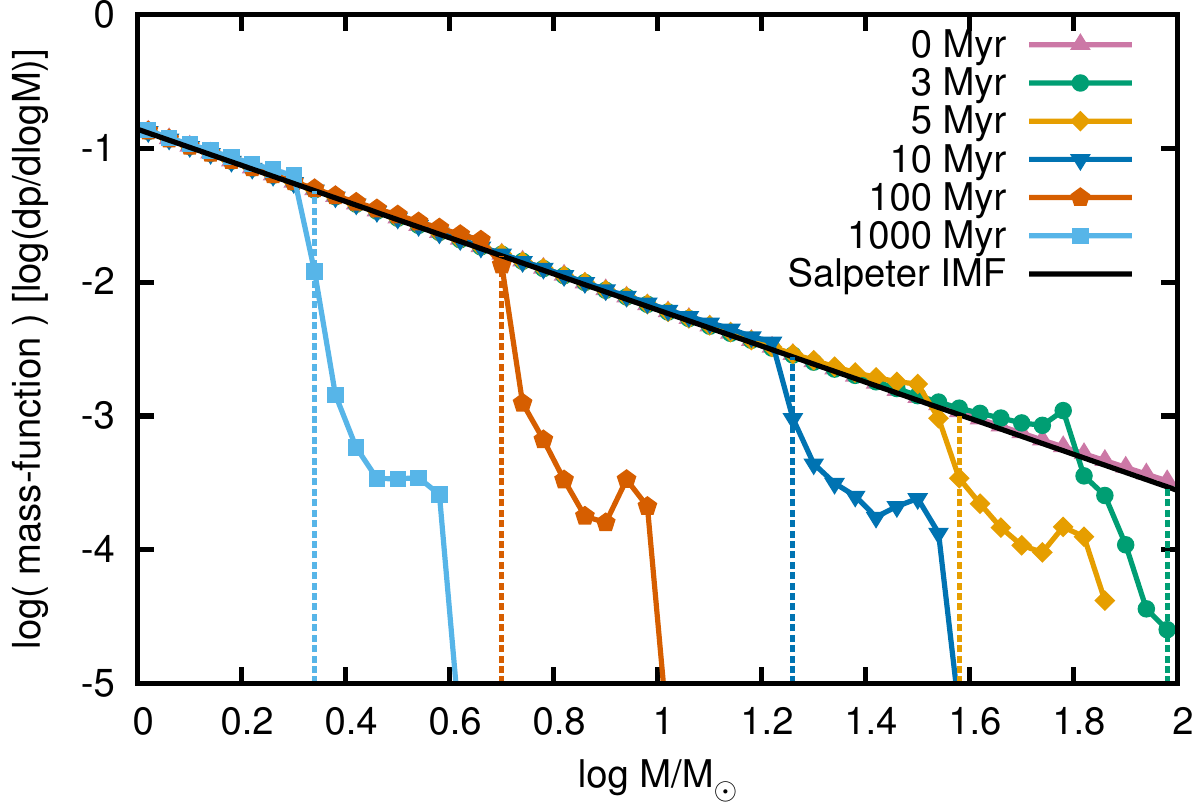}
\caption{As the bottom panel of Fig.~\ref{fig:histo-binaries-no-overspin-protection},
but for a population of stars with a primordial binary fraction of
$f_{\mathrm{B}}=0.5$. As discussed in Sec.~\ref{sec:star-cluster}
the number of stars in the binary tails are halved compared to a pure binary star population.}
\label{fig:histo-star-cluster-f-0.5}
\end{figure}

We use the PDMFs of populations of single and binary stars presented
in Secs.~\ref{sec:single-star-populations} and~\ref{sec:binary-star-populations}
to build stellar populations composed of a mixture of coeval single and binary stars. Let
$f_{\mathrm{B}}$ be the binary fraction at birth, i.e.\ the number of binary
systems divided by the number of total stellar systems ($f_{\mathrm{B}}=N_{\mathrm{B}}/[N_{\mathrm{S}}+N_{\mathrm{B}}]$
with $N_{{\rm B}}$ the number of binary systems and $N_{{\rm S}}$
the number of single stars). The considered binaries have initial orbital 
separations shorter than $10^4\,\rsun$ --- initially wider binaries are treated as single 
stars in our models (Sec.~\ref{sec:pop-syn}). The mass function of a 
population with this binary fraction is then,
\begin{equation}
\frac{\mathrm{d}p}{\mathrm{d}\log m} = (1-f_{\mathrm{B}})\cdot\left(\frac{\mathrm{d}p}{\mathrm{d}\log m}\right)_{\mathrm{s}}+f_{\mathrm{B}}\cdot\left(\frac{\mathrm{d}p}{\mathrm{d}\log m}\right)_{\mathrm{b}}.
\end{equation}
Here $\left(\mathrm{d}p/\mathrm{d}\log m\right)_{\mathrm{s}}$
is the mass function of single stars only and correspondingly
$\left(\mathrm{d}p/\mathrm{d}\log m\right)_{\mathrm{b}}$ is
the mass function of binary stars only. 
In Fig.~\ref{fig:histo-star-cluster-f-0.5} we
show the observed mass functions of stellar populations with a primordial binary 
fraction of $f_{\mathrm{B}}=0.5$ (i.e.\ two out of three stars initially born in binaries).
In comparison to Fig.~\ref{fig:histo-binaries-no-overspin-protection},
the number of stars in the binary tail in the PDMFs (Sec.~\ref{sec:binary-star-populations})
is attenuated. 
The tail of the mass functions of stellar populations 
with a binary fraction of 50\% is 
decreased by $0.3\,\mathrm{dex}$, i.e.\ by a factor of 2, compared to
those with a binary fraction of 100\%.

\subsection{Quantification of evolutionary effects on the PDMF}\label{sec:quantifying-evol-effects}

\begin{figure}
\center
\includegraphics[width=0.46\textwidth]{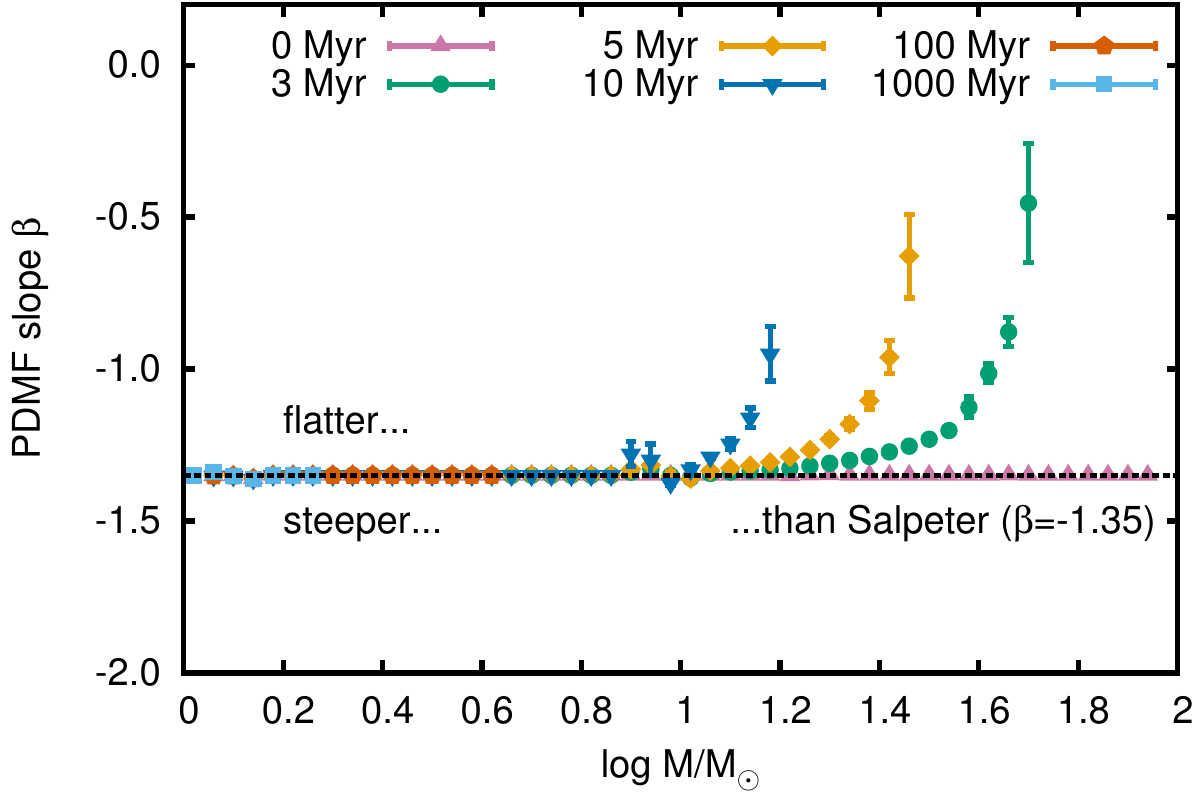}
\caption{Slopes of the PDMF of single stars (Sec.~\ref{sec:single-star-populations})
as a function of logarithmic mass. Stellar wind mass loss
flattens the high mass end of the PDMFs.}
\label{fig:slopes-starburst-singlestars-datapoints-3}
\end{figure}

The stellar wind mass-loss peak at the high mass end flattens
the mass function. We quantify the flattening by computing
the PDMF slopes as a function of stellar mass. We fit straight lines
piecewise to three mass bins at a time by a Levenberg--Marquardt method \citep{levenberg1944,marquardt1963}
and show the slopes $\beta$ of the \emph{single star} PDMFs
in Fig.~\ref{fig:slopes-starburst-singlestars-datapoints-3}.
The errors are statistical $1\sigma$ deviations from the best fit.
The more massive a star the more mass is lost by stellar winds, hence the
accumulation and resulting flattening of the PDMF is strongest in the $3\,\mathrm{Myr}$
population. The slope at the high mass end of the mass function
of young stellar populations ($\lesssim 10\,\mathrm{Myr}$) is much shallower than the Salpeter IMF,
$\Gamma=-1.35$. We find extremes of the PDMF slope of $\beta\approx-0.45\pm0.20$
at the highest masses around $50\,\msun$ ($\log M/\msun \approx1.7\,{\rm dex}$)
in the $3\,{\rm Myr}$ population. The flattening is still significant
for $10\,\mathrm{Myr}$ populations at masses more than $10\,\msun$ for
which the slope flattens to values of up to $\beta\approx-0.95\pm0.09$.
In other words, not accounting for wind mass loss when determining
the slope of the IMF for the most massive stars may lead to IMF slopes 
that are biased by up to 1\,dex.

\begin{figure}
\center
\includegraphics[width=0.46\textwidth]{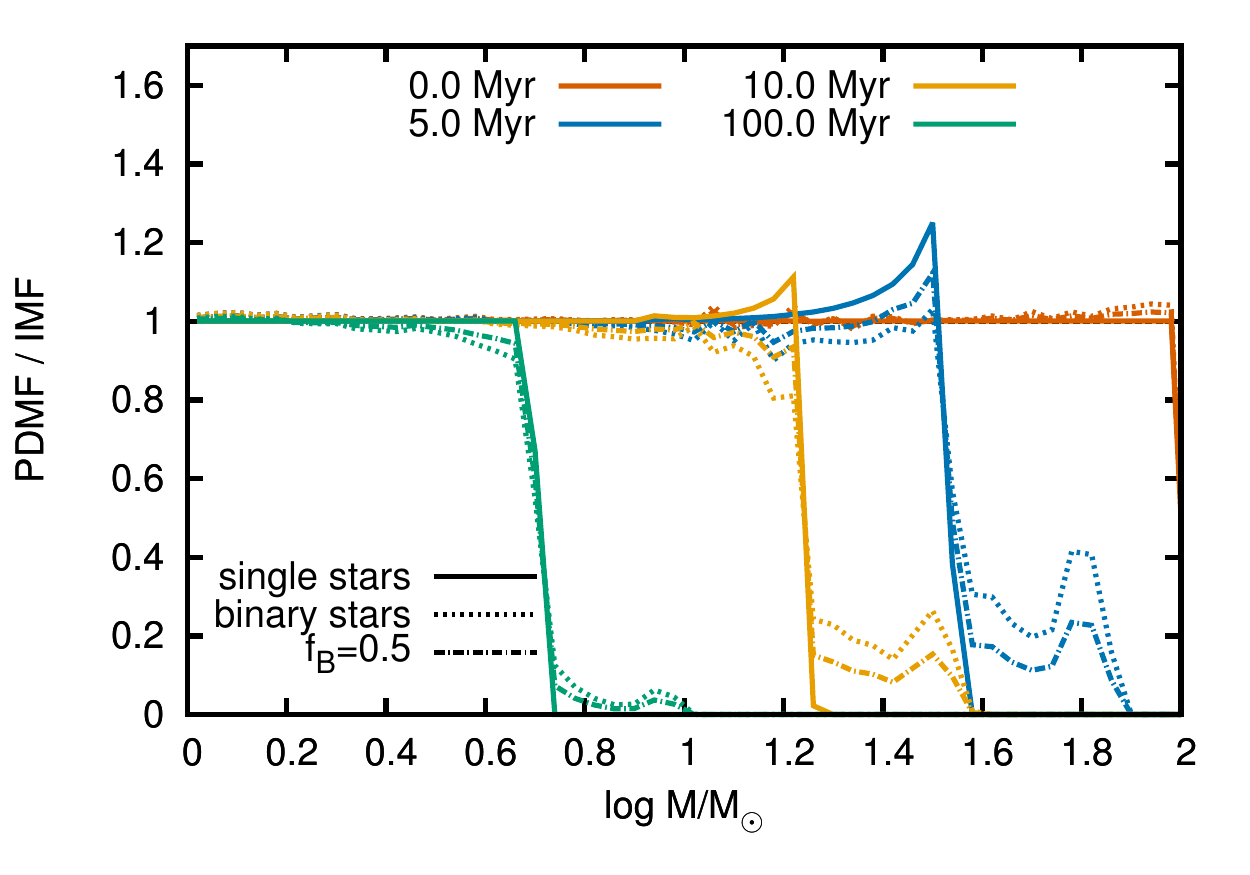}
\caption{Ratio of the PDMFs to IMFs for different ages and primordial binary
fractions. The PDMFs are constructed from the known stellar masses
(approaches 1 and 2 in Sec.~\ref{sec:construct-mf}).}
\label{fig:mf-ratio-allMS}
\end{figure}

We further quantify the wind mass loss peak and the binary tail
by dividing the PDMFs of populations consisting purely of
single stars, purely of binary stars and of a mixture 
of single and binary stars by the initial distribution
functions (Eqs.~\ref{eq:psi} and~\ref{eq:imf-star2-analytic}) in 
order to determine the relative importance of the evolutionary
effects compared to the initial distribution of stellar systems. 
First, we construct PDMFs from the theoretically known stellar masses
and not using the ML inversion method and show the ratio of 
PDMF to IMF in Fig.~\ref{fig:mf-ratio-allMS}. This enables
us to distinguish between the effects of single and binary star physics and
of unresolved binaries on the PDMF. 
Again, the truncations of the PDMFs are due to finite stellar lifetimes.

The PDMF of the $5\,{\rm Myr}$ single star population reaches a level
of more than $120\%$ of the IMF because of an accumulation of stars
caused by wind mass loss (blue solid line close to the turn-off mass of
$\log m \approx 1.5\,{\rm dex}$).
The wind mass loss peak becomes weaker with age until it completely disappears
in stellar populations older than $30\,\mathrm{Myr}$ (at a metallicity 
of $Z=0.02$ with a Salpeter IMF; see also Sec.~\ref{sec:single-star-populations}).

Stellar mergers and mass transfer by RLOF shift stars toward higher masses. 
The number of stars that are slightly 
less massive than the turn-off stars is therefore less than the initial number of stars,
i.e.\ $\mathrm{PDMF}/\mathrm{IMF}<1$ (with no wind mass loss).
The tail of the $5\,{\rm Myr}$
binary population reaches an average level of more than $30\%$ of
the initial distribution function, i.e.\ on average about one third
of the IMF in a mass range from about $40\,\msun$ ($\log m\approx1.6\,{\rm dex}$)
to $80\,\msun$ ($\log m\approx1.9\,{\rm dex}$) is re-populated by
binary evolution. This level gradually decreases the older the population
and the smaller the binary fraction.

\begin{figure}
\center
\includegraphics[width=0.46\textwidth]{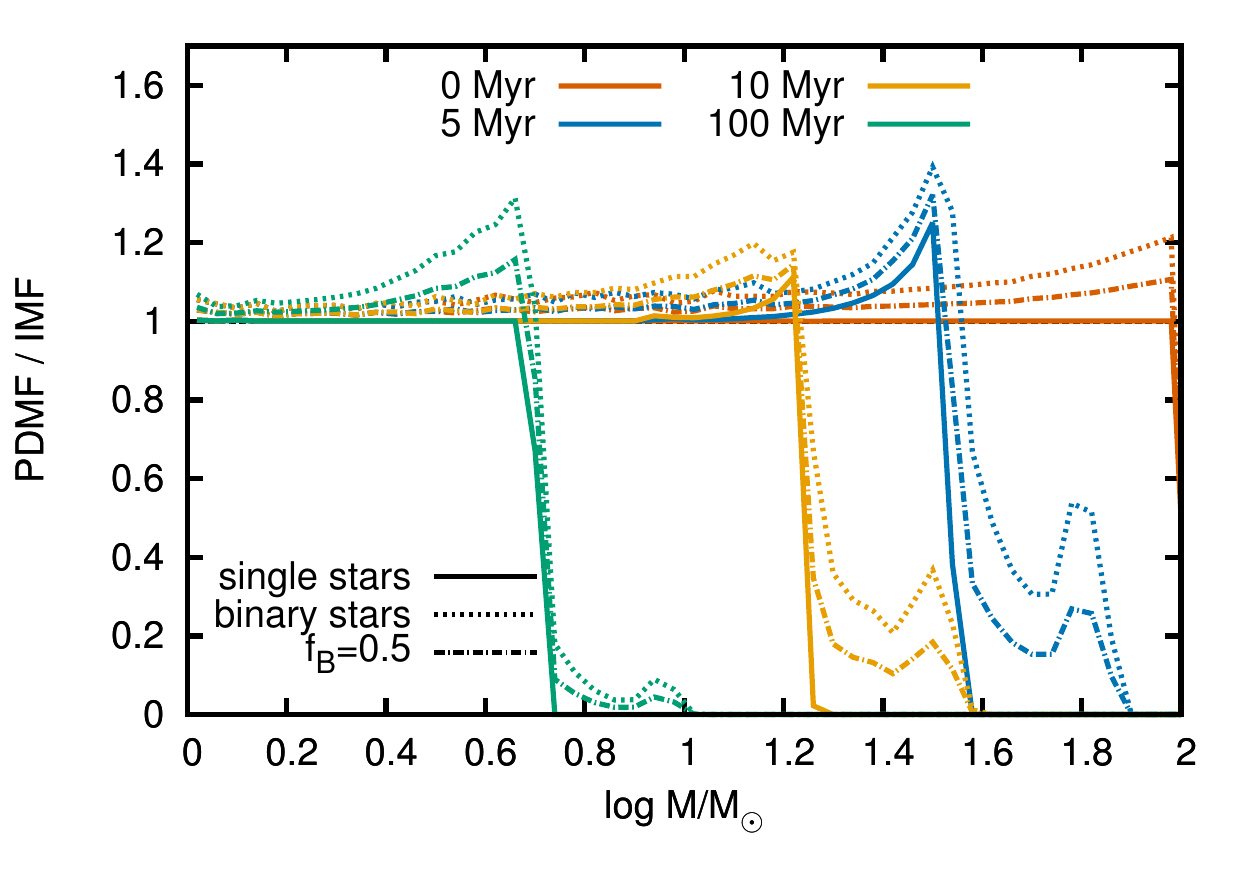}
\caption{As Fig.~\ref{fig:mf-ratio-allMS} but constructed
using the mass-luminosity inversion method (approach~3 in Sec.~\ref{sec:construct-mf}).}
\label{fig:mf-ratio-obs}
\end{figure}

Being unable to resolve binaries shifts the observed mass function
to larger masses (Fig.~\ref{fig:mf-ratio-obs}). This is well known
\citep{1991A&A...250..324S,1993MNRAS.262..545K,2008ApJ...677.1278M,2009MNRAS.393..663W}
and is evident from the initial mass distribution of pure binary
star populations ($0\,{\rm Myr}$ binary population in Fig.~\ref{fig:mf-ratio-obs}).
In a single star population there is no difference between
the observed IMF and the Salpter IMF. With unresolved binaries, 
the PDMF exceeds the IMF and the ratio of PDMF to IMF increases with larger binary fractions.
In stellar populations with $f_{\mathrm{B}}=0.5$
there are $5-10\%$ more stellar systems than expected from the Salpeter IMF 
and up to $\sim 20\%$ more for pure binary populations ($f_\mathrm{B}=1$). The
increase is mass dependent: it is greater for larger masses and is
understood from the ML relation, which can be approximated as $L=L_{0}M^{x}$. In
unresolved binaries with equal masses $M$ the observed mass $M_{\mathrm{obs}}$ is
\begin{equation}
M_{\mathrm{obs}}=\left(\frac{L_{1}+L_{1}}{L_{0}}\right)^{1/x}=2^{1/x}M,\label{eq:mobs}
\end{equation}
i.e.\ larger than $M$ by a factor of $2^{1/x}$. Low mass stars
have a larger exponent $x$ than high mass stars,
hence the larger the mass $M$ the larger the factor by which the observed mass is increased 
(e.g.\ $L\propto M^{4.5}$ for $1\leq M/\msun\leq2$ and $L\propto M^{1.8}$
for $60\leq M/\msun\leq100$ ZAMS stars, \citealp{1996MNRAS.281..257T}).
This translates into an increase of the mass function by a factor of
\begin{equation}
\frac{\psi(\ln M)}{\psi(\ln M_{\mathrm{obs}})}=2^{-\Gamma/x}=1.23\dots1.68,\label{eq:pdmf-shift}
\end{equation}
where $\Gamma$ is the IMF slope. 
Unresolved binaries flatten the
PDMF most strongly at the largest stellar masses (under the assumption
that $\Gamma$ is constant in the considered
mass range). We discuss the effect of unresolved binaries on mass functions 
with respect to previous work in more detail in Appendix~\ref{sec:unresolved-binaries}.

Figure~\ref{fig:mf-ratio-obs} shows that unresolved binaries
flatten the PDMF at masses less than the turn-off mass instead of steepening the PDMF as was
the case for resolved binaries. The wind mass-loss peaks in the PDMF reach a level of up to
$\sim140\%$ of the IMF.
The tail of the $5\,{\rm Myr}$ binary population is re-populated
by on average $\gtrsim40\%$ and with maximum levels of about $55\%$
of the IMF. 
The number of stars in the binary tail is halved in populations with a
binary fraction of $f_{\mathrm{B}}=0.5$.

We have shown above that stellar wind mass loss, binary products and 
unresolved binaries re-shape the high mass end of mass functions which
may complicate IMF determinations. To avoid biased IMF 
slope determinations in young as well as old clusters, we suggest to either exclude 
those parts of the mass function that are expected to be affected by stellar winds, 
binary products and/or unresolved binaries or to correct for these effects. 
As evident from Fig.~\ref{fig:mf-ratio-obs}, this concerns roughly the mass 
ranges from $0.5\text{--}0.6$ to twice the turn-off mass in our models (about 
$\pm0.3\,\mathrm{dex}$ around the turn-off).

\section{Blue straggler stars}\label{sec:blue-stragglers}
The stars in the tail of mass functions are mainly rejuvenated binary products ---
they are classical blue straggler stars.
We define a blue straggler as a star whose mass is larger than that
of the turn-off stars. This definition does not include all blue stragglers
because there are also rejuvenated binary products that are less massive than the turn-off.
In the following sections we characterise the blue straggler stars in the
binary tail and make predictions about their frequencies, binary fractions and ages
as a function of cluster age and compare to observations to test our predictions and
to finally improve our understanding of binary evolution.

Because our stellar evolution code cannot evolve stars more massive than
$100\,\msun$ (Sec.~\ref{sec:bse-code}), boundary effects for ages $\lesssim 3\,\mathrm{Myr}$
are expected. This age range is indicated
by hatched regions in Figs.~\ref{fig:bss-ratios},~\ref{fig:bss-binary-fraction} 
and~\ref{fig:bss-ages}.

\subsection{Expected and observed blue straggler star frequencies}\label{sec:bss-ratios}
Blue straggler stars and their frequencies have been investigated in the past 
using population synthesis calculations including binary stars \citep[e.g.][]{1984MNRAS.211..391C,1994A&A...288..475P,1998A&A...334...21V,2001MNRAS.323..630H,2005MNRAS.363..293H,2009MNRAS.395.1822C,2013ApJ...777..106C}. 
Qualitatively, population synthesis calculations show that binary star evolution
forms stars that appear as blue stragglers in colour--magnitude diagrams through 
binary mass transfer and mergers. Quantitatively however, the picture is more complicated. 
Some models predict blue straggler frequencies 
in agreement with observations \citep[e.g.][]{1984MNRAS.211..391C,1994A&A...288..475P,2001MNRAS.323..630H,2005MNRAS.363..293H,2013ApJ...777..106C}
while others disagree \citep[e.g.][]{2009MNRAS.395.1822C,2013AJ....145....8G}. This problem is not yet fully resolved.

\citet{2007A&A...463..789A} present a catalogue of blue stragglers in Galactic 
open star clusters of various ages. They count the number of blue straggler stars and 
the number of stars down to two magnitudes below the turn-off based on colour--magnitude
diagrams. This is a complicated task because of observational uncertainties 
such as field contamination. In addition, the turn-off and age of the cluster 
are uncertain, in particular for the younger cluster where only a few stars
define the turn-off. 

In our models, the number of blue stragglers, $N_\mathrm{bss}$, is given by the number of stars
more massive than the turn-off mass. The number of stars two magnitudes
below the turn-off, $N_{2}$, is computed from the number of stars with luminosities in the
range $10^{-5/4} L_\mathrm{to}$ to $L_\mathrm{to}$, where $L_\mathrm{to}$ is the
turn-off luminosity (the factor $10^{-5/4}$ corresponds to two magnitudes).

\citet{2013ApJ...777..105S} show that the number of blue straggler stars selected from 
their position in an observed HR diagram following the prescription of \citet{2011MNRAS.415.3771L} 
can be less by a factor of about two than the number of blue stragglers
selected from models based on stellar masses. 
The reason for the difference is that the criteria of \citet{2011MNRAS.415.3771L}
that are used to identify blue stragglers observationally in an HR diagram of a modelled star cluster do not 
cover all blue stragglers produced in that model (cf. Fig.~1 of \citealt{2013ApJ...777..105S}).
\citet{2007A&A...463..789A} identify blue stragglers differently in HR diagrams. They
count stars as blue stragglers that lie between the 
ZAMS and an isochrone appropriate for the cluster. 
It is therefore likely that the number of observationally and theoretically identified blue stragglers
do not differ by a factor of 2 in our case --- we expect the difference to be smaller but can
not rule out systematic differences. \citet{2013ApJ...777..105S} further show
that there is a strong correlation between the number of observationally and theoretically identified
blue stragglers such that the relative number of blue straggler stars as a function of
cluster age may be compared directly.

\begin{figure}
\center
\includegraphics[width=0.46\textwidth]{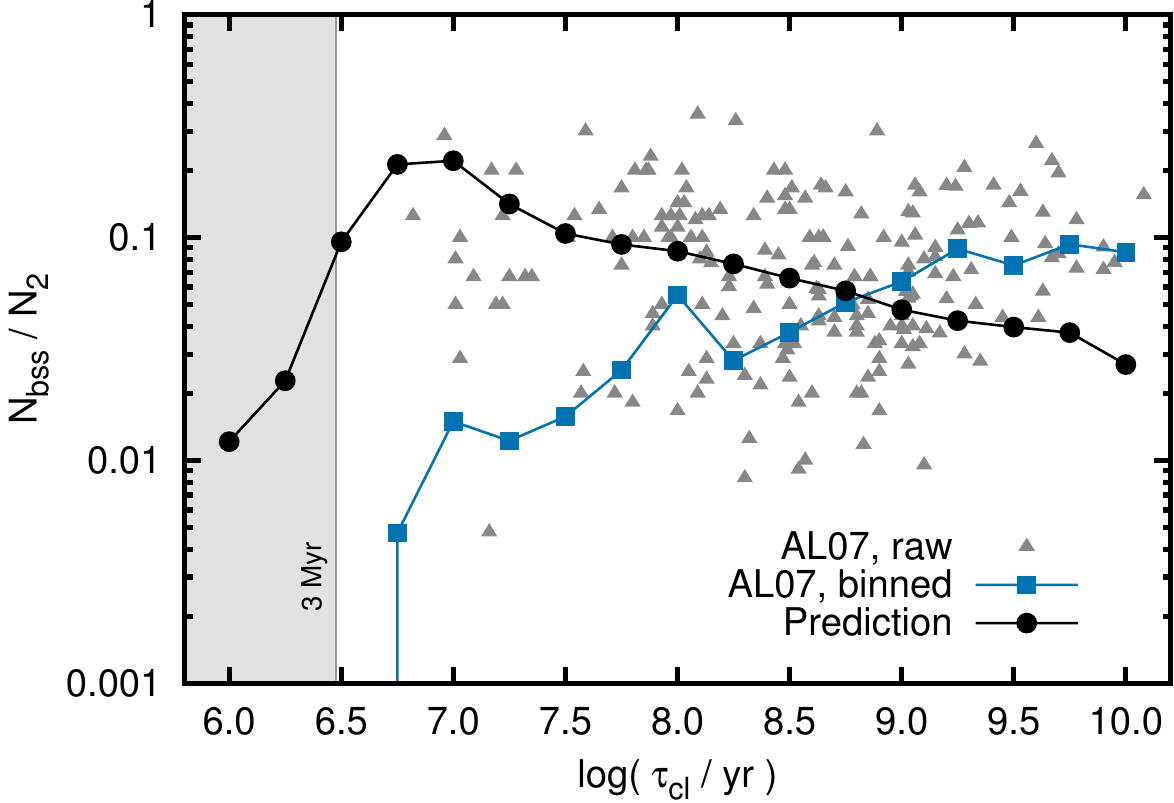}
\caption{Ratio of the number of blue straggler stars, $N_\mathrm{bss}$, to the number 
of stars two magnitudes below the turn-off, $N_2$, as a function of cluster age, $\tau_\mathrm{cl}$.
The raw observational data for each Galactic open cluster of the blue 
straggler catalogue of \citet{2007A&A...463..789A}, AL07, is given by the 
grey triangles, while the same data binned by cluster age is given by
blue squares. Note that no blue straggler is found in $66\%$ of the clusters
younger than $500\,\mathrm{Myr}$. These clusters are not on the logarithmic scale but cause the 
decreasing blue straggler frequency of the binned observational data with younger cluster ages.  
Our model predictions are overlayed in black dots. The grey regions indicate ages for which 
boundary effects play a role.}
\label{fig:bss-ratios}
\end{figure}

From the catalogue of \citet{2007A&A...463..789A}, we put the number of blue 
straggler stars, $N_\mathrm{bss}$, and stars down to two magnitudes below the turn-off, $N_2$, in age bins of 
size $0.25\,\mathrm{dex}$ and plot the ratio $N_\mathrm{bss}/N_2$ as a function of cluster age
in Fig.~\ref{fig:bss-ratios}. 
Additionally, we show the individual data points for every
cluster and our model predictions.
In some clusters, no blue straggler star is found. These clusters are not visible on the 
logarithmic scale used in Fig.~\ref{fig:bss-ratios} but do contribute to the binned data. 

In our models, the ratio $N_\mathrm{bss}/N_2$ decreases in older populations.
Binary evolution is more efficient in producing 
blue stragglers in high mass, young binaries than in low mass, old binaries.
This is because the number of binaries that produce blue straggler stars by MS mergers
and by stable mass transfer, i.e.\ the number of binaries that do \emph{not} go through a common envelope
phase, is larger in high mass than in low mass binaries (cf. Figs.~\ref{fig:phase-space-analysis}
and~\ref{fig:binary-parameter-space-2msun}--\ref{fig:binary-parameter-space-100msun}).
The peak in the predicted ratio $N_\mathrm{bss}/N_2$ around $10\,\mathrm{Myr}$ ($\log \tau_\mathrm{cl}/\mathrm{yr}=7$)
is due to a change in the distribution functions of the initial orbital periods 
from {\"O}pik's law to the results of \citet{2012Sci...337..444S} for O-star binaries (Sec.~\ref{sec:pop-syn}).
The orbital period distribution of \citet{2012Sci...337..444S} favours tight over wide binaries
more than {\"O}pik's law does and thus binaries that form massive blue straggler stars.

A similar decreasing blue straggler ratio $N_\mathrm{bss}/N_2$ in older clusters 
is predicted from binary evolution calculations by \citet{2009MNRAS.395.1822C}
but our ratios are larger by factors of $3$--$5$. \citet{2009MNRAS.395.1822C} 
notice a similar difference (of a factor of at most two) between their predicted blue straggler 
frequencies and those computed with the population synthesis code of \citet{2002MNRAS.329..897H}.
\citet{2009MNRAS.395.1822C} find that the difference is related to
how they treat binaries in counting the number of stars two magnitudes below the turn-off. We count 
each binary as one stellar system because binaries are unresolved in the colour--magnitude diagrams
from which the observed blue straggler frequencies are deduced. We notice
another difference that might explain the discrepancy: we use a maximum initial orbital 
separation of $10^4\,\rsun$ while \citet{2009MNRAS.395.1822C} use $5.75\times 10^6\,\rsun$.
Hence, our models contain more interacting binaries (cf. Fig.~\ref{fig:phase-space-analysis}) that can 
form blue straggler stars. In other words, our effective binary fraction is larger. 

The observations indicate an opposite trend to our models: the ratio $N_\mathrm{bss}/N_2$ 
increases with cluster age. 
There are complications that render a direct comparison between the observed
and our predicted ratio $N_\mathrm{bss}/N_2$ difficult. 
There exist two classes of star clusters in the catalogue of \citet{2007A&A...463..789A}:
clusters with and without blue straggler stars. 
The bi-modality is mainly found in clusters younger than about $500\,\mathrm{Myr}$. 
There is no blue straggler star in $3\%$ of their star clusters older than $500\,\mathrm{Myr}$,
in $38\%$ of star clusters with ages between $100$ and $500\,\mathrm{Myr}$ and
in $80\%$ of star clusters younger than $100\,\mathrm{Myr}$. This bi-modality is not
understood 
but causes the observed decreasing 
blue straggler frequency with younger cluster ages. 

Stochastic sampling (\citealt{2014ApJ...780..117S} and Appendix~\ref{sec:stochastic-sampling}), i.e.\ 
lower number statistics in young clusters compared to 
older clusters because of the IMF, cannot explain this
bi-modality. Stochastic sampling is expected to increase the scatter between the blue
straggler frequencies of clusters of similar ages but not to create
a bi-modal distribution. However, stochastic sampling may
possibly complicate the accurate determination of the turn-off.

Supernova kicks \citep[e.g.][]{2001LNP...578..424L} might be part of the solution 
to this discrepancy. The binary companion of a blue straggler star that 
formed by RLOF, i.e.\ the former mass donor, will explode if it is massive enough
to undergo core collapse. The supernova can break up the binary such that
the blue straggler leaves the cluster as a runaway star, thereby reducing the number
of blue stragglers observed in the cluster. Some clusters might lose all their blue 
stragglers in this way, giving rise to the observed bi-modality. The observed bi-modality 
is found in star clusters younger than $500\,\mathrm{Myr}$. However, supernova explosions 
occur only in young star clusters ($\lesssim 40\text{--}50\,\mathrm{Myr}$) 
with massive stars ($\gtrsim 7\text{--}8\,\msun$). Supernova kicks can therefore 
explain only part of the discrepancy.

The primordial binary fraction in
our models is $100\%$. A smaller binary fraction linearly decreases the number 
of blue stragglers and hence the ratio $N_\mathrm{bss}/N_2$. Our predictions are therefore likely
upper limits, implying that our models underpredict the observed blue straggler frequencies in 
$\gtrsim100\,\mathrm{Myr}$ star clusters and that additional channels for the formation
of blue stragglers are required. \citet{2011Natur.478..356G} find carbon-oxygen white dwarfs as companions
to blue stragglers in NGC~188. The white dwarf mass and period distributions appear to be consistent 
with a Case~C RLOF formation scenario in which an asymptotic giant branch (AGB) star transfers mass to a MS star.
In our models, RLOF from giants typically leads to common envelope
evolution and thus not to the formation of blue stragglers. However,
\citet{2008MNRAS.387.1416C} show that Case~B and~C mass transfer from 
giants to MS stars can form blue stragglers.
Dynamical cluster evolution also produces blue stragglers in
stellar collisions \citep[e.g.][]{2013AJ....145....8G} and is expected to be 
more efficient the higher the density of the star cluster \citep[e.g.][]{2013ApJ...777..106C}. 
Another form of mass transfer, wind RLOF from AGB stars, 
might be efficient enough to also form blue straggler stars \citep[e.g.][]{2007ASPC..372..397M,2013A&A...552A..26A}.
All three contributions, the Case~B and~C formation scenarios, mergers due to collisions 
and wind RLOF, are missing in our predictions
and might potentially help explaining the too low BSS predictions in $\gtrsim100\,\mathrm{Myr}$ 
star clusters \citep[see also][]{2013AJ....145....8G}.

\subsection{Binary fraction of blue straggler stars}\label{sec:bss-binary-fraction}
A further testable prediction from our models is the binary fraction among
blue straggler stars. The blue stragglers that form from stable mass transfer by RLOF
can potentially be observed as binary stars whereas those that 
form from stellar mergers are observed as single stars.
It is, however, difficult to observationally find the companions of blue stragglers that formed by stable 
mass transfer because the companions can be much fainter than the blue straggler (or even be a compact object)
and the binary orbits are often wide such that most searches for radial velocity 
variations in blue stragglers cannot detect them \citep[e.g.][]{1984MNRAS.211..391C,1994A&A...288..475P}.
The blue straggler binary might even be disrupted if the companion star explodes.
In general, post-interaction binaries can often not be identified as such from 
radial velocity variations and mostly appear to be single stars \citep{2014ApJ...782....7D}.

In our models, the binary fraction of blue straggler stars is independent of the primordial
binary fraction in star clusters because only the sub-sample of
primordial binaries can produce blue stragglers. The binary fraction among 
blue straggler stars is thus purely determined through binary evolution.

\begin{figure}
\center
\includegraphics[width=0.46\textwidth]{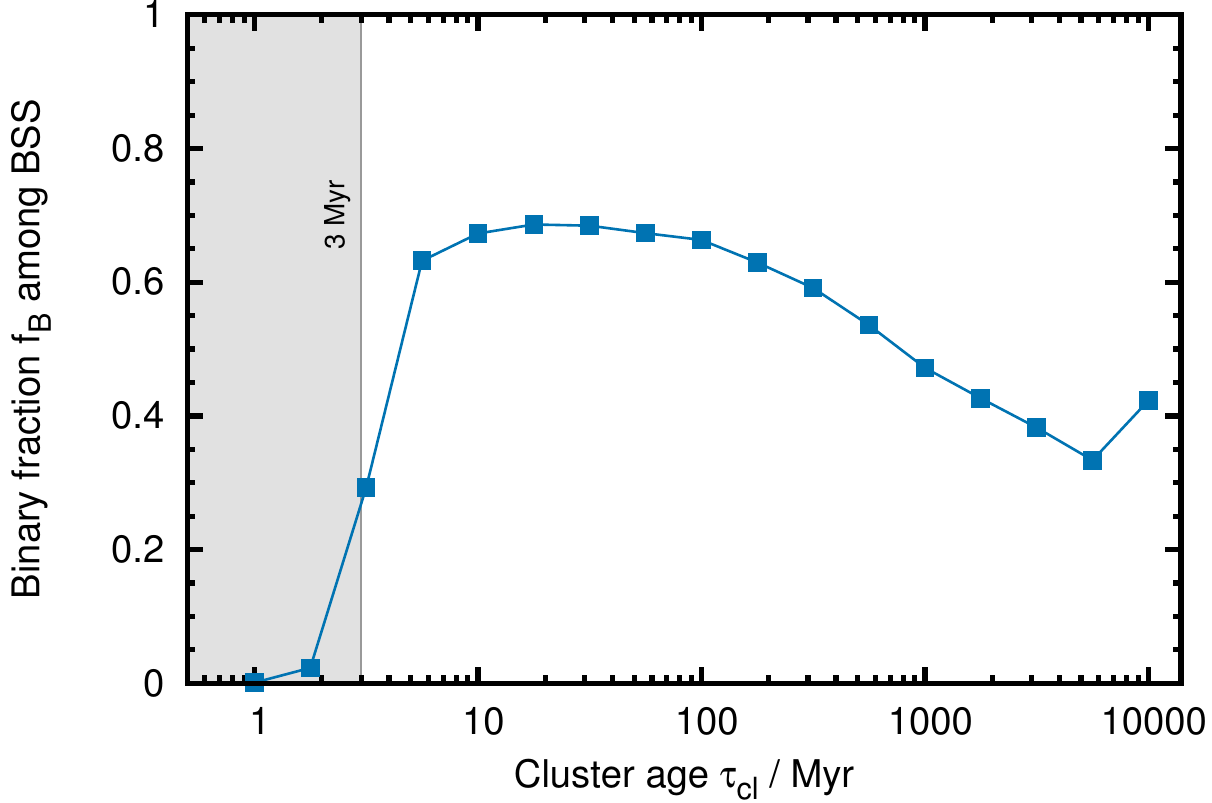}
\caption{Binary fraction among blue straggler stars as a function of cluster age, 
$\tau_\mathrm{cl}$, corresponding to stars in the tail of our PDMFs.
The given binary fractions are upper limits because we neglect 
supernova kicks that may disrupt binaries. The grey regions indicate ages for which boundary effects play a role.}
\label{fig:bss-binary-fraction}
\end{figure}

In Fig.~\ref{fig:bss-binary-fraction} we show the binary fraction $f_\mathrm{B}$ of blue 
straggler and hence tail stars in our PDMFs as a function of
cluster age. In our simulations we assume that all binaries with a mass ratio less than $0.56$ 
merge at the onset of RLOF during Case~A mass transfer (Sec.~\ref{sec:bse-code}), 
i.e.\ mass transfer from a MS star. Because of the assumed flat mass ratio distribution, at least $56\%$
of all Case~A binaries merge and are thus observed as single stars ($f_\mathrm{B}\leq44\%$). 
Our analysis of the binary parameter space 
(Figs.~\ref{fig:binary-parameter-space-2msun}--\ref{fig:binary-parameter-space-100msun})
reveals that more than $56\%$ of all binaries merge during Case~A mass transfer
because even binaries with a mass ratio greater than $0.56$ come into contact if 
they are initially in close orbits \citep[e.g.][]{2001A&A...369..939W}.
In binaries younger than $2\,\mathrm{Myr}$, only the primary stars of very 
short orbit binaries overfill their Roche lobes. 
Most of them, regardless of the mass ratio, merge, leading to binary fractions $\leq10\%$.
Later, the binary fraction increases with time because also binaries in initially wider
orbits interact, producing blue straggler stars by stable Case~A and~B mass transfer without merging,
i.e.\ leaving behind a blue straggler in a binary. 

Case~A mass transfer predominantly forms blue stragglers by merging in more than 
$56\%$ of all Case~A binaries in our simulations. Contrarily, Case~B mass transfer purely creates blue 
straggler star binaries (neglecting binary disruption by supernova explosions). 
The binary fraction in Fig.~\ref{fig:bss-binary-fraction} 
therefore reaches a maximum of $\sim60$--$70\%$ and stays at this level 
after all Case~B binaries have enough time to interact to contribute blue straggler binaries. 

In our models, Case~B mass transfer in binaries with primary stars 
more massive than $5\,\msun$ leads to fewer common envelope phases 
and consequently more blue stragglers than in binaries with less massive 
primary stars (cf. Case~B regions of binaries with $2\,\msun$ and 
$5\,\msun$ primary stars in Figs.~\ref{fig:binary-parameter-space-2msun} 
and~\ref{fig:binary-parameter-space-5msun}, respectively).
The binary fraction of blue straggler stars therefore decreases in populations 
older than the lifetime of $5\,\msun$ stars, 
i.e.\ older than $\sim 100\,\mathrm{Myr}$, and is given by the fraction of stars that merge 
during Case~A (and early Case~B) mass transfer.

Our blue straggler star binary fractions are upper limits because
we neglect supernova kicks that might disrupt binaries and because the companions 
might be hard to detect observationally. We exclude supernova kicks because they
introduce a random process that can only be fully taken into account by
many repeated calculations which is impractical in our approach.

\subsection{Apparent ages of blue straggler stars}\label{sec:bss-ages}
Mass gainers and mergers appear to be younger than other cluster members because they are
rejuvenated by mass accretion (Sec.~\ref{sec:bse-code}). The age of a cluster is given by the MS
lifetime of the turn-off stars, $\tau_\mathrm{MS}(M_\mathrm{to})\propto M_\mathrm{to}^{1-x}$,
where $x$ is the exponent of the ML relation, $L\propto M^x$. The most massive 
blue straggler stars have a mass of about twice the mass of the turn-off stars
in our models. Their apparent age is thus a factor 
$\tau_\mathrm{MS}(2M_\mathrm{to})/\tau_\mathrm{MS}(M_\mathrm{to})=2^{1-x}$
smaller than their true age if we neglect mixing of fresh fuel into convective 
cores which makes stars look even younger than the analytic approximation given here. The ML
relation of MS stars is flatter, i.e.\ has a smaller exponent $x$, at high 
masses. The exponent approaches $x=1$ in stars close to the Eddington 
limit and is as large as $x=4$--$5$ in low mass stars ($\sim1$--$2\,\msun$).
The most massive blue straggler with exponent $x=4$, i.e.\ a blue straggler of 
lower mass in older clusters, appears to be rejuvenated by a factor $1/8$ (the cluster
appears eight times as old as the blue straggler). Contrarily, the most 
massive blue straggler with an exponent $x=2$, i.e.\ a blue straggler of higher 
mass in younger clusters, appears to be rejuvenated by only a factor of $1/2$ 
(the cluster appears twice as old as the blue straggler).

\begin{figure}
\center
\includegraphics[width=0.46\textwidth]{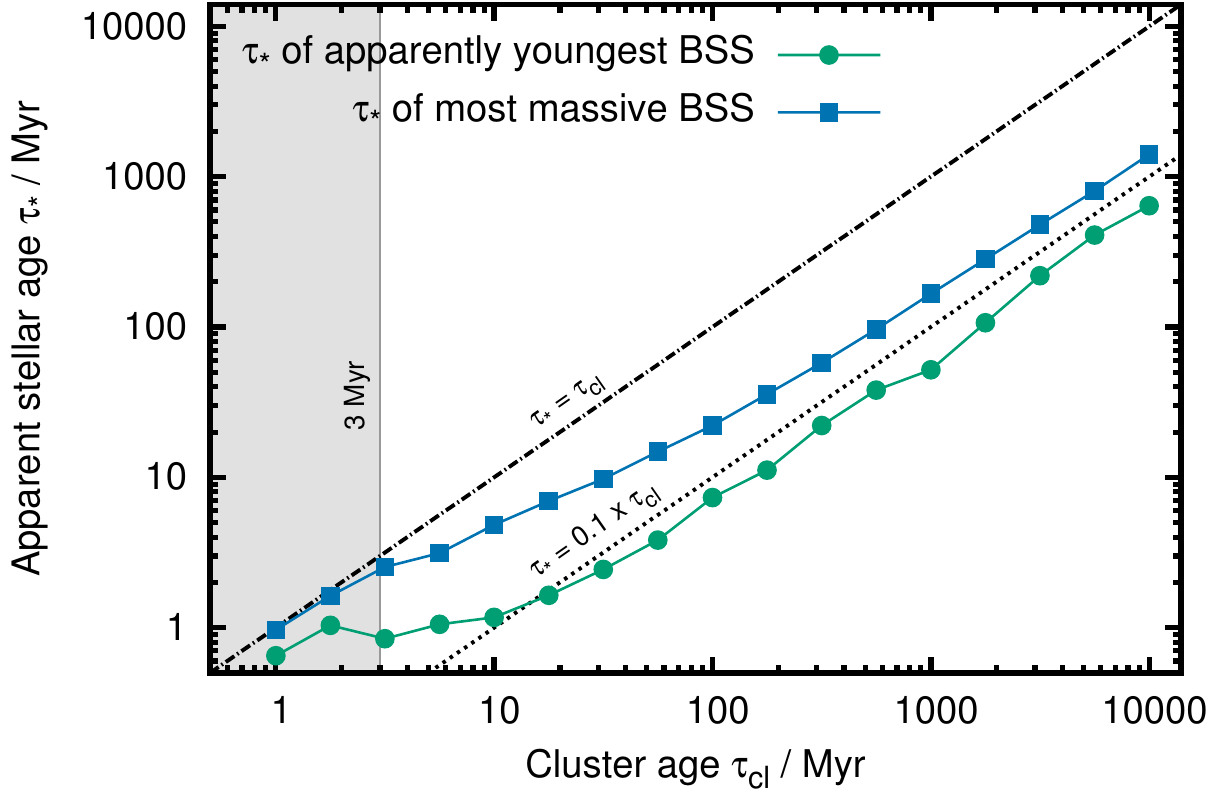}
\caption{Apparent stellar age, $\tau_{*}$, of the most massive and the apparently youngest blue straggler star 
as a function of cluster age, $\tau_\mathrm{cl}$. The grey regions indicate ages for which 
boundary effects play a role.}
\label{fig:bss-ages}
\end{figure}

In Fig.~\ref{fig:bss-ages} we show the apparent stellar age, $\tau_{*}$, of the most massive and of the 
apparently youngest blue straggler star in our simulations as a function of cluster age, $\tau_\mathrm{cl}$.
The above mentioned trend for the most massive blue straggler is recovered, i.e.\
the most massive blue straggler in young star clusters appears to be less rejuvenated
than in old ones: in a $10\,\mathrm{Myr}$ star cluster, the most massive blue straggler
appears to be younger than the cluster by a factor of $0.48$ while this factor is smaller,
only $0.17$, at $1\,\mathrm{Gyr}$.

The most massive blue straggler is not the one that appears youngest (Fig.~\ref{fig:bss-ages}). 
The apparently youngest blue straggler forms from accretion onto a relatively 
unevolved star which then becomes a blue straggler. In our models, the apparently 
youngest blue straggler is about $40\%$ more massive than the turn-off and forms
from late Case~A/early Case~B mass transfer in a binary with a mass ratio $0.5$--$0.6$. The mass ratio
ensures that the secondary star is relatively unevolved. 
Mass accretion then brings the secondary mass above the mass of
the turn-off, forming a blue straggler close to the ZAMS, i.e.\ apparently very young --- the
apparent age is of the order of $7\%$ of the cluster age (cf. Fig.~\ref{fig:bss-ages}).

The apparent age of the most massive blue straggler depends on how massive blue stragglers
can get through mergers and stable RLOF, and on how much fresh fuel is mixed
into convective cores during these processes. The apparent age of the most massive 
blue straggler is younger the more massive the star gets and the more mixing occurs.
The apparent age of the seemingly 
youngest blue straggler depends on the choice of the critical mass ratios, $q_\mathrm{crit}$, 
and the mass transfer efficiency because these parameters
determine how small the initial mass ratio can be in order to accrete enough mass to form 
a blue straggler star. The smaller the mass ratio, the less evolved is the progenitor 
of the blue straggler and hence the younger the blue straggler appears.
Note that the mass ratio can also not be too small: there has to be enough transferred mass
such that the secondary mass can exceed the turn-off mass.

\section{Determination of star cluster ages}\label{sec:determination-cluster-age}

As shown in Sec.~\ref{sec:bss-ages}, a stellar population can be 10 times
older than some of its most massive stars appear to be. Such
rejuvenated stars can bias the age determination of stellar populations
because it is not always clear whether a star was influenced 
by binary mass transfer in the past or not. Especially in young clusters with OB stars, 
the most luminous and hence most massive
stars are often investigated in detail because they are easiest to observe.
Their ages are then sometimes interpreted as the cluster age although
the most massive stars are likely rejuvenated binary products.

To avoid potential biases and confusion of stars with rejuvenated binary 
products such as stellar mergers,
stars at least half the mass of the most massive cluster members should 
be used to determine ages of stellar populations. This rule of thumb is 
implicitly taken into account when dismissing blue stragglers from fitting 
the turn-off of well-populated star clusters in colour--magnitude or HR diagrams.

The turn-off may be blurred by binary products and it is sometimes, 
especially in young star clusters, challenging to determine its location.
In such cases, the mass function may provide a promising alternative because
possibly rejuvenated binary products can be identified by the binary tail and
the age of the stellar population can be directly determined by reading-off 
the turn-off mass.

First, we consider single star mass functions to clarify the general approach 
to determine cluster ages from mass functions.
Single star mass functions are truncated at the present-day turn off mass, $M_{{\rm to,p}}$,
because of finite stellar lifetimes. The present-day and initial mass of the turn off stars
in old stellar populations are the same because stars have negligible stellar winds.
The age of old populations therefore follows directly from the MS age
of stars with an initial mass of that of the turn-off. In young stellar populations, 
stars lose mass by stellar winds and a peak forms in the mass function.
The present-day mass of the turn off stars that is read-off from the truncation 
of the mass function is no longer equal to their initial mass 
because of stellar winds. However, without further modeling or the need of stellar models, we can correct
for wind mass loss by redistributing the number of excess stars in the wind mass loss peak, $N_1$, 
such that the mass function is filled up to the initial mass of the turn off stars, 
$M_{{\rm to,i}}$ (cf. Fig.~\ref{fig:pdmf-bump-imf-z} and Sec.~\ref{sec:single-star-populations}).
Rewriting the mass lost by stellar winds on the MS, 
$\Delta M = M_\mathrm{to,i}-M_\mathrm{to,p}$, in Eq.~(\ref{eq:excess-stars-def}), 
we have for the number of excess stars in the wind mass loss peak,
\begin{equation}
N_{1}=\frac{A}{\Gamma}\left(M_{{\rm to,i}}^{\Gamma}-M_{{\rm to,p}}^{\Gamma}\right)\,,
\label{eq:excess-stars-def2}
\end{equation}
from which we find the initial mass of the turn-off stars, 
\begin{equation}
M_{{\rm to,i}}=\left(\frac{N_{1}\Gamma}{A}+M_{{\rm to,p}}^{\Gamma}\right)^{1/\Gamma},\label{eq:turn-off-mass}
\end{equation}
in the cluster. The age of the star cluster follows from the MS lifetime
of the turn off stars with initial mass $M_{{\rm to,i}}$.

The mass function of binary stars allows us to determine the
age of the stellar population in a similar way. The difference is that 
the binary star mass functions are not truncated at the turn-off mass but 
rather at about twice this mass. In practice the mass function may be truncated at less than twice the turn-off mass
because of stochastic sampling. Consequently, another indicator of the turn-off mass
than the truncation of the mass function is
required to determine the age of the stellar population. We use the onset of
the binary tail for that purpose. In old stellar populations, 
the onset of the binary tail, i.e.\ the turn-off mass, is indicated by a steep 
decrease in the number of stars. In young stellar populations, the number of stars in 
our models do not change that strongly at the onset of the binary
tail. In such cases, we propose the use of the wind mass 
loss peak to indicate the onset of the binary tail and hence to determine the turn-off 
mass and the age of the stellar population. The number of excess stars
in the peak can be influenced by unresolved binaries and binary evolution 
(Sec.~\ref{sec:binary-star-populations}). The correction
for stellar wind mass loss to derive the initial mass of the turn off stars 
may therefore not be possible without the use of stellar models that provide the mapping of
masses at the end of the MS to initial masses.

The advantage of this over other methods to determine stellar ages is that it is not biased by 
apparently younger, rejuvenated binary products. Instead, the proposed method identifies and explicitly uses
rejuvenated binary products to determine the turn-off mass and hence to derive the cluster age.

This new method is used by \citet{2014ApJ...780..117S} to determine the ages of the young Arches 
and Quintuplet star clusters in the Galactic centre. The mass functions of Arches and Quintuplet show
a wind mass loss peak from which it is possible to identify the binary tail and derive an unambiguous age.
The age from the wind mass loss peak results in older cluster ages 
than previously derived for these clusters from the most luminous stars. The most luminous stars
belong to the binary tail of the mass function and are thus likely rejuvenated binary products.

\section{Conclusions}\label{sec:conclusions}

We use a rapid binary evolution code
to investigate how single and binary star evolution shape PDMFs 
with time. To that end, we set up coeval populations of single and binary stars,
follow their evolution in time and construct mass functions at
ages ranging from $\mathrm{Myr}$ up to $\mathrm{Gyr}$. Our code incorporates 
all the relevant single and binary star physics that directly alters stellar masses ---
wind mass loss, mass transfer in binaries by RLOF 
and by winds, stellar mergers and rejuvenation of stars. 

Finite stellar lifetimes truncate the mass functions and
wind mass loss results in an accumulation of stars in the PDMF creating
a peak at the high mass end. The magnitude of the peak depends on the
strength of stellar winds, i.e.\ on stellar mass and metallicity,
and the slope of the IMF: the flatter the IMF the stronger the peak.
The peak can thus be used to constrain stellar wind mass loss.
We investigate the age and mass ranges of stellar populations for which
we expect a wind mass loss peak in their mass functions. Typically,
the peak is present in stellar populations younger than about 
$10\,\mathrm{Myr}$ at $Z=0.02$ corresponding to stars initially more massive
than about $18\,\msun$ (Figs.~\ref{fig:pdmf-bump-imf-z} and~\ref{fig:visibility-mdot-peak}).
Less massive stars have too weak stellar winds.
The peak flattens the PDMF slopes at the high mass end by up to $60\%$ 
in $5\,\mathrm{Myr}$ old stellar populations with a Salpeter IMF
(IMF slope $\Gamma=-1.35$; Fig.~\ref{fig:slopes-starburst-singlestars-datapoints-3}).

Binary interaction, i.e.\ mass transfer and stellar mergers,
reshape the PDMF at the high mass end, forming a tail which extends 
the PDMF of single stars by a factor of 2 in mass. 
The PDMF tail consists of rejuvenated binary products that are not expected to exist from
single star evolution and are better known as blue stragglers.
Binary interactions are more efficient in producing the binary tail 
at high masses (Sec.~\ref{sec:binary-parameter-space}).
The number of rejuvenated binary products in young star clusters ($\sim 5\,{\rm Myr}$)
reaches more than $30\%$ of the initial number of stars in the mass range corresponding
to the tail (Figs.~\ref{fig:mf-ratio-allMS} and~\ref{fig:mf-ratio-obs}).
Binary interactions are therefore efficient in repopulating the high
mass end of PDMFs even in star clusters that are only a couple of million years old.

Unresolved binaries flatten the slope of the PDMF. The slope is flattened by about
$0.1$ in our zero-age populations, which is in agreement with previous work on unresolved multiple systems. 
Altogether, stellar winds, mass exchange in binary systems and unresolved binaries reshape
the high mass end of mass functions within $\pm0.3\,\mathrm{dex}$ of the turn-off mass 
and can therefore bias IMF determinations.

We compare our predicted blue straggler frequencies to those from the 
blue straggler catalogue of Galactic open star clusters
\citep{2007A&A...463..789A}. Our models predict a decreasing blue straggler
frequencies with increasing cluster age --- in contrast to the observations.
The observed blue straggler frequency drops with younger ages because there 
are no blue stragglers identified in two-thirds 
of the open clusters younger than $500\,\mathrm{Myr}$ but only in $3\%$ of 
clusters older than that. 
Our models predict about the right amount of blue stragglers in young clusters 
($\lesssim100\,\mathrm{Myr}$) but too few in older clusters. Additional 
blue straggler formation channels, such as mergers resulting from stellar collisions 
and wind Roche-lobe overflow from AGB stars, are likely required to explain the observed 
frequency of blue stragglers in older clusters. Also, the treatment of Case~B and~C 
mass transfer may need revision in our models to allow for the formation of blue stragglers from 
RLOF of giants to MS stars.

The binary fraction among the blue straggler stars in the tail of our PDMFs
varies between $40\%$ and $70\%$, depending on cluster age.
It is largest in young ($\sim 10\,\mathrm{Myr}$)
and smallest in old stellar populations ($\sim 5000\,\mathrm{Myr}$).
Our binary fractions are upper limits because we neglect supernova kicks that may disrupt binaries.

The most massive blue stragglers in the binary tail have apparent ages that are younger by factors of
$0.17$--$0.48$ than the real cluster age, i.e.\ cluster ages inferred
from the most massive stars would be too young by factors of $2$--$6$.
Some of the less massive blue stragglers may show apparent ages that are even younger by a
factor of 10.
Cluster ages derived from the most luminous stars should therefore be treated with caution
because of a likely confusion with rejuvenated binary products. 
If in doubt, and to avoid potential confusion, cluster ages should not be based on the most luminous cluster members.
Instead, we propose the use of mass functions to identify rejuvenated binary products and
to derive cluster ages.
This is possible because the turn-off mass and hence the cluster age can be directly determined 
from mass functions.

In old stellar populations, the turn-off mass
is best indicated by a steep decrease of the number of stars at the onset of
the binary tail. In younger populations, this decrease of the number of stars is less and 
the wind mass loss peak may be used instead to determine the turn-off mass. 
Stellar mass functions and especially the wind mass loss peak constitute a new and unambiguous clock 
to age-date star clusters. This technique is applied by \citet{2014ApJ...780..117S}
to the Arches and Quintuplet star clusters to determine unambiguous cluster ages
by identifying likely binary products and thereby resolved the apparent age discrepancies 
among the most luminous members in both clusters.

The binary products in the tail of young PDMFs have potentially far-reaching consequences. So far, they have 
mostly been neglected when computing the feedback from stellar populations. 
However, the stars in the tail are the most massive stars in a stellar population and
can therefore contribute significantly to the ionising radiation, the mechanical 
feedback from stellar winds and supernovae explosions and to the chemical enrichment.
In the youngest star clusters, binary products might even become so massive that they
explode as pair-instability supernovae, thereby contributing significantly to the 
metal production in the Universe \citep{2002ApJ...567..532H,2009Natur.462..624G,2012ARA&A..50..107L}.

\begin{acknowledgements}
F.R.N.S. acknowledges the fellowships awarded by the German National Academic Foundation (Studienstiftung)
and the Bonn-Cologne Graduate School of Physics and Astronomy.
R.G.I. would like to thank the Alexander von Humboldt foundation.
S.d.M. acknowledges support by the Einstein Fellowship program 
through grant PF3-140105  awarded by the Chandra X-ray Center, which is 
operated by the Smithsonian Astrophysical Observatory for NASA under the contract NAS8-03060.
\end{acknowledgements}

\appendix

\section{Binary parameter space continued}\label{sec:appendix-binary-parameter-space-cont}

In Sec.~\ref{sec:binary-parameter-space} we describe how
much mass is transferred and accreted in our binary models with $10\,\msun$
primary stars to understand quantitatively
how binary evolution shapes the high mass end of PDMFs.
Here, we continue this description by providing figures equivalent to
Figs.~\ref{fig:m1-10-beta-no-op}, \ref{fig:m1-10-dMtrans-nop-tot}
and \ref{fig:m1-10-dMacc-nop} but for primary masses of $2$, $5$, $20$, $50$, $70$ and $100\,\msun$ 
(Figs.~\ref{fig:binary-parameter-space-5msun}--\ref{fig:binary-parameter-space-100msun}, respectively).
These analyses enable us to fully understand the quantitative results 
presented in this paper. The top panels (a) contain the mass transfer
efficiency $\beta$ as defined in Eq.~\eqref{eq:beta-code-practice}, the
middle panels (b) the mass transferred from the primary to the secondary
stars during stable RLOF and the bottom panels (c) the mass accreted by
the secondary stars during stable RLOF.

\clearpage

\begin{figure}
\center
$\mathbf{M_1=2\,\msun}$
\includegraphics[width=0.46\textwidth]{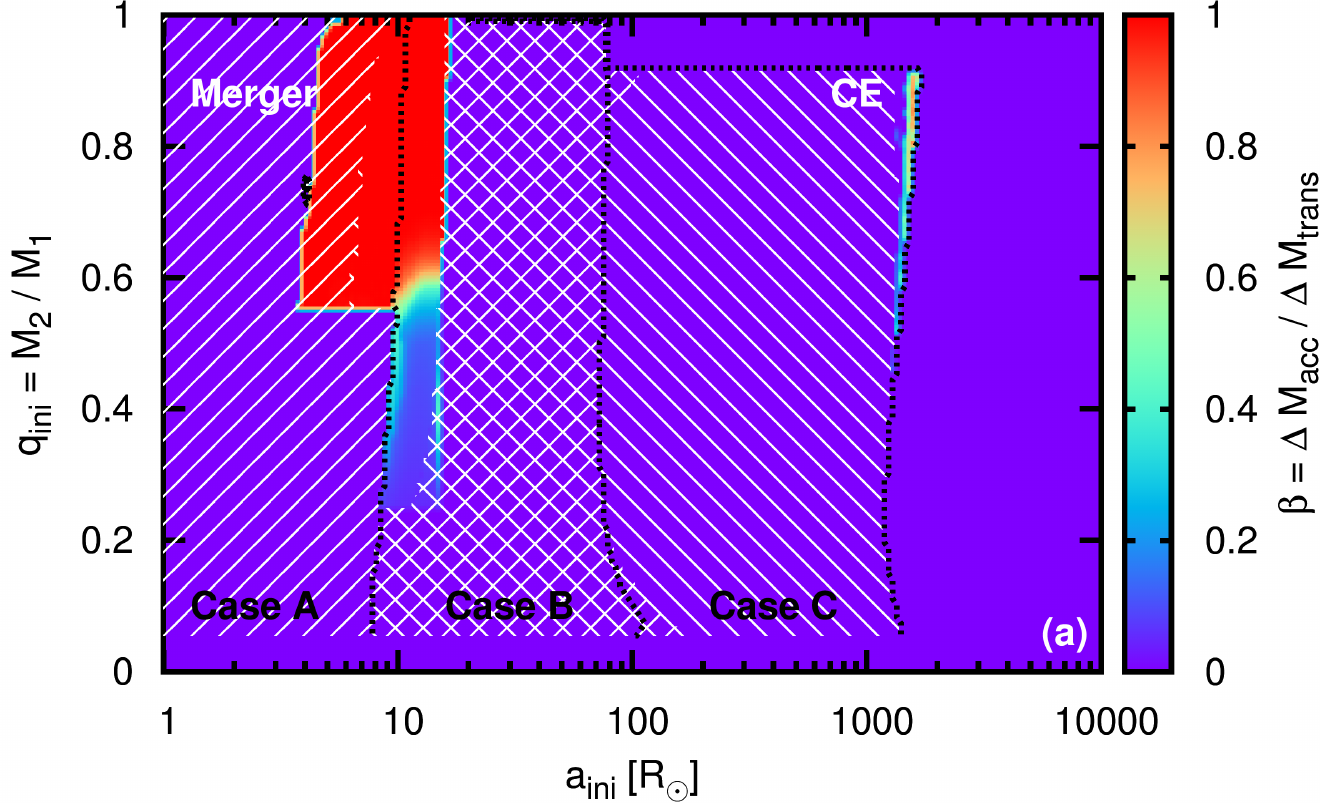}
\includegraphics[width=0.46\textwidth]{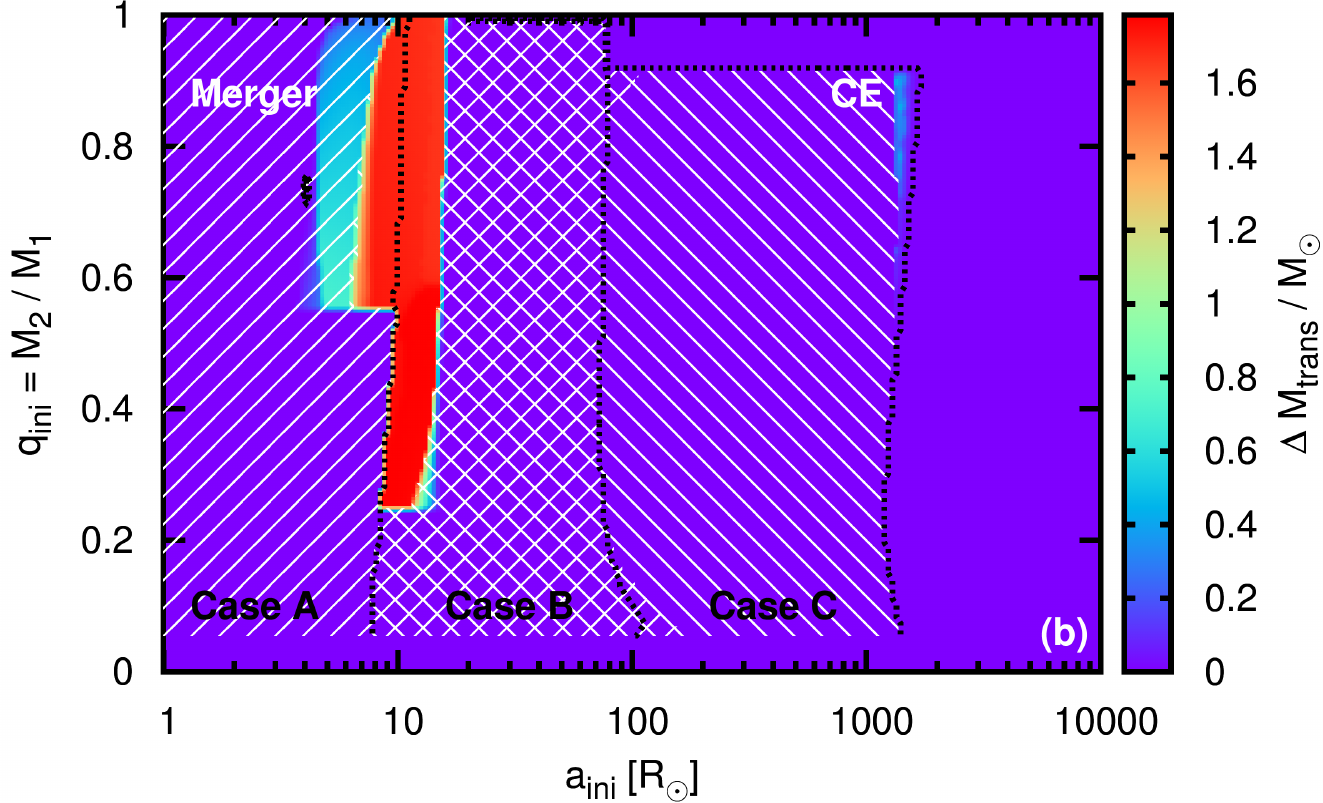}
\includegraphics[width=0.46\textwidth]{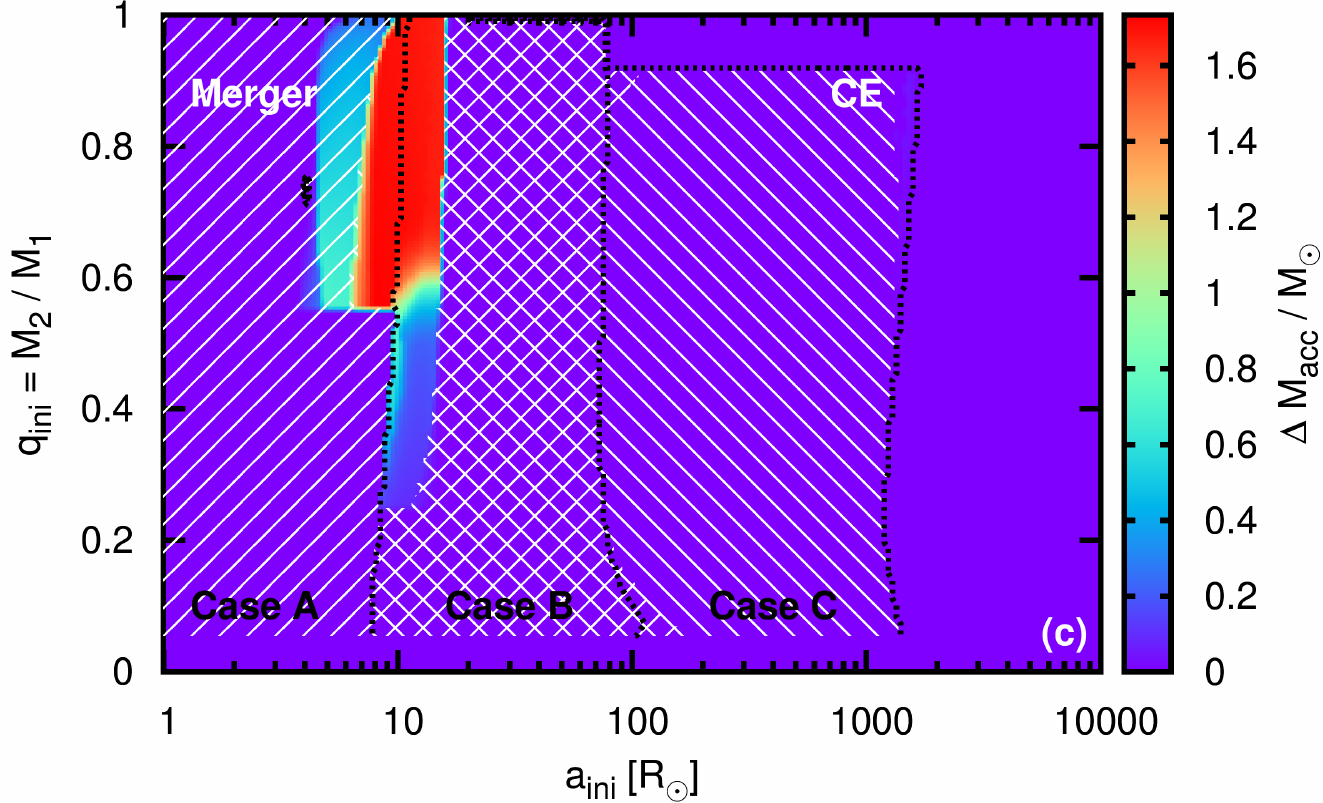}
\caption{Mass transfer efficiency $\beta$ 
(top panel (a)), transferred mass from primary to secondary during stable
RLOF (middle panel (b)) and mass accreted by secondary star during stable
RLOF (bottom panel (c)) as functions of the initial mass ratio $q_\mathrm{ini}$ (i.e.\ initial
secondary mass) and initial orbital separation $a_{\mathrm{ini}}$
for $2\,\msun$ primary stars. The shaded regions have the same meaning
as in Fig.~\ref{fig:m1-10-beta-no-op} and indicate binaries which merge and/or go through a
common envelope phase.}
\label{fig:binary-parameter-space-2msun}
\end{figure}

\begin{figure}
\center
$\mathbf{M_1=5\,\msun}$
\includegraphics[width=0.46\textwidth]{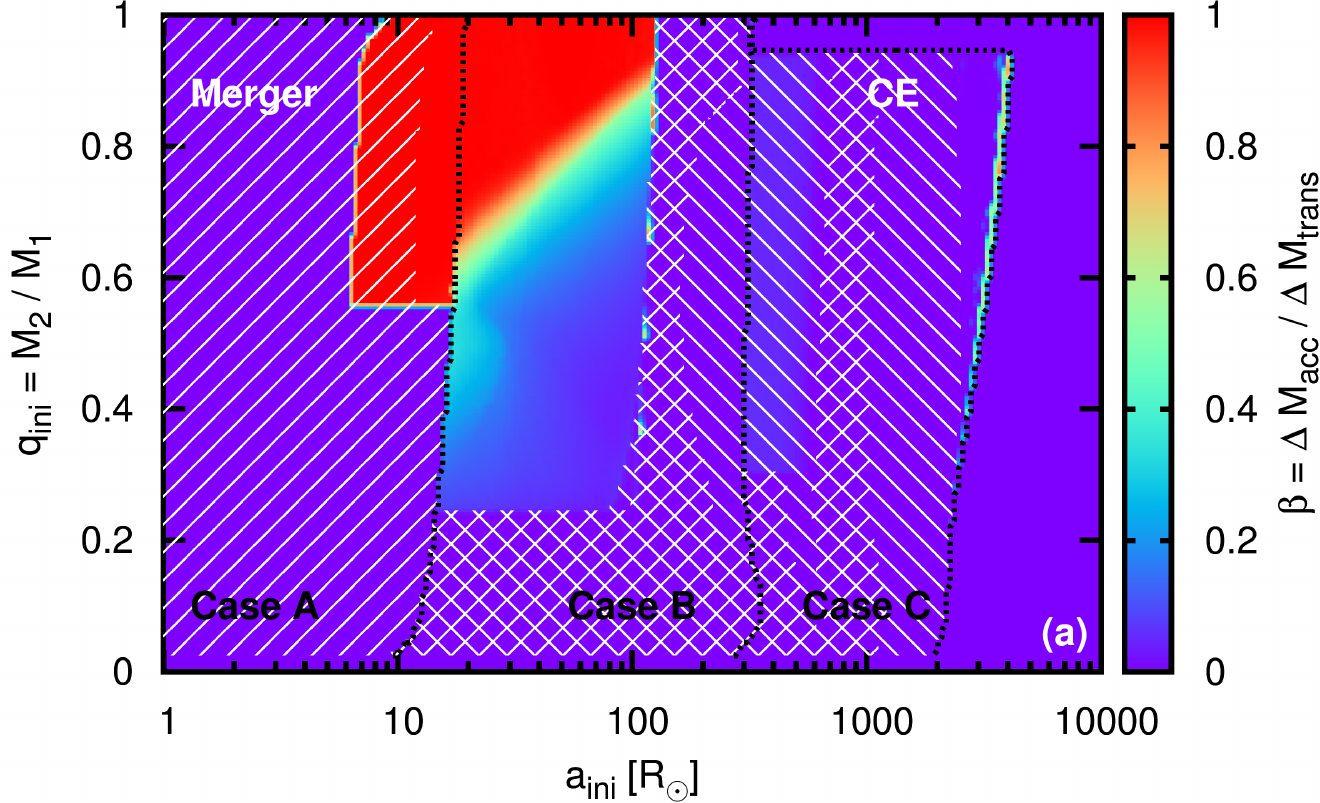}
\includegraphics[width=0.46\textwidth]{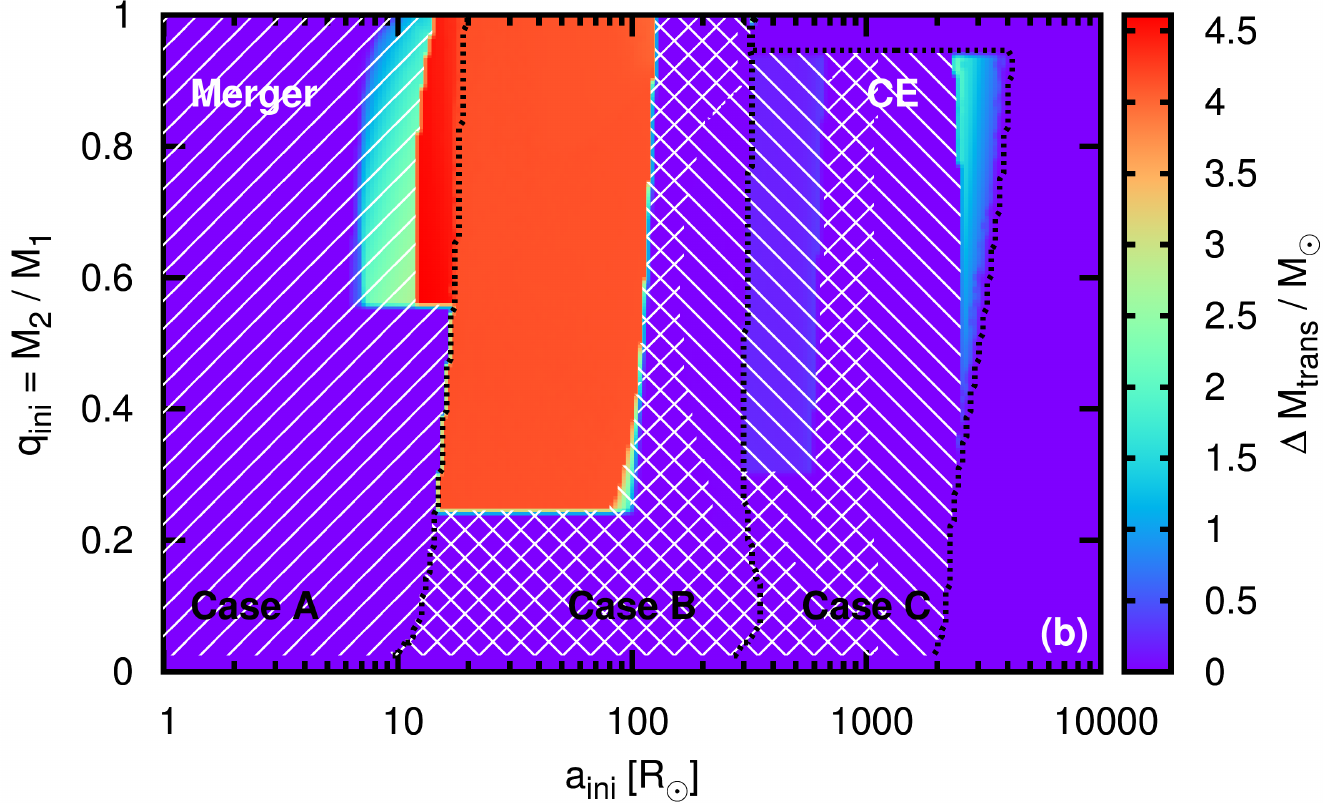}
\includegraphics[width=0.46\textwidth]{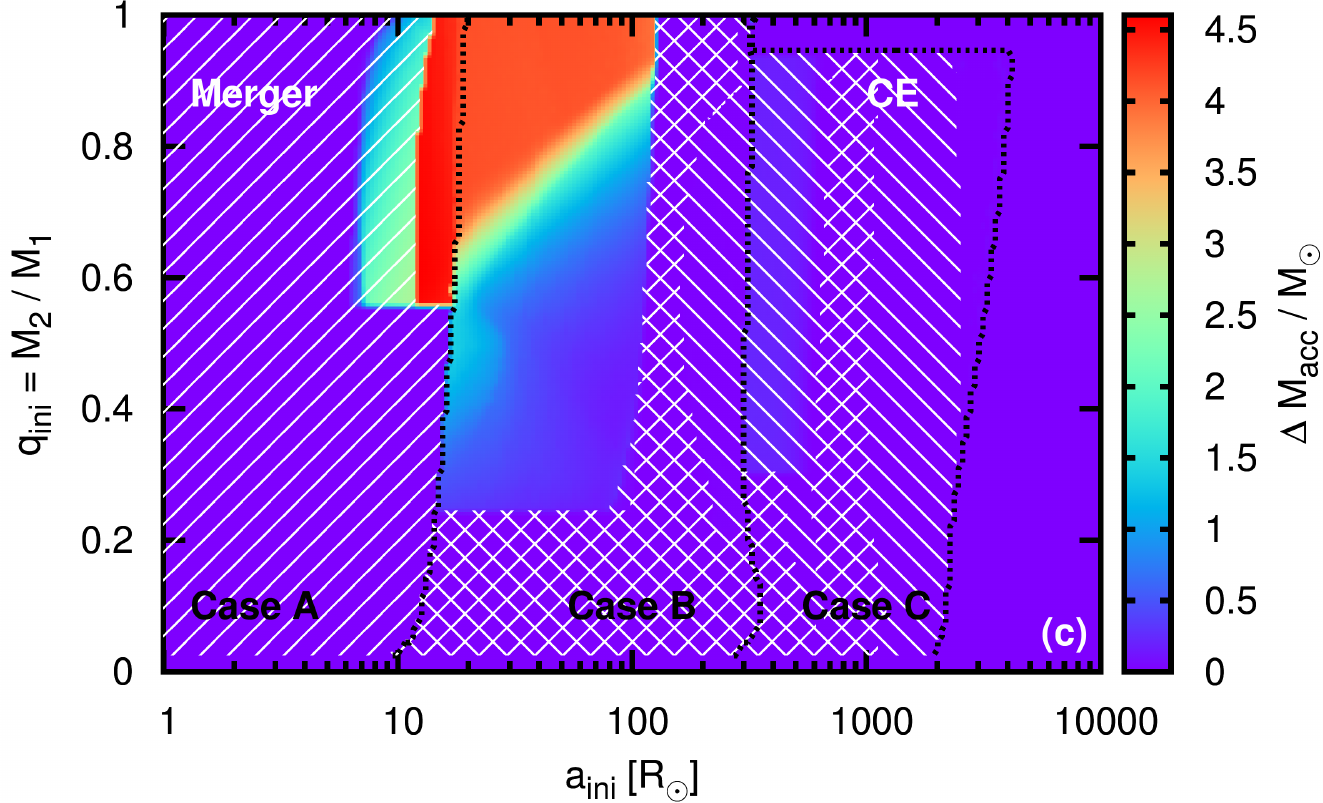}
\caption{As Fig.~\ref{fig:binary-parameter-space-2msun} but for $5\,\msun$
primary stars.}
\label{fig:binary-parameter-space-5msun}
\end{figure}

\clearpage

\begin{figure}
\center
$\mathbf{M_1=20\,\msun}$
\includegraphics[width=0.46\textwidth]{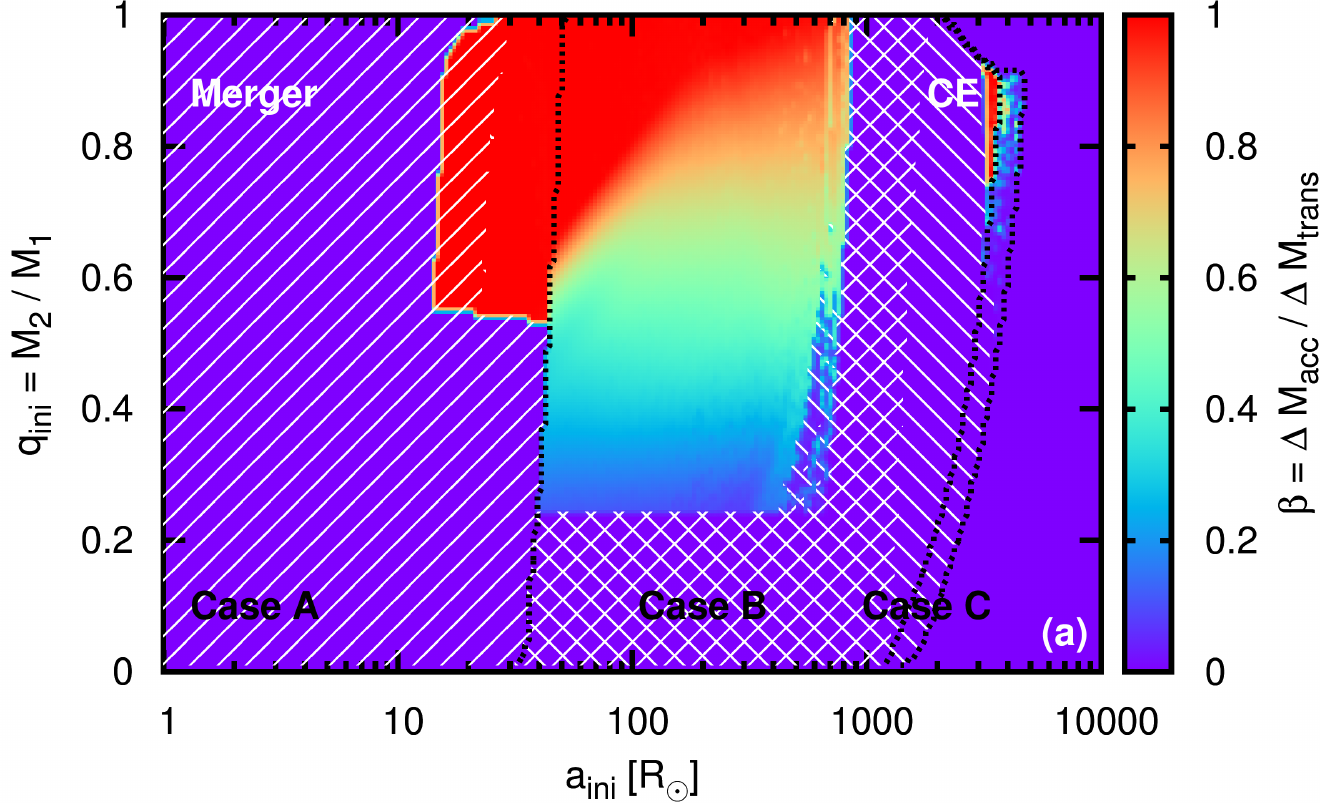}
\includegraphics[width=0.46\textwidth]{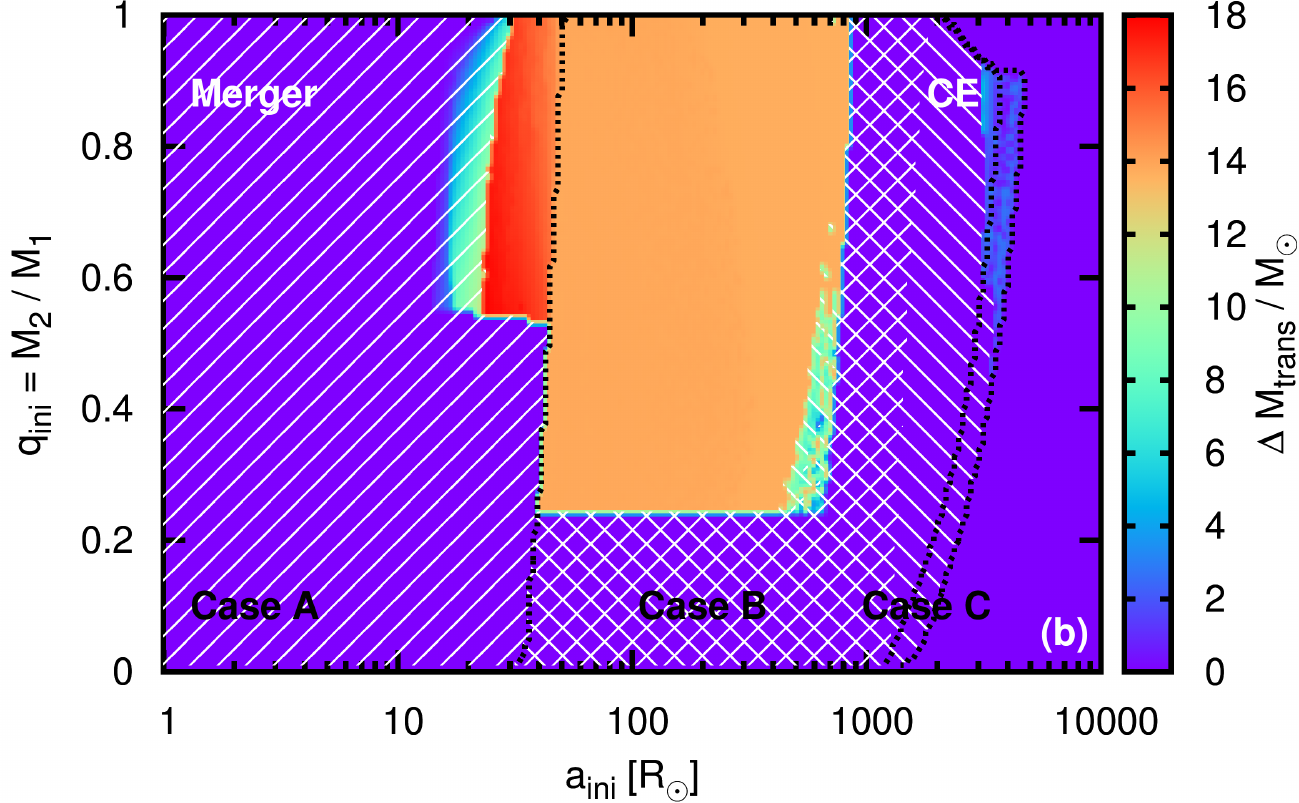}
\includegraphics[width=0.46\textwidth]{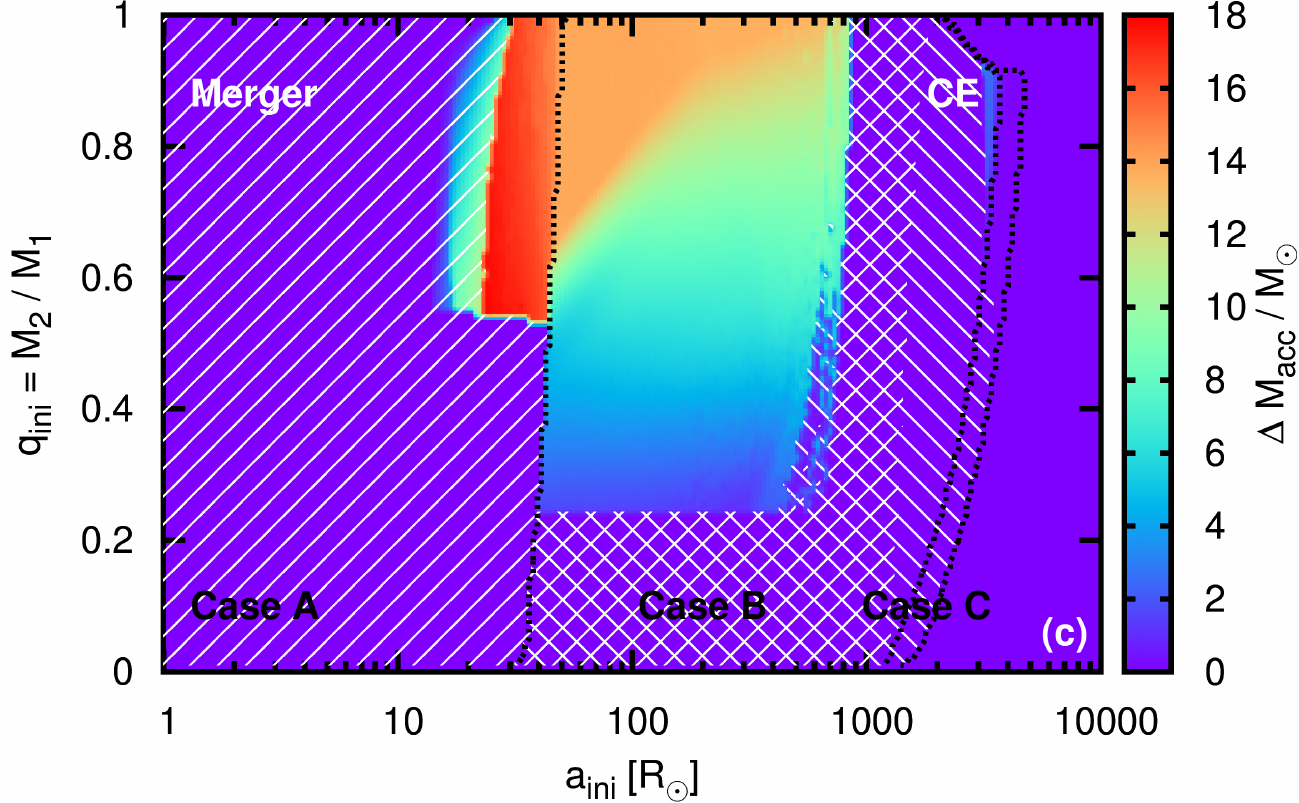}
\caption{As Fig.~\ref{fig:binary-parameter-space-2msun} but for $20\,\msun$
primary stars.}
\label{fig:binary-parameter-space-20msun}
\end{figure}

\begin{figure}
\center
$\mathbf{M_1=50\,\msun}$
\includegraphics[width=0.46\textwidth]{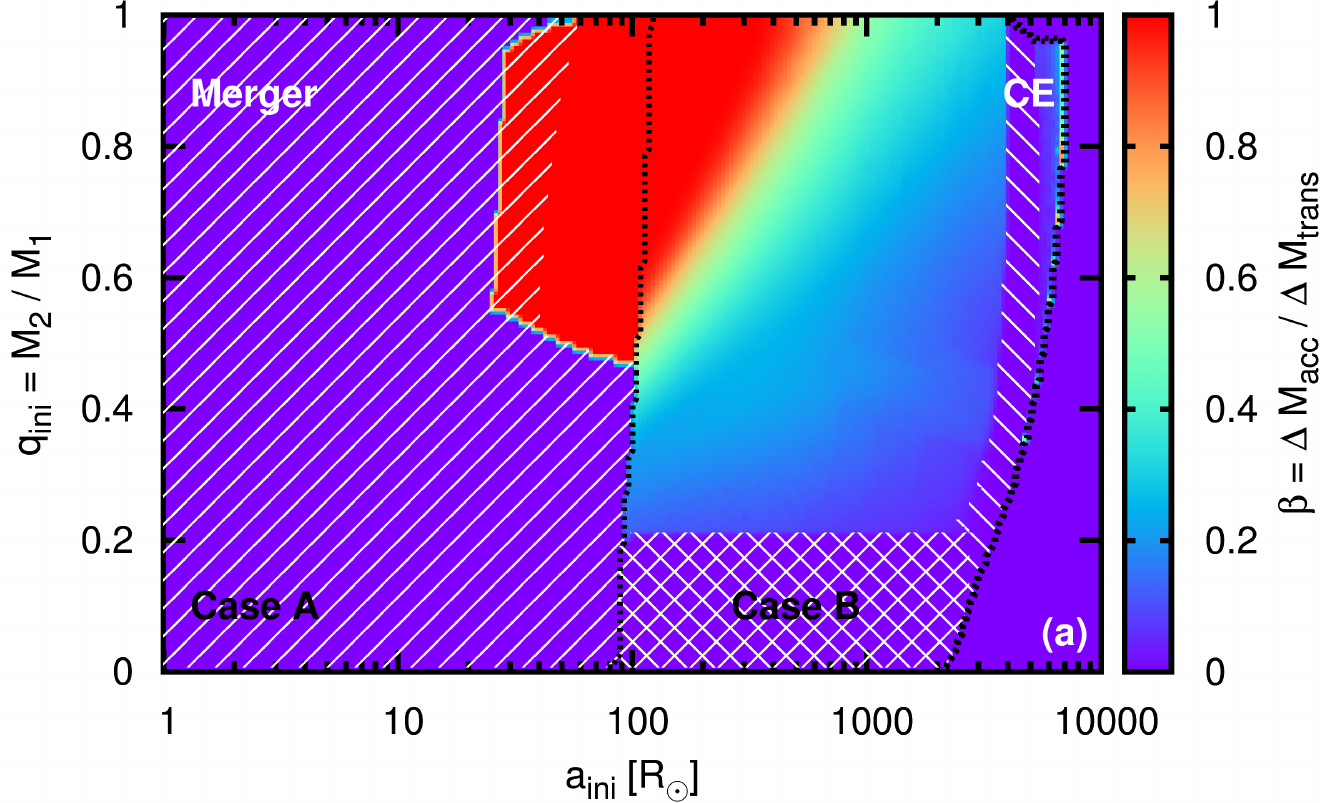}
\includegraphics[width=0.46\textwidth]{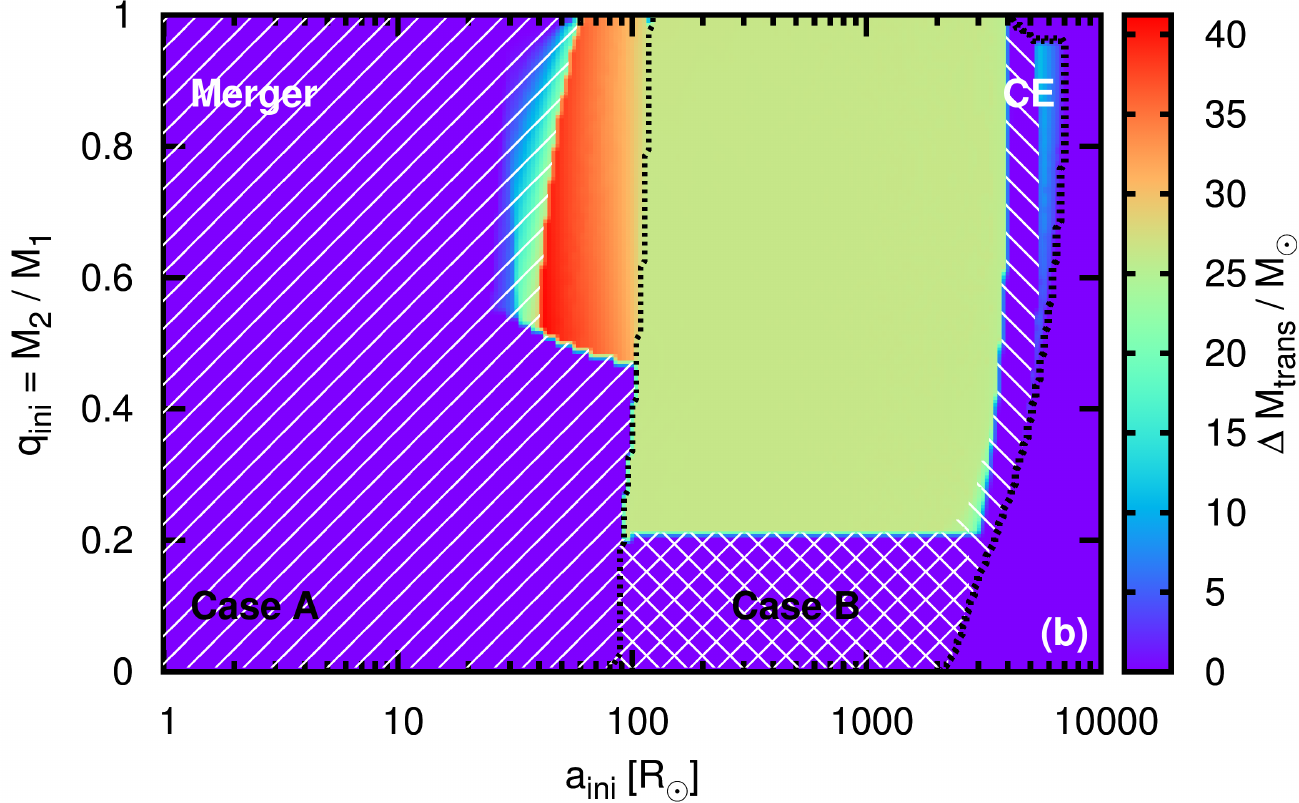}
\includegraphics[width=0.46\textwidth]{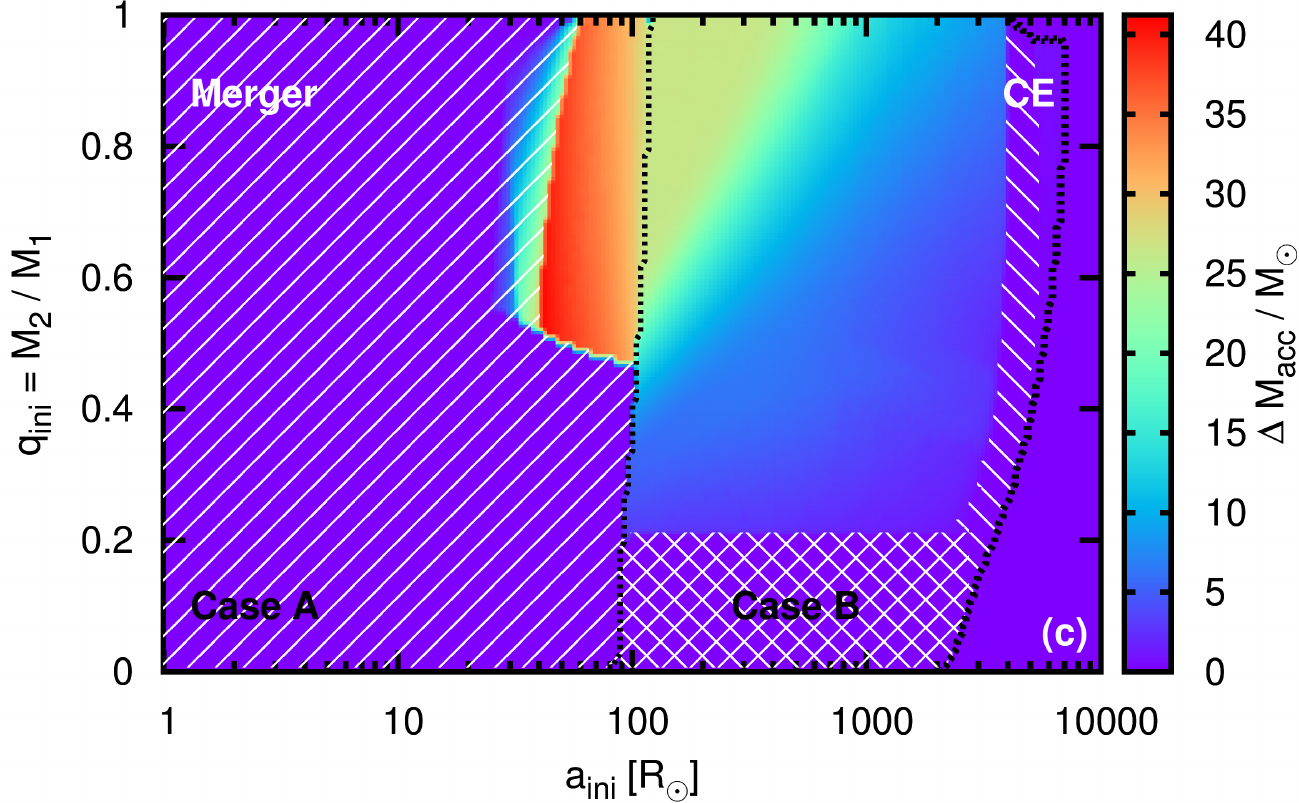}
\caption{As Fig.~\ref{fig:binary-parameter-space-2msun} but for $50\,\msun$
primary stars.}
\label{fig:binary-parameter-space-50msun}
\end{figure}

\clearpage

\begin{figure}
\center
$\mathbf{M_1=70\,\msun}$
\includegraphics[width=0.46\textwidth]{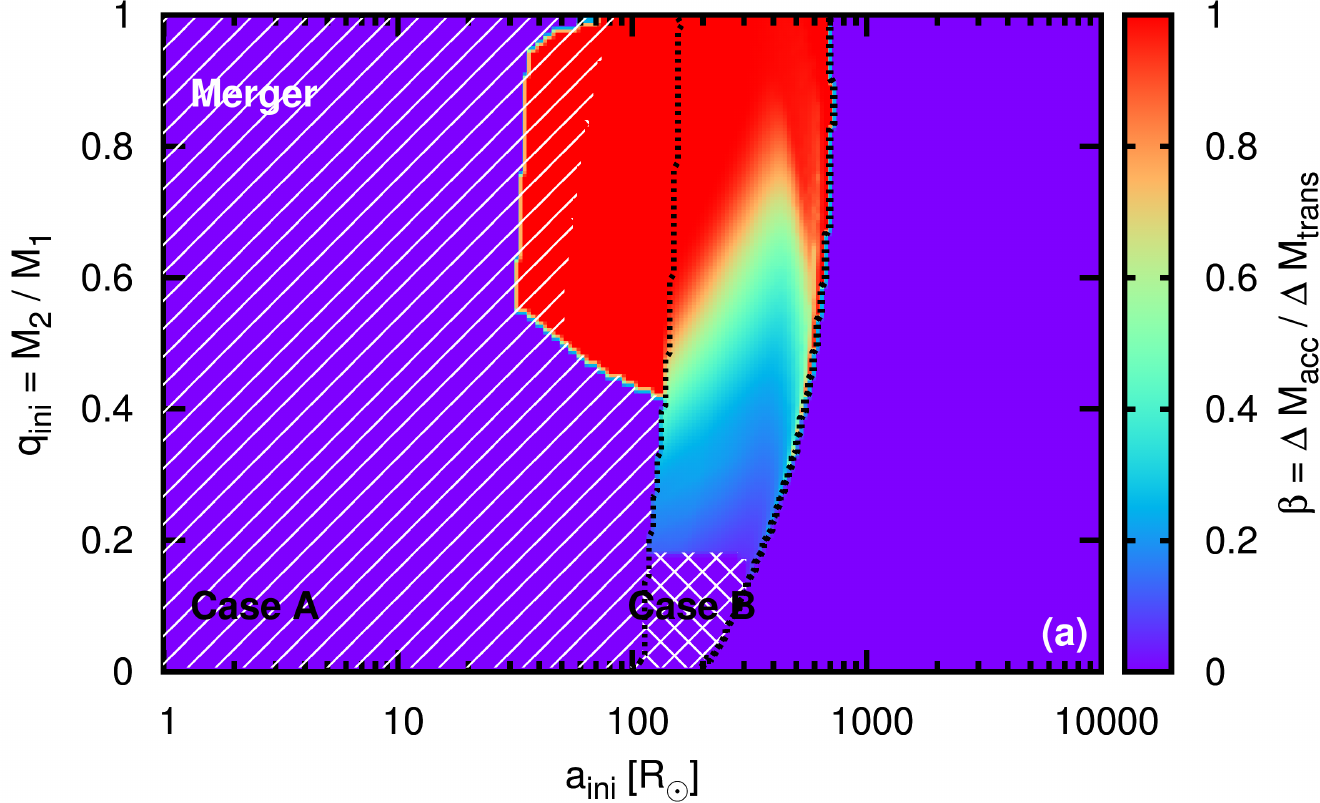}
\includegraphics[width=0.46\textwidth]{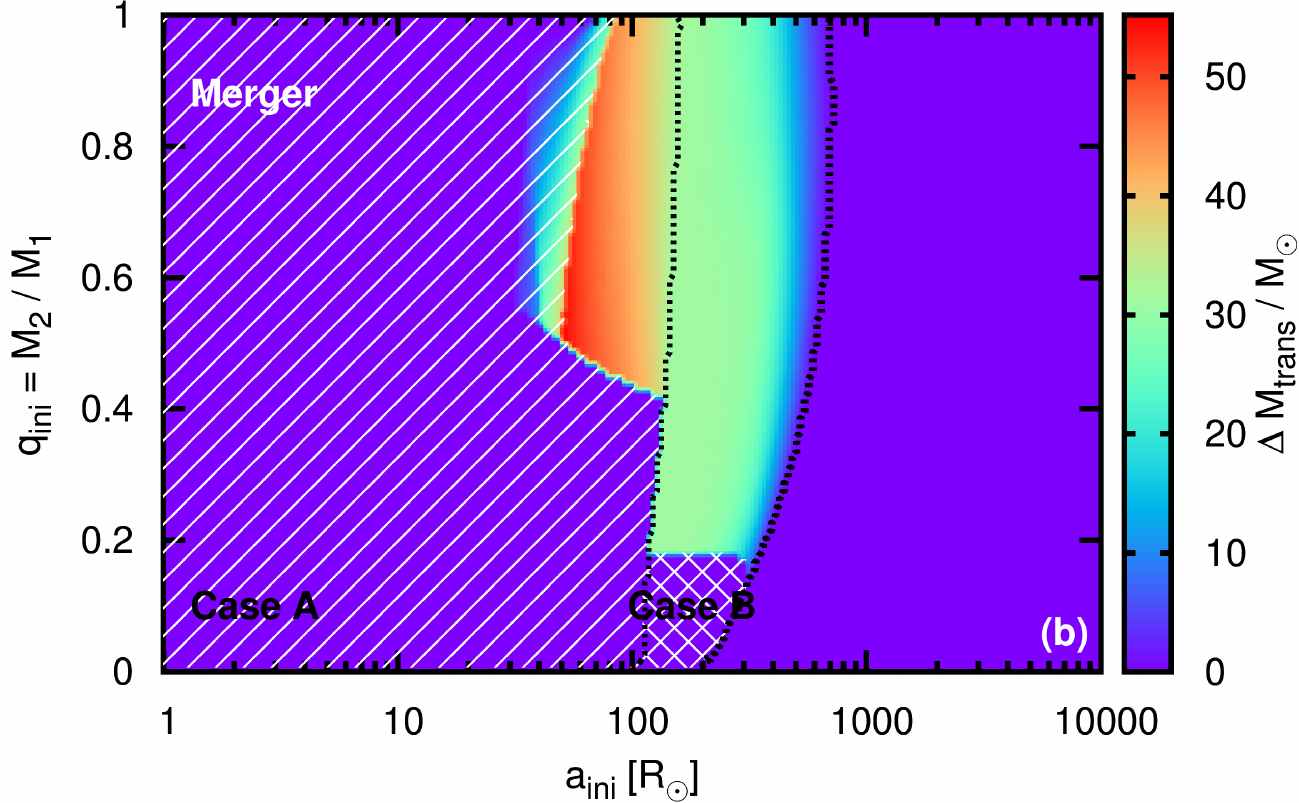}
\includegraphics[width=0.46\textwidth]{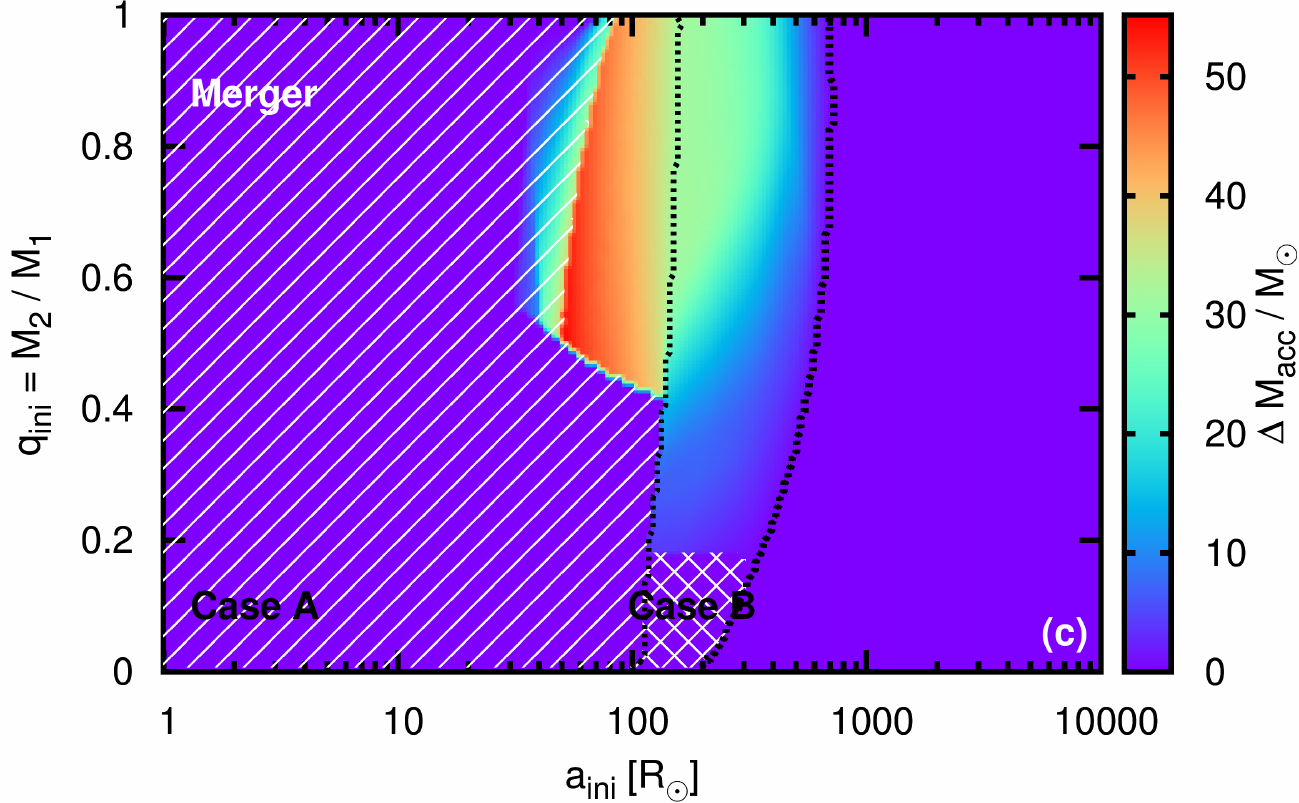}
\caption{As Fig.~\ref{fig:binary-parameter-space-2msun} but for $70\,\msun$
primary stars.}
\label{fig:binary-parameter-space-70msun}
\end{figure}

\begin{figure}
\center
$\mathbf{M_1=100\,\msun}$
\includegraphics[width=0.46\textwidth]{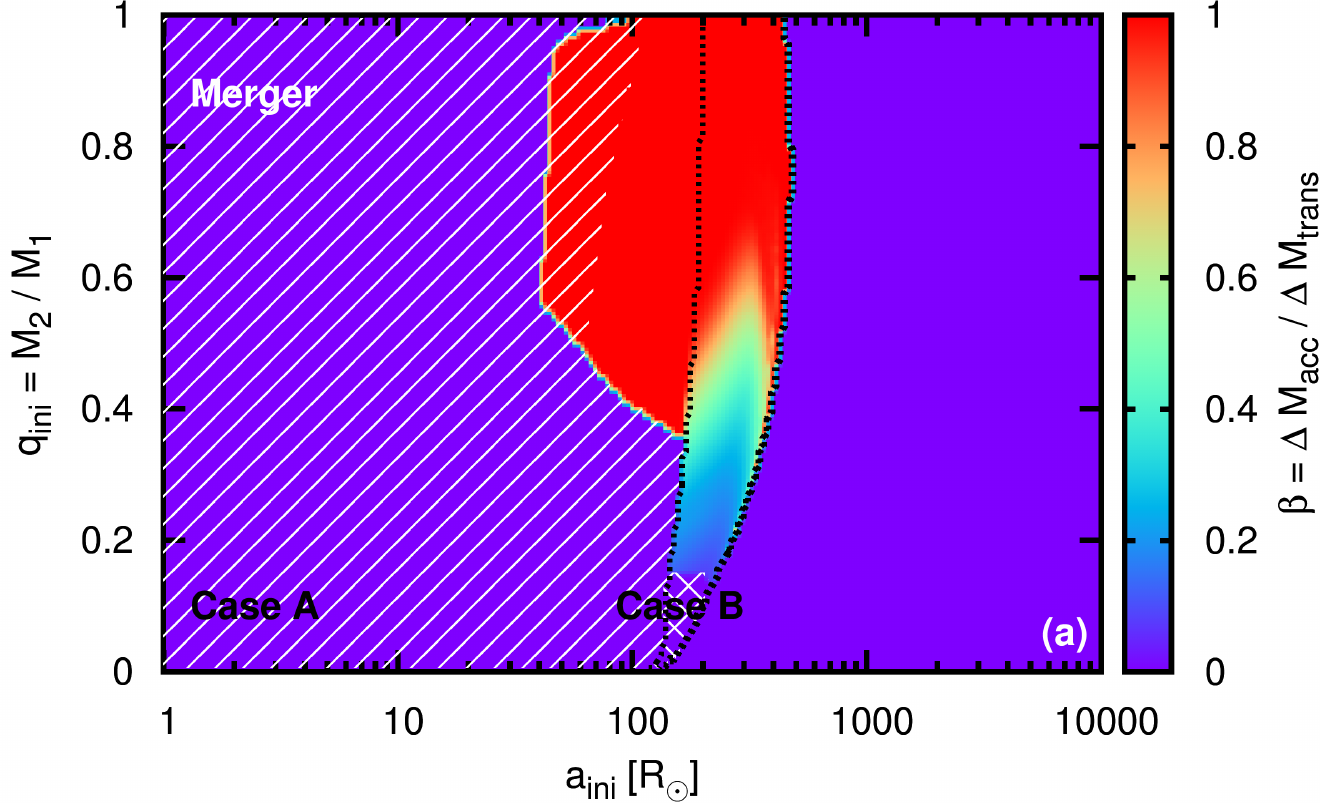}
\includegraphics[width=0.46\textwidth]{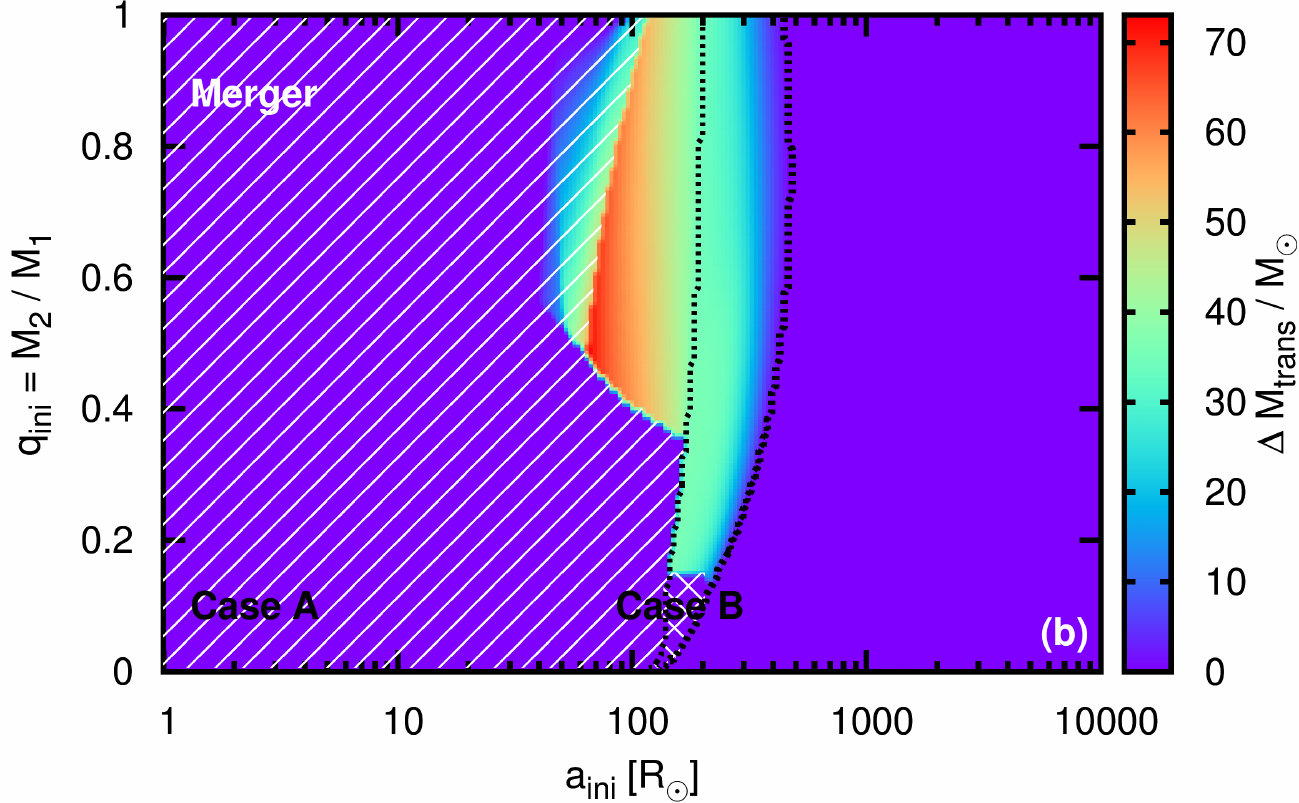}
\includegraphics[width=0.46\textwidth]{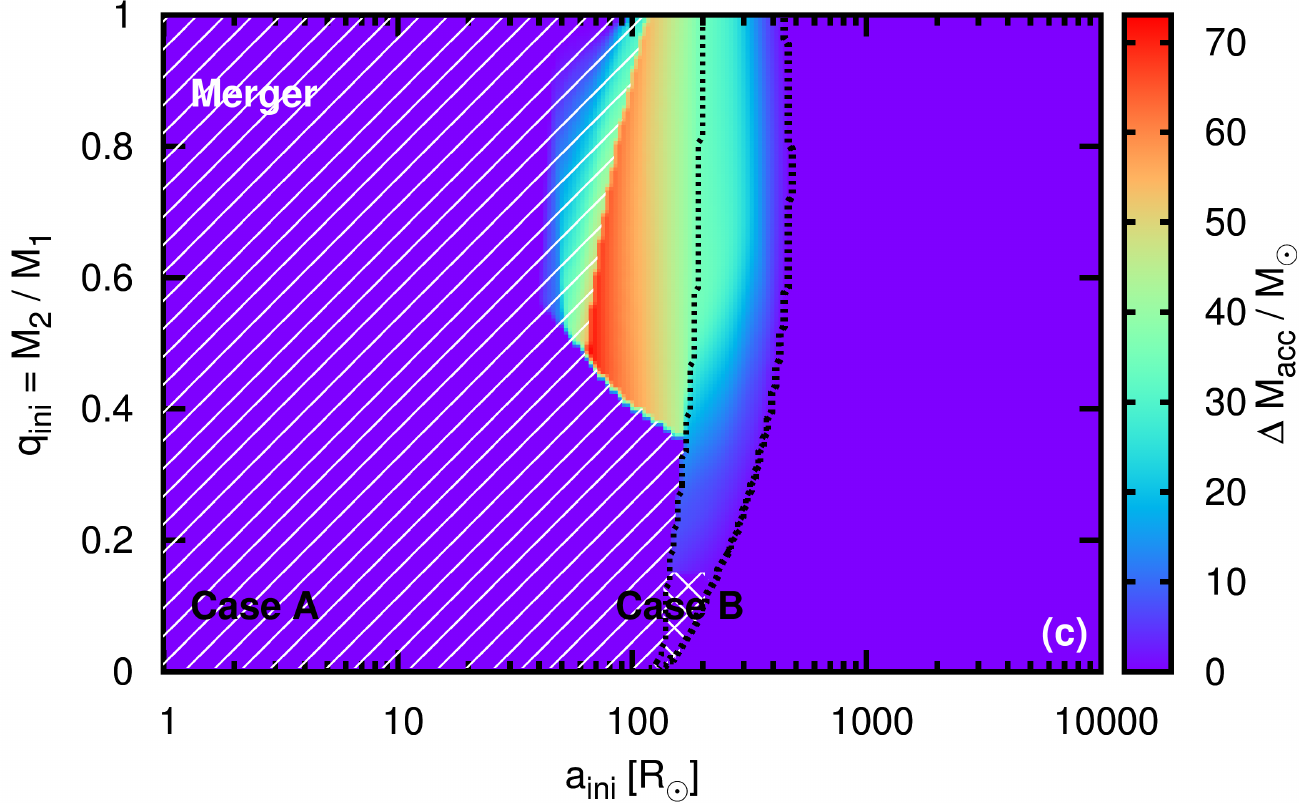}
\caption{As Fig.~\ref{fig:binary-parameter-space-2msun} but for $100\,\msun$
primary stars.}
\label{fig:binary-parameter-space-100msun}
\end{figure}

\clearpage

\section{Uncertainties in the models}\label{sec:uncertainties-models}

\subsection{Single star evolution}\label{sec:uncertainties-single-stars}

Stars more massive than $10\,\msun$ accumulate in a peak at the high
mass end of PDMFs because of stellar wind mass loss (Sec.~\ref{sec:single-star-populations}).
The magnitude of the accumulation depends on the IMF slope $\Gamma$
and on the strength of stellar winds, i.e.\ on the wind mass loss prescription
and the metallicity. Our wind mass loss prescription for MS stars
\citep{1990A&A...231..134N,1989A&A...219..205K} tends to slightly
underestimate stellar wind mass loss compared to \citet{2000A&A...362..295V,2001A&A...369..574V}.
We compare our initial and end-of-MS stellar masses to the latest
non-rotating stellar models of \citet{2011A&A...530A.115B} and \citet{2012A&A...537A.146E} and
find that the turn-off masses agree within $1$--$3\%$ for stars
less massive than about $50\,\msun$ and that we overestimate the
turn-off masses by up to $20\%$ for more massive stars. This deviation
is primarily because of the Wolf--Rayet wind mass loss rates used in
\citet{2011A&A...530A.115B} and \citet{2012A&A...537A.146E} for massive MS stars which we only
apply for post-MS stars. Increasing the wind mass loss for
stars $\geq50\,\msun$ to match the turn-off masses of the latest detailed stellar
models results into stronger peaks at the high mass end of PDMF younger
than about $4.3\,\mathrm{Myr}$. Older populations are not affected.

The wind mass loss peak flattens PDMFs (Sec.~\ref{sec:quantifying-evol-effects}).
Whether this effect needs to be taken into account when deriving the
IMF from observations depends on how mass functions are constructed
from a measured sample of stars. If, on the one hand, measured stellar
luminosities are converted to masses by means of a suitable ML
relation, the flattening of the PDMF by wind mass loss needs to
be taken into account. If, on the other hand, the
observed stars are compared individually to stellar evolution tracks
to find their initial masses, this effect does not need to be corrected
for because stellar tracks usually include wind mass loss. Alternatively,
the mass function slope can be determined omitting the high mass end
and therefore the wind mass loss peak.

\subsection{Uncertain binary physics}\label{sec:uncertain-physics}

In this section we discuss changes in the treatment of binary 
interactions and their importance for the PDMFs.
Besides mass, angular momentum is transferred during RLOF.
\citet{1981A&A...102...17P} found
that a uniformly rotating star needs to accrete only 5--10\% of
its initial mass through a disc to reach a critical (Keplerian) velocity
at the equator such that the outermost layers are no longer bound
to the star \citep[see also][]{2005A&A...435.1013P}.
This point of view is debated and there are several arguments regarding
accretion and decretion disks which might be able to dissipate angular
momentum such that stars stay always below critical rotation and thus
can accrete much more mass \citep{1976IAUS...73..237L,1991ApJ...370..604P,2011A&A...527A..84K}.
If a star is spun up to over-critical rotation it probably sheds as
much mass as is needed to rotate below critical. Mass is lost from
the system and takes away angular momentum from the star. In the case
of RLOF this means that not all transferred mass is accreted, mass
transfer is non-conservative and typically only a few percent of the
transferred mass is accreted \citep{2012ARA&A..50..107L}. Low mass transfer efficiencies
reduce the PDMF binary tail. For stellar
mergers we assume that $10\%$ of the system mass is lost and takes
away the excess angular momentum. This is likely rather
an upper limit because detailed collision simulations of massive MS stars show that
$<10\%$ is lost even in equal-mass merger events \citep{2013MNRAS.434.3497G}.

Detailed binary models find mass transfer efficiencies of about
$100\%$ for Case~A and of the order of $40$--$100\%$ for early
Case~B mass transfer \citep{2001A&A...369..939W}. Also, tides in
short period binaries and magnetic fields in combination with mass
loss (e.g., by a stellar wind), i.e.\ magnetic braking, are efficient in dissipating angular
momentum such that stars do not reach critical rotation during
RLOF. Magnetic fields seem to be even generated by strong shear because
of mass accretion (see e.g., Plaskett's star, \citealp{2013MNRAS.428.1686G})
and stellar mergers \citep{2008MNRAS.387..897T,2009MNRAS.400L..71F,2012ARA&A..50..107L}.

It is mentioned by \citet{2002MNRAS.329..897H} that the critical
mass ratio $q_{{\rm crit}}=m_{2}/m_{1}=0.25$ that determines whether
stars come into contact during thermal timescale mass transfer, e.g.,
when the primary star is a Hertzsprung gap star, is rather approximate.
Case~B mass transfer can lead to a contact phase for extreme mass ratios 
because the mass transfer timescale becomes faster than the thermal 
timescale of the accretor 
\citep[e.g.][]{1976ApJ...206..509U,1977A&A....54..539K,1977PASJ...29..249N,1994A&A...288..475P,2001A&A...369..939W}.
We implement a similar criterion with the consequence that nearly
all binary systems undergoing Case~B mass transfer go through a contact
phase in which both stars merge or eject their common envelopes. Merging
stars results in a post-MS object and hence a reduction of MS secondary
stars. Ejecting envelopes terminates any mass transfer such that the 
initial secondary stars cannot accrete any matter ---
in other words the mass transfer efficiency is $0\%$ for most
Case~B binaries. The resulting changes to our PDMFs are small,
because we already limit mass accretion to the thermal timescale of
the accretors. 

The initial distribution functions of secondary masses (or 
mass ratios) and orbital periods are also important for our results. We
assume an initially flat mass ratio distribution, i.e.\ all mass ratios
are equally probable. Small initial mass ratios often lead to contact
phases (see the discussion of the binary parameter space in Sec.~\ref{sec:binary-parameter-space}).
A distribution function of initial mass ratios which favours equal
mass binaries enhances the effect of mass transfer and reduces the
effect of MS coalescence. Also the Case~B contribution, which is
the dominant mass transfer channel at later times
(Fig.~\ref{fig:mt-cases}), is increased because the Case~B mass transfer
efficiency is higher for larger mass ratios. If the initial orbital
separation distribution favours small initial orbital separations
compared to {\"O}pik's law, Case~A and~B mass transfer (including
stellar mergers) occur more frequently which increases the magnitude
of the tail of the PDMFs. 

Different IMF slopes $\Gamma$
enhance or reduce the importance of binary physics. Binary evolution
usually increases the mass of stars: let this increase be given by
a multiplicative factor $\alpha$, i.e.\ the increased mass is $\alpha M$.
If we increase all stellar masses $M$ by the same factor $\alpha$,
we enhance the mass function at the increased mass $\alpha M$ by the factor,
\begin{equation}
\frac{\psi(\ln M)}{\psi(\ln\alpha M)}=\alpha^{-\Gamma}.
\end{equation}
The steeper the slope of the IMF, i.e.\ the more negative $\Gamma$,
the bigger the effect of any mass gain. Also the flattening of the PDMF because
of unresolved binaries is more important for steeper IMF slopes and
less important for flatter IMF slopes (Sec.~\ref{sec:quantifying-evol-effects}).
This trend is opposite if mass is lost e.g., by stellar winds: then, 
a flatter IMF enhances and a steeper one reduces the accumulation of stars.

\section{Unresolved binary stars}\label{sec:unresolved-binaries}

Binaries in photometric studies are usually unresolved which leads to an overestimation of
their masses (see Eq.~\ref{eq:mobs}). A mass function, even at zero
age, thus looks different compared to the IMF. We find changes in
the slope of the mass function of the order of $\Delta\beta\approx0.1$
toward flattened mass functions for zero age populations.
The flattening because of unresolved binaries is more important for larger masses
because of the mass dependence of the power law index of the ML
relation (Sec.~\ref{sec:quantifying-evol-effects}).
The problem of unresolved binaries is well known
\citep{1991A&A...250..324S,1993MNRAS.262..545K,2008ApJ...677.1278M,2009MNRAS.393..663W}. 
To compare with their results it is important to know how they distributed
stars in binaries \citep[see e.g., the review by][]{2010ARA&A..48..339B}.

\citet{1991A&A...250..324S}, for example, do not use a flat mass ratio
distribution, but they draw both binary components randomly from one
IMF. For an initial IMF slope of $\Gamma=-1.5$ (our initial IMF slope
is Salpeter, i.e.\ $\Gamma=-1.35$) and a binary fraction of 50\% they
derive a PDMF slope of $\beta=-1.16$; for a binary fraction of 100\%
they arrive at $\beta=-1.10$. These changes occur according to their
analysis in a mass range of $2$--$14\,\msun$ and are larger than
we find. The major difference between their calculation and ours is
the assumed distribution function of stars in binaries (mass ratios
and IMF slope).

\citet{2008ApJ...677.1278M} not only investigates
unresolved binary stars but also higher order multiples and chance
superpositions in dense clusters. Taking only the effect of binary
stars into account they conclude ``that for most cases the existence
of unresolved binaries has only a small effect on the massive-star
IMF slope'' of the order of $|\Delta\beta|\approx0.2$.
They use the same mass ratio distribution as in our analysis but slightly
different IMF slopes.

\citet{2009MNRAS.393..663W} investigate the effect of unresolved
binaries and higher order multiples for different pairing methods.
None of their pairing methods correspond to our flat mass ratio distribution.
A steepening of the observed PDMF by $\Delta\beta=0.1$ is reported.
Their PDMF is steeper because they take a certain number of stars
given by their IMF and pair them randomly into binaries (their RP
method). The probability of finding a massive star grouped with another
massive star is therefore less than to find it grouped with a lower
mass companion, hence estimated system masses do not change significantly
for massive stars. At lower masses it is opposite: relatively more
binaries with stars of similar masses are found and the companions
lead to higher system mass estimates. In total this leads to a steepening
of the PDMF.

All in all it seems that unresolved binaries have limited
influence on the PDMF of zero age populations. Pairing stars which
are randomly sampled from an IMF steepens the mass function,
whereas a flat mass ratio distribution flattens the mass function
as is shown in our analysis.
Observations of massive, i.e.\ O-type binaries favour a flat mass ratio 
distribution \citep{2012Sci...337..444S}, so the high mass 
end of PDMFs of young stellar populations with O-type
stars is expected to be flattened if binaries are unresolved.

\section{Stochastic sampling}\label{sec:stochastic-sampling}
The statement that the most massive stars are likely blue stragglers 
has to be treated with caution, because it
is only valid in ``rich'' (i.e.\ massive) clusters where there are enough
stars to sample the binary parameter space well --- this problem is known 
as stochastic sampling. 
Binary evolution e.g., re-populates the $5\,\mathrm{Myr}$ PDMF on
average by more than $40\%$ of the IMF above $40\,\msun$ ($\log m\approx1.6$).
If there is initially only one star in this mass range, we expect about
$0.4$ stars to be re-populated by binary evolution --- if there are
initially 10 stars (if the cluster is a factor of 10 more massive),
we expect to find about four binary products. So if the cluster is not
massive enough, there might be no blue straggler star at a certain
age. It will also take some time until binary star evolution produces
the first blue straggler stars. This again depends on the cluster
richness and on how binaries are distributed: the shorter the initial
period, the earlier the binary interaction. When the initially most
massive stars in a cluster evolve toward the end of core hydrogen
burning we know that blue stragglers can be produced by MS mergers
and Case~A mass transfer. So the whole contribution of MS mergers 
and Case~A mass transfer products of binaries with initial $100\,\msun$ primary stars
to the blue straggler star population is expected to be present after 
about $3\,\mathrm{Myr}$. Blue stragglers can be even present at younger ages
--- this is just a matter of stochastic sampling 
which is investigated in more detail in \citet{2014ApJ...780..117S}.

\clearpage

\bibliographystyle{apj}

\end{document}